\documentclass[%
reprint,
superscriptaddress,
 amsmath,amssymb,
 aps,
]{revtex4-2}

\usepackage{graphicx}
\usepackage{dcolumn}
\usepackage{bm}
\usepackage{xcolor}

\usepackage{booktabs}

\usepackage[utf8]{inputenc}
\usepackage[T1]{fontenc}
\usepackage{mathptmx}
\usepackage{etoolbox}

\usepackage{enumitem}

\usepackage{cleveref}
\crefname{figure}{Fig.~}{Figs.~}
\Crefname{figure}{Figure~S}{Figures~S}

\begin{document}

\preprint{APS/123-QED}

\title{Differentiable Cardiac Electrophysiology Simulations for Dynamical State and Parameter Estimation}

\author{Adarsh Pashikanti}
\altaffiliation{These authors contributed equally to this work.}
\affiliation{Cardiovascular Research Institute, University of California, San Francisco, San Francisco, USA}
\affiliation{Department of Physics, University of California, Berkeley, Berkeley, USA}
\author{Shrey Chowdhary}
\altaffiliation{These authors contributed equally to this work.}
\affiliation{Cardiovascular Research Institute, University of California, San Francisco, San Francisco, USA}
\author{Alex Ho}
\altaffiliation{These authors contributed equally to this work.}
\affiliation{Cardiovascular Research Institute, University of California, San Francisco, San Francisco, USA}
\author{Weizhen Li}
\affiliation{Department of Biomedical Engineering, The George Washington University, Washington D.C., USA}
\author{Emilia Entcheva}
\affiliation{Department of Biomedical Engineering, The George Washington University, Washington D.C., USA}
\author{Jan Christoph}
 \email{jan.christoph@ucsf.edu}
 \homepage{http://cardiacvision.ucsf.edu}
 \affiliation{Cardiovascular Research Institute, University of California, San Francisco, San Francisco, USA}
 \affiliation{Division of Cardiology, University of California, San Francisco, San Francisco, USA}
 \affiliation{Department of Biomedical Engineering \& Therapeutic Sciences, University of California, San Francisco, USA}

\newcommand{\stimes}{{\times}}

\begin{abstract}

The heart's contractions are triggered by action potential waves,
which propagate rapidly through the cardiac muscle and exhibit diverse spatio-temporal dynamics during different heart rhythms. 
The dynamics are governed by biophysical laws arising in reaction-diffusion systems and modeled with partial differential equations (PDEs) in cardiac electrophysiology simulations.
However, fitting such models to measurement data to develop digital twins or patient-specific computer models is challenging.
Here, we introduce differentiable cardiac electrophysiology simulations that can be fitted automatically to spatio-temporal measurement data of action potential waves in cardiac tissue. 
By comparing the simulated dynamics with the observation data, we define a loss function that is minimized via gradient-based optimization. 
Backpropagating the loss gradient through the differentiable PDE solver with respect to the model parameters and initial condition enables us to learn the parameters and recover the full dynamics, even with sparse, noisy, or partial observations.
Implemented using both the finite-difference and smoothed particle hydrodynamics (SPH) methods, our simulation framework can be applied to pixel-, voxel-, or point-based data, such as 2D or 3D slabs, or arbitrary anatomical shapes, such as the heart's ventricles.
For instance, we recover spiral wave dynamics from observations across a 2D grid, scroll waves from observations across a 3D bulk's surface, or early activation sites inside a 3D bi-ventricular geometry from epicardial observations.
We also fit one phenomenological model to another, and to imaging data of a voltage spiral wave in a cardiac monolayer cell culture.
With experimental data, we employed a perceptual loss based on the Video Joint-Embedding Predictive Architecture, which enables fitting to noisy imaging data, and a generative diffusion model to estimate initial conditions and constrain solutions.
Our results demonstrate that differentiable physics simulations are a powerful methodology for fitting cardiac electrophysiology models to data.
Utilizing such techniques could improve the diagnosis of rhythm abnormalities in patients and facilitate the development of personalized models or digital twins of the heart.

\begin{description}
\item[Keywords]
Digital Twin, Cardiac Electrophysiology, Differentiable Physics Simulations, Machine Learning
\end{description}

\end{abstract}

\maketitle


\section{Introduction}
\label{sec:introduction}

Computer simulations of the heart are increasingly used for disease modeling, arrhythmic risk stratification, drug or medical device evaluation, treatment planning, and procedural guidance. 
These developments have fueled growing interest in the concept of a digital twin of the heart: a patient-specific computational model that integrates anatomy, physiology, and measurement data to reproduce and predict cardiac function.
However, personalizing such models remains a major challenge because the heart's dynamical state and physiological parameters are difficult to obtain.
From a dynamical systems perspective, not only can some dynamical variables, such as the transmembrane voltage, only be observed indirectly or incompletely, but other hidden variables cannot be observed at all.
While it is possible to measure the heart's anatomy using magnetic resonance imaging (MRI) or ultrasound, functional measurements of the heart's electrophysiology rely on sparse, indirect, or incomplete data.
For instance, the 12-lead electrocardiogram (ECG) is a reflection of the heart's electrical state on the body surface. It is an indirect, low-resolution measurement.
Direct approaches, such as catheter-based electrode mapping~\cite{Gepstein1997} or voltage-sensitive optical mapping~\cite{Chowdhary2026}, provide higher resolutions, however, only on the heart's surface, not from within the heart muscle. 
Further, only optical mapping provides measurement data of transmembrane voltage, and is a truly panoramic high-resolution measurement.
Catheter mapping is a sparse, point-by-point measurement that yields electrograms, rather than a physiological variable.
Accordingly, the complete set of variables of a biophysical model describing the heart's dynamics must be inferred using state and parameter estimation techniques.
Together, this challenge is particularly pronounced in cardiac electrophysiology simulations, where nonlinear reaction-diffusion dynamics govern the propagation of action potential waves through a complex tissue anatomy and produce highly intricate spatio-temporal dynamics during arrhythmias~\cite{Davidenko1992, Pertsov1993, Alonso2016, Rappel2022, Christoph2018}, yet, the full dynamics must be inferred from limited observation data.

A wide range of techniques has been developed to address state and parameter estimation in cardiac systems and related nonlinear dynamical models,
ranging from data assimilation~\cite{Berg2011, Hoffman2016, LaVigne2017, Hoffman2020, Marcotte2021, Marcotte2023, Mendez2024}, Kalman filters~\cite{LaVigne2017, Marcotte2023}, particle swarm optimization~\cite{Seemann2009, Chen2012, Loewe2015, Rheaume2023, Cairns2025}, gradient-based optimization~\cite{Loewe2015}, least squares~\cite{Dokos2004}, stochastic optimization approaches such as simulated annealing~\cite{Lombardo2016} or genetic algorithms~\cite{Syed2005, Bot2012, Cairns2017}, physics-informed neural networks \cite{HerreroMartin2022}, and Bayesian approaches~\cite{Pezzutto2022}, among others.
As much as the techniques vary in their mathematical foundations, from an application perspective, their respective strengths and limitations make them better suited for certain applications than others.
Particle swarm optimization and genetic algorithms were primarily used to fit action potentials in single cells, while data assimilation was primarily used to reproduce wave dynamics in tissues.
Bayesian and iterative approaches with reaction-Eikonal models~\cite{Neic2017} were used with 12-lead ECG and 3D MRI data for personalizing bi-ventricular electrophysiology~\cite{Pezzutto2020, Gillette2021, Camps2025} and electromechanics~\cite{Doste2026} simulations of sinus rhythm.
Together, these techniques have enabled parameter inference and dynamical state reconstructions, including reconstructions of activation sequences, repolarization times, or wave patterns at the tissue level. 
However, many existing techniques struggle to fully exploit spatio-temporal data, particularly when fitting models to irregular, arrhythmic wave dynamics.

\begin{figure}[htb]
  \centering
  \includegraphics[clip, trim=0.0cm 0.0cm 0.0cm 0.0cm, width=0.48\textwidth]{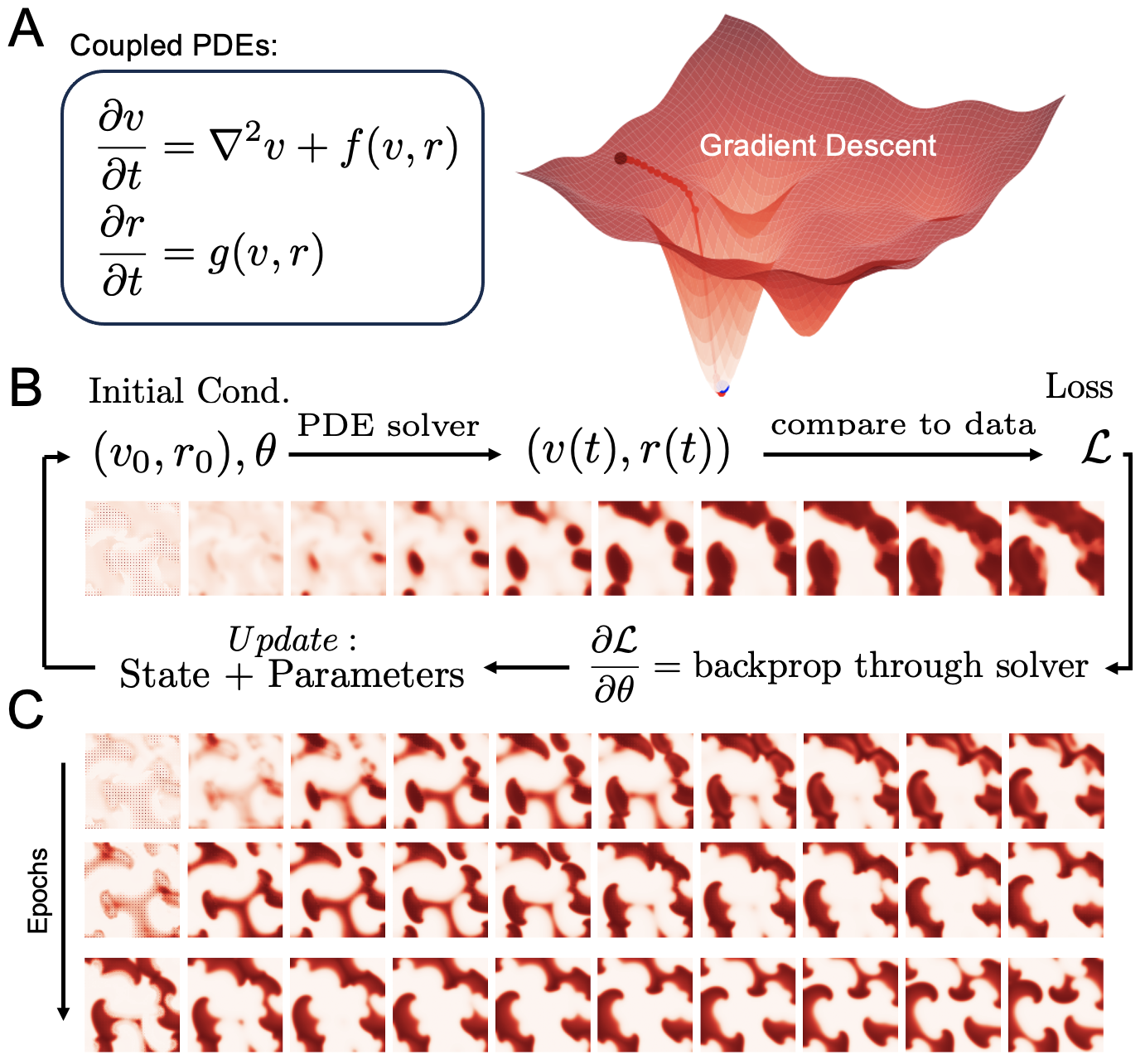}
  \caption{
Gradient-based fitting to spatio-temporal data using differentiable cardiac electrophysiology simulations.
\textbf{A} Coupled partial differential equations (PDEs) as a biophysical model for excitable wave dynamics in a reaction-diffusion system. 
Gradient descent finds the optimal simulation (initial state and matching parameters) that minimizes the loss $\mathcal{L}$ between data and model (illustration).
\textbf{B} Initial state $(v_0,r_0)$ and trajectory $(v(t), r(t))$ solved with differentiable solver. Loss gradient backpropagated through solver to update the model's state and parameters.
\textbf{C} Iterative refinement of simulation and minimization of loss over optimization steps (epochs).
 }
  \label{fig:1}
\end{figure}

Recent advances in automatic differentiation, differentiable programming, and machine learning have given rise to the emerging field of differentiable physics, which offers a unifying framework for integrating physical models with gradient-based optimization~\cite{Hu2020, Thuerey2021}. 
By constructing simulation pipelines that are fully differentiable, it becomes possible to propagate gradients through numerical solvers and directly optimize simulation parameters using backpropagation, see Fig.~\ref{fig:1}. 
This paradigm has been enabled by modern automatic differentiation tools such as JAX~\cite{jax} and PyTorch~\cite{pytorch}, and has already shown promise in various areas, such as fluid and molecular dynamics~\cite{Winchenbach2026, jax-md}.
Differentiable physics solvers were introduced for simulating cardiac mechanics~\cite{Thomas2025}, parameter estimation of spiral wave dynamics in excitable media~\cite{Lettermann2024}, and, in conjunction with neural networks, for fitting cardiac action potentials~\cite{Kashtanova2023}.
In cardiac electrophysiology, differentiable simulations offer a powerful alternative to traditional fitting procedures, enabling integration of biophysical models, irregular geometries, and diverse data modalities.

In this work, we build on these developments and introduce differentiable cardiac electrophysiology reaction-diffusion simulations that enable efficient, automated fitting to spatio-temporal data, including, for instance, simulated data in slab- and heart-shaped tissue models and action potential spiral waves in a petri dish.
We used several complementary techniques in conjunction with the gradient-based optimization that facilitate learning, such as iterative multi-horizon fitting.
In particular, with experimental data, we employed a perceptual loss based on the Video Joint-Embedding Predictive Architecture (V-JEPA) that overcomes limitations associated with a pixel-based loss and enables fitting to noisy, artifact-afflicted data, and a denoising diffusion probabilistic model (DDPM) to estimate initial conditions and further constrain the solution space towards a particular biophysical model.
We assess the efficacy of our approach under various conditions with partial, sparse, or noisy observations.
For instance, we fit a 3D model to epicardial observations of a simulated focal wave pattern originating in the septum of a bi-ventricular geometry, recover simulated spiral wave dynamics from sparse observations across a grid, scroll waves inside a bulk from surface observations, and fit a 2D model to noisy imaging data of an action potential spiral wave in a monolayer cell culture.

\begin{figure}[htb]
  \centering
  \includegraphics[clip, trim=0.0cm 0.0cm 0.0cm 0.0cm, width=0.48\textwidth]{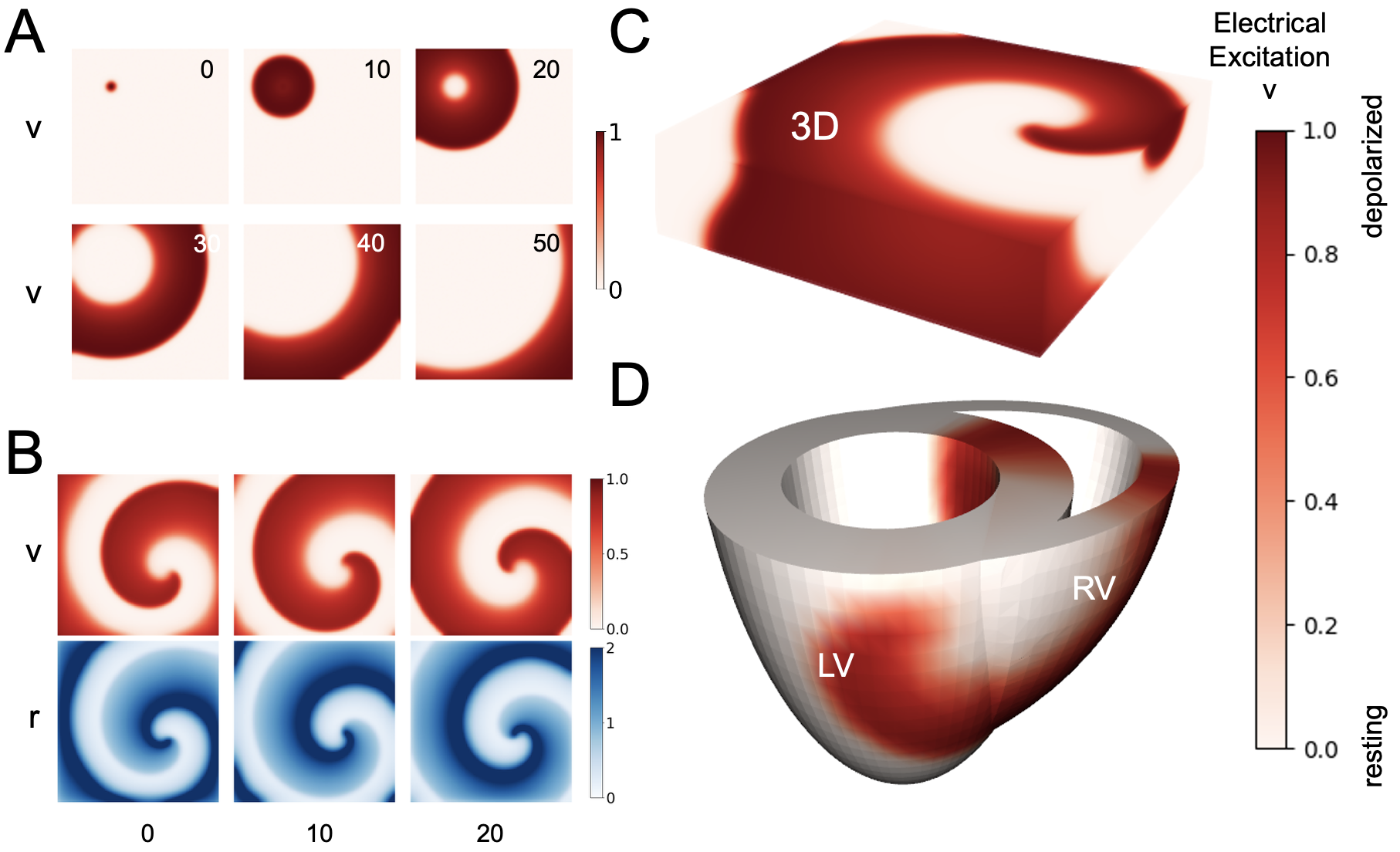}
  \caption{
Simulations of electrical action potential waves (red: depolarized tissue, white/gray: resting tissue) in 2D and 3D tissues. 
\textbf{A} 2D focal wave originating from a point source (voltage variable $v$ only, timesteps 0-50). 
\textbf{B} 2D spiral wave (red: voltage variable $v$, blue: refractory variable $r$, 1 rotation occurs within approx. 20-50 timesteps).
\textbf{C} 3D scroll wave dynamics inside slab/bulk.
\textbf{D} 3D scroll wave in idealized bi-ventricular geometry (LV: left ventricle, RV: right ventricle).
Simulations were performed using the finite differences (A-C) and smoothed particle hydrodynamics (SPH) methods (D) with several different models (Aliev-Panfilov, Mitchell-Schaeffer, and others).
All dynamics were subsequently learned using differentiable simulations.
 }
  \label{fig:methods:simulations}
\end{figure}

\section{Methods}
\label{sec:methods}
We employed the finite differences and smoothed particle hydrodynamics (SPH) methods to simulate nonlinear waves of electrical excitation in two-dimensional (2D) sheets, three-dimensional (3D) slab/bulk tissues, and 3D idealized bi-ventricular-shaped heart models, see Fig.~\ref{fig:methods:simulations}.
Further, we imaged voltage spiral waves in cardiac monolayer cell cultures, see Fig.~\ref{fig:results:cellculture}.
We then fitted differentiable electrophysiology simulations to the spatio-temporal data, starting with rough initial guesses of the model parameters and dynamical states.
In sections \ref{sec:results:2Dfocus}-\ref{sec:results:sparse}, we assumed that the model equations are known, in section \ref{sec:results:crossmodel}, we learned the dynamics with different model equations than those used for data generation, and with the experimental data in section \ref{sec:results:experiments}, we did not know and had to assume the model equations.

\subsection{Biophysical Modeling of Action Potential Waves}
\label{sec:methods:models}
We simulated electrical action potential waves in cardiac muscle tissue using two distinct two-variable phenomenological reaction-diffusion models~ \cite{AlievPanfilov1996,MitchellSchaeffer2003}.
We used the Aliev-Panfilov (AP) model~\cite{AlievPanfilov1996} with 2 state variables and 6 parameters, and the Mitchell-Schaeffer (MS) model~\cite{MitchellSchaeffer2003} with 2 state variables and 5 parameters, 
to produce spatio-temporal action potential wave data. 
Generally, the models comprise coupled partial differential equations (PDEs) describing the excitable and refractory kinetics, along with a diffusion term that supports the propagation of nonlinear waves of electrical excitation through the tissue.
While more detailed ionic models describe richer or finer features of action potential dynamics, with some of their variables and parameters relating to sodium, potassium, and calcium ion channels, simpler phenomenological models (AP, MS) capture the essence of the dynamics.
We primarily used the AP model~\cite{AlievPanfilov1996}: 
\begin{eqnarray} 
\label{eq:modelu}
\frac{\partial v}{\partial t} & = & \nabla \cdot (D \nabla v) - k v (v-a) (v-1) - v r \\
\label{eq:modelr}
\frac{\partial r}{\partial t} & = & \left(\varepsilon_0 + \frac{\mu_1 r}{u+\mu_2}\right) (k v(a+1-v)-r)
\end{eqnarray}
Here, the 6 parameters $\{ D, a, k, \epsilon_0, \mu_1, \mu_2 \}$ determine the properties of the waves (e.g. wave propagation speed, wavelength, excitation threshold, distance between waves, number of waves, etc.).
The AP parameter values used in the different parts of the study are shown in table \ref{tab:parameters}.
We performed simulations in 2D sheets, 3D bulks / slabs, and 3D heart-shaped geometries.
The simulations in the regular geometries, see Fig.~\ref{fig:methods:simulations}A-C), were integrated using the finite differences method.
The simulations in the heart-shaped, bi-ventricular geometries, see Fig.~\ref{fig:methods:simulations}D), were performed using the smoothed particle hydrodynamics (SPH) method using the SPHinXsys library~\cite{Zhang2021,Zhang2021b}.
The 2D simulations were performed in simulation domains with a size of $128 \times 128$ pixels.
The 3D bulk / slab simulations were performed in a simulation domain with a size of $64 \times 64 \times 16$ voxels.

\subsection{Differentiable Simulations of Cardiac Electrophysiology}
\label{sec:methods:diffsim}
We developed differentiable cardiac electrophysiology simulations using the JAX library~\cite{jax} and related packages from the JAX ecosystem.
JAX is a Python library for accelerator-oriented array computation and program transformation, designed for high-performance numerical computing and large-scale machine learning. 
In particular, JAX provides the environment for gradient-descent-based optimization and differentiable physics simulations as it can automatically differentiate native Python and Numerical Python (NumPy) functions.
We developed two versions of differentiable cardiac electrophysiology simulations: one based on the finite differences method and the other on the smoothed particle hydrodynamics (SPH) method, see sections \ref{sec:methods:diffsim:finitedifferences} and \ref{sec:methods:diffsim:SPH}, which allows us to perform simulations on regular and irregular geometries.

\subsubsection{Finite Differences Method}
\label{sec:methods:diffsim:finitedifferences}
The finite differences simulations were implemented as a fully differentiable pipeline utilizing the JAX ecosystem. 
The spatial domain was discretized using finite difference stencils (e.g., a 9-point Laplacian) implemented natively in JAX.
This allowed the operations to be Just-In-Time (JIT) compiled via the Accelerated Linear Algebra (XLA) compiler, allowing highly efficient, parallelized execution on graphics processing units (GPUs).
For temporal integration, we utilized Diffrax~\cite{kidger2021}, a JAX-based library for differentiable differential equation solvers. 
The coupled system of partial differential equations (PDEs) was integrated using the \texttt{Tsit5} solver, an explicit Runge-Kutta method of order 5(4). 
To maintain numerical stability, we employed a Proportional-Integral-Derivative (PID) controller for adaptive step sizing with both relative and absolute tolerances set to $10^{-4}$. 
Crucially, exact gradients of the observation loss with respect to the biophysical parameters and initial dynamical state were computed using reverse-mode automatic differentiation. 
To mitigate the massive memory overhead typical of computing gradients via the adjoint method in long spatio-temporal sequences~\cite{Lettermann2024}, we applied the \texttt{RecursiveCheckpointAdjoint()} method. 
Doing so keeps the gradient computation exact while maintaining a computationally feasible memory footprint that scales as $\mathcal{O}(N \log N)$ in the number of solver steps $N$.
Finally, the reaction-diffusion parameters and spatial fields were parameterized using the JAX-based Equinox library~\cite{equinox}. 
By encapsulating the model parameters and spatial discretizations as PyTrees (\texttt{eqx.Module}), Equinox provides an object-oriented architecture that integrates with JAX's functional programming constraints. 
The parameter recovery and state estimation were subsequently driven by the Optax library. 
We utilized the Adam optimizer~\cite{Adam} coupled with either a constant learning rate or an exponential decay learning rate schedule across our multi-horizon training blocks described in section \ref{sec:methods:learn:multi-horizon}.

\subsubsection{Smoothed Particle Hydrodynamics (SPH) Method}
\label{sec:methods:diffsim:SPH}
\begin{figure}[htb]
  \centering
  \includegraphics[clip, trim=0.0cm 0.0cm 0.0cm 0.0cm, width=0.48\textwidth]{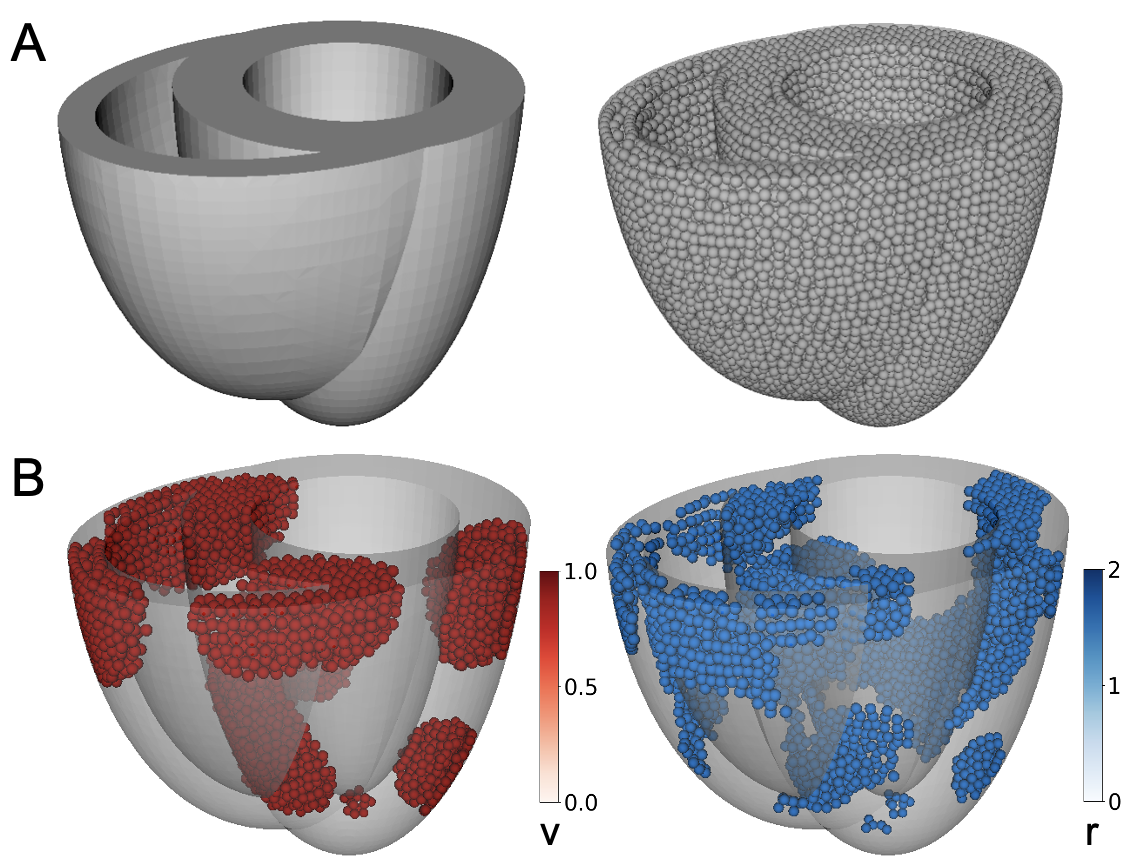}
  \caption{
Bi-ventricular simulations of cardiac electrophysiology using the smoothed particle hydrodynamics (SPH) method.
\textbf{A} Surface geometry and approximation of volume with particles. We simulated focal and reentrant rhythms, see Figs.~\ref{fig:methods:simulations}D), \ref{fig:results:SPH1}-\ref{fig:results:SPH4}.
\textbf{B} Initialization used in section \ref{sec:results:SPH} to begin the gradient descent-based learning. The initial learned voltage and refractory variables $v'_0(\vec{x})$ and $r'_0(\vec{x})$ were set to a random Perlin noise field (here shown with $v,r>0.1$ thresholding), while observations of the voltage variable $\bar{v}$ were obtained only from the epicardium.
 }
  \label{fig:methods:sph}
\end{figure}

The SPH simulations were implemented using JAX-MD~\cite{jax-md}, a differentiable molecular dynamics framework, reimplementing the cardiac electrophysiology simulation part of the open-source SPHinXsys library~\cite{Zhang2021,Zhang2021b} as a fully differentiable pipeline.
The simulations were initialized using a bi-ventricular template geometry loaded from a surface mesh, see Fig.~\ref{fig:methods:sph}.
Particles were placed on a regular grid inside the mesh, and a 50-step iterative relaxation procedure was applied to achieve a near-uniform particle distribution, in which particles repel one another via a pressure-like force derived from the SPH kernel gradient.
Spatial interpolation was performed with the Wendland C2 kernel (5th-order, compact support radius $2h$, $h = 1.2\,\Delta x$), and neighbor lists were managed using JAX-MD's sparse neighbor list with a cutoff of $2h$.
A kernel correction (renormalization) matrix $\mathbf{B}$ was computed per particle to enforce first-order consistency of the SPH gradient operator.
A transmural scalar field $\psi \in [0,1]$ was computed by solving a diffusion equation on the particle set, with Dirichlet boundary conditions of $\psi = 1$ on the epicardium and $\psi = 0$ on the endocardium.
From $\psi$, a fiber direction $\mathbf{f}_0$ was assigned to each particle by rotating a circumferential base direction about the local sheet normal, with the helix angle varying linearly from $-70^\circ$ at the endocardium to $+80^\circ$ at the epicardium, following standard rule-based fiber assignment methods~\cite{Doste2019}.
Action potential propagation was modeled using the AP model, see section \ref{sec:methods:models}, augmented with an anisotropic diffusion tensor $\mathbf{D} = D_\mathrm{iso}\mathbf{I} + D_\mathrm{ani}\mathbf{f}_0\mathbf{f}_0^\top$ aligned with the local fiber direction.
The reaction-diffusion system was integrated using a Strang splitting scheme: a half-step of AP reaction, a full diffusion step, and a second half-step of AP reaction, which yields second-order accuracy in time.
The reaction terms for the voltage $v$ and refractory variable $r$ were integrated using a quasi-steady-state (QSS) exponential integrator, which is analytically exact for linear-coefficient ODEs and improves stability at larger time steps.
Diffusion was advanced using a second-order Runge-Kutta scheme with the SPH Laplacian corrected by the anisotropy tensor and the per-particle correction matrix $\mathbf{B}$.
As with the finite differences method, the entire simulation pipeline was implemented in JAX and is end-to-end differentiable.
Gradients of the observation loss with respect to the initial dynamical state and model parameters were computed using \texttt{jax.value\_and\_grad} with gradient checkpointing applied to the diffusion integration steps to manage memory.
Parameter recovery was performed using the Adam optimizer~\cite{Adam} with a learning rate of $10^{-4}$ via Optax.

\subsection{Gradient-based Learning of Spatio-Temporal Cardiac Action Potential Wave Dynamics}
\label{sec:methods:training}

In this study, we aim to learn spatio-temporal action potential wave dynamics inside an excitable tissue.
We treat the excitable tissue as a dynamical system $\mathbf{S}$.
The system's configuration is defined by a dynamical state $\mathbf{y}$, which is a vector of state variables $\{ y_1, y_2, \dots \}$ (e.g. excitability, refractoriness, etc.), and the system's dynamics correspond to the evolution of this state vector over time.
We assume a system of coupled partial differential equations (PDEs) that describe or approximate the dynamics, see eqs.~(\ref{eq:modelu}) and (\ref{eq:modelr}) for an example.
Accordingly, the dynamics are described by:
\[
\begin{aligned}
\frac{\partial \mathbf{y}}{\partial t} & = f \!\left( \mathbf{y}, \nabla^2 \mathbf{y}, \ldots, \mathbf{x}, t; \boldsymbol{\theta}
\right),
&& \mathbf{x}\in\Omega,\; t\in(0,T]
\end{aligned}
\]
where $f$ are a specific set of equations of motion, see eqs.~(\ref{eq:modelu})-(\ref{eq:modelr}), $\boldsymbol{\theta}$ is a set of model parameters, and $\Omega$ is the spatial domain of a spatially extended system.
The initial condition corresponds to the configuration of all state vectors throughout the spatial domain of the system at time $t=0$:
\[
\begin{aligned}
\mathbf{y}(\mathbf{x},t=0)
&= \mathbf{y}_0(\mathbf{x}),
&& \mathbf{x}\in\Omega
\end{aligned}
\]
where $\mathbf{y}_0 = (v_0,r_0)$ for two-variable models.
Here, we aim to estimate the initial condition $\mathbf{y}_0 (\mathbf{x})$ and a set of parameters $\{ \theta_1 , \dots , \theta_n \}$ that minimize a loss function:
\begin{align} 
\label{eq:loss-parameters}
\mathcal{L} \left ( \mathbf{y}_{t_0}, \dots , \mathbf{y}_{t_n} \right )
\end{align}
evaluated over a series of subsequent observations of a subset of the dynamical state variables (the ones that can be observed) at observation time points $t_0, \dots , t_n$, see section \ref{sec:methods:learn:loss}.
In particular, we assume that the model parameters and the dynamical state $\mathbf{y}$ are unknown at the beginning of the learning.
The problem corresponds to a state and parameter estimation task.

In this study, the learning was performed by observing the excitatory or voltage variable $v$, see section \ref{sec:methods:learn:loss}, and by subsequently continuously adjusting the initial condition and model parameters using gradient-based optimization until the loss is minimized.
In some tasks, we assumed that the voltage variable could only be observed in certain locations of the spatial domain, e.g. on parts of the surface or in sparse locations.
The dynamics were learned using the Adam~\cite{Adam} optimizer with either a constant learning rate of $10^{-3}$, or a linearly or exponentially decreasing learning rate, see also section \ref{sec:methods:learn:multi-horizon}.
Learning was performed on NVIDIA RTX A5000 and A6000 graphics processing units (GPUs), and took between 4 and 24 hours depending on the task.

\subsubsection{Loss Function}
\label{sec:methods:learn:loss}
In all simulations, we computed at a minimum a total loss comprising an observation loss and a regularization loss :
\begin{eqnarray} 
\label{eq:loss}
\mathcal{L} = & \mathcal{L}_{obs} + \mathcal{L}_{reg}
\end{eqnarray}
The observation loss $\mathcal{L}_{obs}$ measures the discrepancy or mean squared error between the observed excitatory or voltage variable $\bar{v}$ and the learned variable $v'$:
\begin{eqnarray} 
\label{eq:loss-data}
\mathcal{L}_{obs} = & \frac{1}{N} \sum (\bar{v}(\mathbf{x},t) - v'(\mathbf{x},t))^2
\end{eqnarray}
over all observed data (pixels, points) within a specified time frame (e.g. horizon, see section \ref{sec:methods:learn:multi-horizon}).
The regularization loss $\mathcal{L}_{reg}$ is a penalty term that activates when the learned dynamic variables $v'$ and $r'$ leave their typical range, and enforces them to be $0 \le v' \le 1$ and $r' \ge 0$:
\begin{eqnarray} 
\label{eq:loss-reg}
\mathcal{L}_{reg} = & \sum \max(-u,0) + \sum \max(-r,0) +  \sum \max (v-1,0)
\end{eqnarray}
For example, if $v' < 0 , r = 1 \rightarrow \mathcal{L} = - v'$ with the AP model.
In addition to $\mathcal{L}_{obs}$ and $\mathcal{L}_{reg}$, we introduced other losses whenever necessary.
With sparse data, see section \ref{sec:results:sparse}, we introduced a smoothness loss $\mathcal{L}_{s}$, which corresponds to the mean squared gradient magnitude of the learned refractory variable:
\begin{eqnarray} 
\label{eq:loss-smooth}
\mathcal{L}_{s} = \alpha \cdot \| \nabla r' \|_2^2
\end{eqnarray}
where $\nabla$ denotes the discrete spatial gradient computed using forward differences, and $\alpha$ is a weighting factor controlling the strength of the contribution of this loss to the overall loss.
The smoothness loss penalizes large spatial gradients in the field of the learned refractory variable $r'$, which arise, for instance, with sparse observations, see section \ref{sec:results:sparse}.
We only used the smoothness loss with sparse data in addition to the total loss $\mathcal{L}$ and set $\alpha = 0$ otherwise.

With experimental data, see section \ref{sec:results:experiments}, we augmented the total loss $\mathcal{L}$ with a perceptual loss:
\begin{eqnarray} 
\mathcal{L} = & \mathcal{L}_{obs} + \mathcal{L}_{reg} + \mathcal{L}_p
\end{eqnarray}
the perceptual loss being:
\begin{eqnarray} 
\label{eq:loss-perceptual}
\mathcal{L}_p = \frac{1}{T}\sum_{t} \left\| \frac{f_\theta(v'_t)}{\|f_\theta(v'_t)\|} - \frac{f_\theta(\bar{v}_t)}{\|f_\theta(\bar{v}_t)\|} \right\|^2
\end{eqnarray}
where $v'_t$ and $\bar{v}_t$ denote the simulated and observed frames at time $t$, respectively.
The perceptual loss $\mathcal{L}_p$ is computed in the feature space of a frozen vision transformer encoder $f_\theta$ pre-trained via a Video Joint-Embedding Predictive Architecture (V-JEPA)~\cite{Eing2026}.
We pre-trained the encoder on simulated single- and multi-spiral AP dynamics, and held the encoder weights fixed throughout optimization.
Both the simulated and observed frames were z-score normalized using the global mean and standard deviation of the synthetic training dataset, and the perceptual loss was computed as the mean squared error (MSE) between $\ell_2$-normalized per-frame feature vectors.
We also introduced a phase loss $\mathcal{L}_{\phi}$ defined as:
\begin{eqnarray} 
\label{eq:loss-smooth}
\mathcal{L}_{\phi} = \alpha_{\phi} \cdot \| \phi - \phi ' \|
\end{eqnarray}
where $\phi$ and $\phi '$ are the ground-truth and learned phase angles and $\alpha_{\phi}$ is a weighting factor.
However, we abandoned the phase loss because we obtained better performance with the perceptual loss.

\subsubsection{Initial Parameter Guesses}
Initial guesses for the learned parameters $\theta'_i = \{ \theta_1', \theta_2', ... \theta_i'\}$ were set to random values within certain reasonable parameter-specific bounds, often resulting in substantial mismatches between the initial guesses $\theta'_i$ and the ground-truth values $\theta_i$ of at least 20\% and in some cases more than 50-100\% per parameter.
Fig.~\ref{fig:barplots} shows true parameters (black bar, left), initial parameter guesses (center, gray), and learned parameters for 3 different cases (2D focus, 2D spiral, sparse observations).
In the lower panel in Fig.~\ref{fig:barplots}A), the initial value for the diffusion coefficient was $D=0.01$, while the true value was $D=0.0011$.
Fig.~\ref{fig:results:parameter_history} shows that the parameters can fluctuate wildly across a wide range of values throughout learning, possibly diverging further from the initial guess before converging back to the ground-truth.
With cross-model learning, see section \ref{sec:results:crossmodel}, we first learned approximations of the parameters in a single cell (0D), then continued with the learned parameters in a 1D ring, and finally used those approximations to initialize the 2D learning.

\subsubsection{Multi-Horizon Learning Schedule}
\label{sec:methods:learn:multi-horizon}
Learning was performed sequentially across multiple segments of the observed dynamics.
For example, if the entire observation data shows 3 rotations of a spiral wave, we divided the 3 rotations into segments corresponding to the first, second, and third rotations, or into finer divisions, such as quarter rotations.
We refer to each segment as a 'horizon' and the learning over multiple segments as multi-horizon learning.
Importantly, horizons can overlap.
We empirically found that multi-horizon learning outperforms learning the dynamics all at once in a single horizon, even when learning it over many epochs, see Fig.~\ref{fig:supplement:learning-setup}B,C).
Further, we found that dividing the dynamics into many horizons dramatically outperforms selecting fewer horizons.
If the horizon is too long, then the loss quickly saturates, because small changes in the parameters and initial condition lead to large deviations in the dynamical trajectories, whereas short horizons yield much smaller deviations and non-saturating losses.
As a result, we abandoned using long single or few horizons and instead used rolling multi-horizon learning schedules with many (e.g. $30-60$) overlapping subsequent horizons for learning a brief period of the spatio-temporal dynamics (e.g. a focal wave or a few spiral wave rotations).
In the above example with 3 spiral rotations, a simple multi-horizon schedule with few horizons could consist of 6 non-overlapping horizons, each covering half a rotation, and a rolling multi-horizon schedule with many horizons could consist of 60 overlapping horizons, each covering half a rotation as well, but with 95\% overlap between subsequent horizons.

Generally, the multi-horizon learning schedule $S$ comprises $k$ horizons, starts with the first horizon, and continues with subsequent horizons:
$$ S = \{ \mathcal{H}^1,  \mathcal{H}^2,  \mathcal{H}^3, ... ,  \mathcal{H}^k \}$$
Each horizon comprises $n$ time steps $\{t_1, t_2, ... , t_n\}$ at which we obtain observation data:
$$ \mathcal{H}^i = \{ \bar{v}(t_1), \bar{v}(t_2), ..., \bar{v}(t_n) \}$$ 
of the voltage variable $v$. 
Each subsequent horizon is shifted by one time step or stride $s=1$.
The number of time steps or observations together with the stride $s$ equals the horizon duration $h$.
Throughout this study, we used equidistant temporal sampling $\Delta t = t_{i+1} - t_{i} = t_{i+2} - t_{i+1}$ etc. for the observations.
Accordingly, with a rolling multi-horizon learning schedule composed of horizons with duration $h=10$, we learned over the following sequence of horizons and observations:
\begin{eqnarray} 
\label{eq:horizons}
\mathcal{H}^1 = & \{ \bar{v}(t_1), \bar{v}(t_2), ..., \bar{v}(t_{10}) \} \nonumber \\ 
\mathcal{H}^2 = & \{ \bar{v}(t_2), \bar{v}(t_3), ..., \bar{v}(t_{11}) \} \nonumber \\
\mathcal{H}^3 = & \{ \bar{v}(t_3), \bar{v}(t_4), ..., \bar{v}(t_{12}) \} \nonumber \\ 
\mathcal{H}^4 = & ...
\end{eqnarray}
The horizon duration can vary across the sequence of horizons, see tables \ref{tab:learning-schedule} and \ref{tab:learning-schedule3Dscroll} and Fig.~\ref{fig:supplement:learningschedule}.
In each horizon, we continue to estimate the parameters and the dynamical state using the estimates from the previous horizon.
With rolling overlapping horizons and $s=1$, we used the second estimated state of the previous horizon to construct the initial state of the next horizon: 
$$ (v',r')^{i-1}_2 \rightarrow (\bar{v},r')^i_1 $$
where $r'$ is the estimated refractory variable and $\bar{v}$ is the current observation of the voltage variable (their respective spatial fields).
In the case of non-overlapping horizons, we used the last state to initiate the first state of the next horizon.
The loss $\mathcal{L}$ is calculated per horizon, and learning is performed over a number of epochs $n_e$ specified for each horizon.
We varied the number of epochs $n_e$ over the course of the learning, see tables \ref{tab:learning-schedule} and \ref{tab:learning-schedule3Dscroll}.
A simple rolling multi-horizon schedule would be to learn 3 spiral wave rotations, observed across 80 time steps, using $k = 60$ horizons, each with a duration of $n = 20$ time steps, and a stride $s = 1$, for example.
However, we used learning schedules that increased the horizon duration in stages, see table \ref{tab:learning-schedule}, as this promoted parameter convergence, see Fig.~\ref{fig:results:parameter_history}, and improved long-term predictions.

\subsubsection{Initialization}
\label{sec:methods:initialization}
We found that choosing the right initial condition or initial dynamical state $(v_0',r_0')$ before learning can strongly affect the learning, and developed initialization schemes based on empirical testing.
Throughout this study, we assumed that only the voltage variable $v$ could be observed, and that it could be only partially observed in some situations.
For instance, in section \ref{sec:results:SPH}, it could only be observed across the epicardium, in section \ref{sec:results:3Dbulk} on two opposing surfaces of a bulk, and in section \ref{sec:results:sparse} in sparse locations.
Accordingly, the initial state $(v_0',r_0')$ used in the first epoch had to be constructed from the observation $\bar{v}_0$ at time $t=0$ and guesses for the remaining missing data: the refractory field had to be guessed entirely, and the voltage variable had to be guessed in locations without observations.
At the same time, we aimed to initialize the learning without making too many assumptions about the dynamics.

\begin{figure}[htb]
  \centering
  \includegraphics[clip, trim=0.0cm 0.0cm 0.0cm 0.0cm, width=0.46\textwidth]{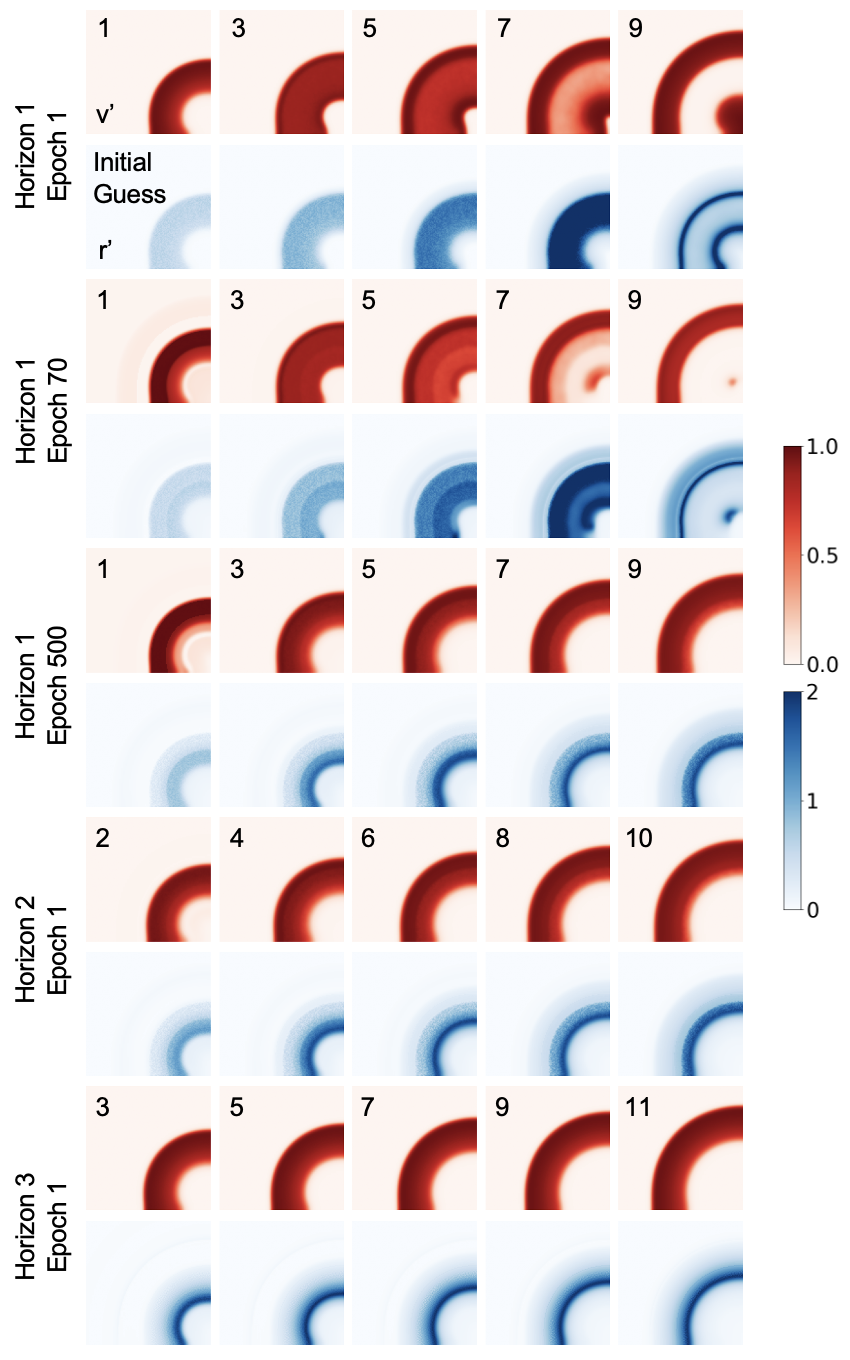}
  \caption{
Learning the initial condition of a focal wave during the early phase of the rolling multi-horizon learning schedule shown in Fig.~\ref{fig:supplement:learningschedule}D).
The maps show the learned voltage field $v'(x,y)$ (red) and the learned refractory field $r'(x,y)$ (blue) of the AP model.
Initially, the true refractory field $r(x,y,0)$ is unknown and must be guessed.
We set $r_0' = \bar{v_0} \cdot (1 + \sigma )$ as an approximation of the observation $\bar{v}_0$, see eq.~(\ref{eq:init2D}).
With this guess, the learned voltage wave propagates both forward and backward (top row).
Within the first 100 epochs, the refractory field is modified such that the voltage wave propagates only forward.
The initial state eventually saturates and can only be refined further towards ground-truth when moving on to the next horizon
(horizon 1: time steps 1-10, horizon 2: time steps 2-11, horizon 3: time steps 3-12, etc.).
In Fig.~\ref{fig:results:sparsity}B), the initial condition is learned within the first 3-4 horizons.}
  \label{fig:methods:initialcondition}
\end{figure}

With the 2D simulations in sections \ref{sec:results:2Dfocus} and \ref{sec:results:2Dspiral}, we chose the following initial guess for the refractory variable:
\begin{eqnarray} 
\label{eq:init2D}
r'_o = & \bar{v}_0 \cdot (1 + \sigma)
\end{eqnarray}
where $\bar{v}_0$ is the observation of the voltage wave at time $t=0$ in horizon 1 and $\sigma$ is Gaussian noise, see also Fig.~\ref{fig:supplement:2D-initialcondition}A). 
This initialization scheme effectively creates a noisy copy of the voltage wave as the refractory wave.
Other initializations, such as $r'_0=0$, $r'_0 = 1$, or $r'_0 = \sigma$ led to less effective learning, see Figs.~\ref{fig:supplement:2D-initialcondition} and \ref{fig:supplement:learning-setup}A), or numerical instability.
In the sparse 2D simulations in section \ref{sec:results:sparse}, we used the same initialization as in the continuous 2D case, but applied eq.~(\ref{eq:init2D}) only in the electrode locations on the grid.
With the 3D bulk simulations, we used the observations $\bar{v}(z=0),\bar{v}(z=16)$ of the voltage variable across the two opposing observed top and bottom surfaces, see Fig.~\ref{sec:results:3Dbulk}, together with a weighted average of these two surface patterns as an initialization for $v$ in the depth of the bulk. 
Accordingly, we used eq.~(\ref{eq:init2D}) for $r'_0$ across the two surfaces and their weighted average for the bulk.
In the bi-ventricular simulations in section \ref{sec:results:SPH}, we chose random spatial patterns for the initial $v'_0$ and $r'_0$ fields throughout the ventricles, see Fig.~\ref{fig:methods:sph}B), and the observation $\bar{v}_0$ at $t=0$ across the epicardium.
The random pattern was generated using Perlin noise and thresholding, see Fig.~\ref{fig:methods:sph}B).
With the 2D cross-model learning, see section \ref{sec:results:crossmodel}, we used a denoising diffusion probabilistic model (DDPM) trained on 2D simulation data, similarly as described in Baranwal et al.~\cite{Baranwal2024}, simulated with the AP model to generate such data when conditioned with MS spiral wave data. 
Subsequently, the DDPM could translate MS voltage patterns into AP spirals with both excitatory and refractory components, which then served as the initial conditions for the learning.
Moreover, the DDPM could translate any spiral-shaped pattern, e.g. those obtained in imaging experiments, and produce a corresponding AP initial condition.
The translation of an MS excitation variable to an AP excitation variable was done via an SDEdit~\cite{Meng2022} style approach, and the generation of the associated refractory variable was done via a RePaint~\cite{Lugmayr2022} style approach.

\subsubsection{Cardiac Monolayer Cell Culture}
\label{sec:methods:cellculture}
We fitted simulations to two different cell cultures, see section \ref{sec:results:experiments} and Fig.~\ref{fig:results:cellculture}.
The data shown in Fig.~\ref{fig:results:cellculture}A) was obtained from Monteiro da Rocha et al.~\cite{MonteiroDaRocha2016}. 
It shows a calcium spiral wave imaged with fura-2.
We used the calcium wave as a proxy for the voltage wave. 
The data was cropped from the corresponding Supplementary Video (.mp4 fileformat) and converted into a numpy array with floating point precision values normalized between $[0,1]$.
The data shown in Fig.~\ref{fig:results:cellculture}B) was generated specifically for this study.
Human iPSC-CM (iCell2 from CDI/Fujifilm) were grown in glass-bottom $35 mm$ dishes (with $14 mm$ circular area) and treated as previously described~\cite{Heinson2023, Liu2023}. 
Briefly, the glass bottom was coated with 50 µg/ml fibronectin, and cells were plated at high density ($270,000$ cells per dish). Cells were maintained in a humidified CO2 incubator at $37^{\circ} C$, with regular exchange of culture medium. 
Measurements were done on day 6 after plating, in OptiMem medium. 
To promote spiral induction, samples were treated with 3nM dofetilide for 30min. 
Spirals were induced through rapid pacing.
Fluorescent voltage recordings were obtained after labeling with 1 µm Berst1 voltage-sensitive dye~\cite{Huang2015,Klimas2020}. 
The optical mapping system used to collect data was described previously~\cite{Heinson2023, Liu2023}. 
Briefly, a Basler camera (Basler acA720-520um, Ahrensburg, Germany) was used to record videos at 100 Hz. 
The field of view was $1.86 cm^2$. 
A 660-nm LED (M660L4, Thorlabs, Newton, NJ, USA) provided oblique trans-illumination and light was collected after an emission filter (ET595/40m+700LP, Chroma, Bellows Falls, VT, USA). 
Signals were acquired with an USB3.0-based Pylon Viewer software.

\section{Results}
\label{sec:results}

We found that differentiable cardiac electrophysiology simulations can be used to estimate states and parameters of 2D and 3D action potential wave dynamics across various tissues and configurations. 
The dynamics and parameters can be learned from sparse, partial, and noisy observations of the voltage variable, yielding full numerical reconstructions of the reaction-diffusion dynamics.
For instance, we were able to locate focal waves originating in the septum from epicardial observations, see Fig.~\ref{fig:results:SPH1}, scroll waves inside a bulk from observing two opposing surfaces of the bulk, see Fig.~\ref{fig:results:scroll-top-bottom}, recover full-resolution 2D reentrant spiral waves measured with a sparse grid of electrodes, see Fig.~\ref{fig:results:sparsity}, and fit a model to an action potential spiral wave imaged in a cardiac monolayer cell culture, see Fig.~\ref{fig:results:cellculture}.
In particular, we found that the method is best suited for learning chaotic spiral or scroll wave dynamics.

\subsection{2D Focal Waves}
\label{sec:results:2Dfocus}

Fig.~\ref{fig:methods:initialcondition} shows the learning process of a simple focal wave propagating through a simulated 2D tissue with isotropic conduction.
The action potential wave was simulated using the AP model, see section \ref{sec:methods:models}, and propagates from the lower right to the upper left.
We employed a rolling multi-horizon learning schedule, as shown in Fig.~\ref{fig:supplement:learningschedule}D and table \ref{tab:learning-schedule}, and observed the voltage variable $v$ at full resolution and without noise across all snapshots within each horizon.
We assumed that the model equations are known.
Overall, the wave is observed within 70 time steps, and, at the beginning of the learning, it has already traveled away from its origin.
The first observation is the snapshot on the left in the top row in Fig.~\ref{fig:methods:initialcondition}.
As only the voltage variable $v$ can be observed, the true initial dynamical state $(v_0,r_0)$ is unknown, and an initial state $(v'_0,r'_0)$ must be guessed to initiate the learning, see section \ref{sec:methods:initialization}.
The first row (Horizon 1, Epoch 1) shows the first epoch, which starts with the guessed initial state $(\bar{v}_0,r'_0)$ composed of the observation $\bar{v}_0$ (first row) and the guess for $r'_0$ (second row).
We used $r'_0 \sim \bar{v}$, see eq.~(\ref{eq:init2D}) for details.
With this initialization, the action potential lacks a refractory tail and propagates both forward and backward during the first epochs.
Within the first 100 epochs, the initial refractory field $r'$ is varied via gradient-based optimization,
such that the learned voltage wave eventually sheds the backward-propagating part and subsequently propagates only forward, as it should.
By the end of Horizon 1, comprising 600 epochs, the learned dynamics exhibit an action potential wave that propagates roughly as expected, from the lower right to the upper left.
At the same time, the parameters are continuously varied during learning to further reduce the loss, see also Fig.~\ref{fig:results:parameter_history}.

\begin{figure*}[htb]
  \centering
  \includegraphics[clip, trim=0.0cm 0.0cm 0.0cm 0.0cm, width=0.95\textwidth]{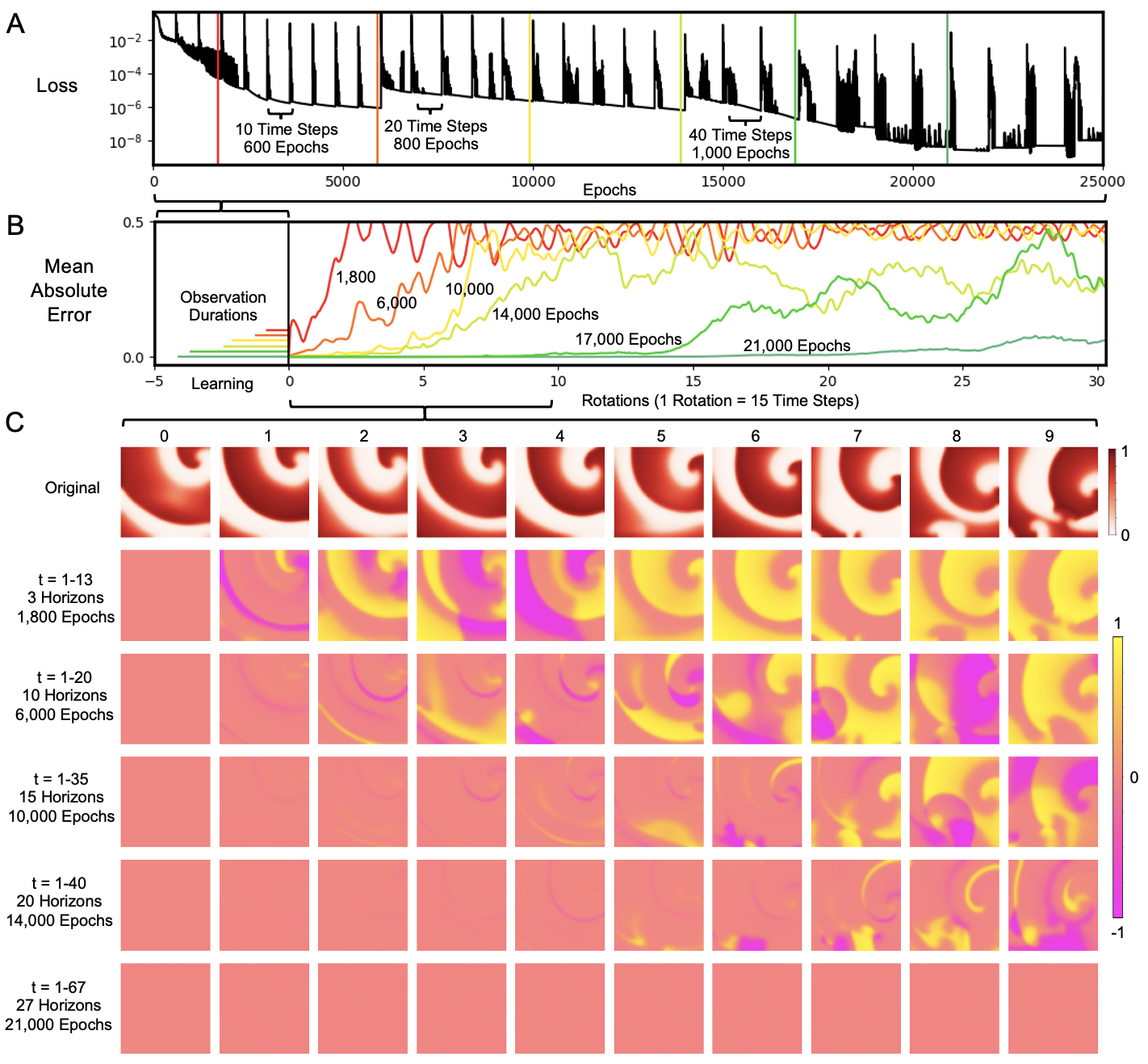}
  \caption{
Reconstruction of spiral wave dynamics using multi-horizon learning schedule with a differentiable simulation. 
Effect of observation and learning duration on state and parameter reconstruction accuracy,
The accuracy is measured as the agreement between original and learned spiral wave dynamics when the two dynamics are evolved into the future and compared to each other, see also Fig.~\ref{fig:results:2D:spiral-divergence-timeseries}. 
Learning ends at 0. Afterwards, the fully learned dynamics co-evolve precisely with the original dynamics for over 20 rotations.
\textbf{A} Loss curve showing multi-horizon learning schedule (10 horizons with 10 time steps each, 10 horizons with 20 time steps each, 10 horizons with 40 time steps each). At the beginning of each new horizon, the loss peaks.
\textbf{B} Mean absolute pixel-wise error between original and learned dynamics over time with different learning schedules: 3, 10, 15, 20, and 27 horizons as shown in
Time units normalized and specified in number of spiral rotations. 
Learning ends at  $t=0$ and the learned state and parameters are used to simulate the dynamics over 30 rotations into the future.
Our multi-horizon gradient-descent-based learning scheme recovers model parameters and the dynamical state so well that, if the learned dynamical state is evolved into the future, it matches the original dynamics precisely over many rotations (green line).
With 27 horizons, the dynamics are identical over 20 rotations. 
or 67 time steps or 20,000 epochs, the dynamics is learned over 
\textbf{C} Top row: Original spiral wave dynamics with alternans, meandering and wave break. 
With 27 horizons learning over 60 time steps or about 4 rotations, we achieve very high reconstruction accuracies. The learned dynamics co-evolve for over 20 rotations, despite alternans, breathing, and spiral wave break-up.
 }
  \label{fig:results:2D:spiral-divergence-loss}
\end{figure*}

\begin{figure*}[htb]
  \centering
  \includegraphics[clip, trim=0.0cm 0.0cm 0.0cm 0.0cm, width=0.8\textwidth]{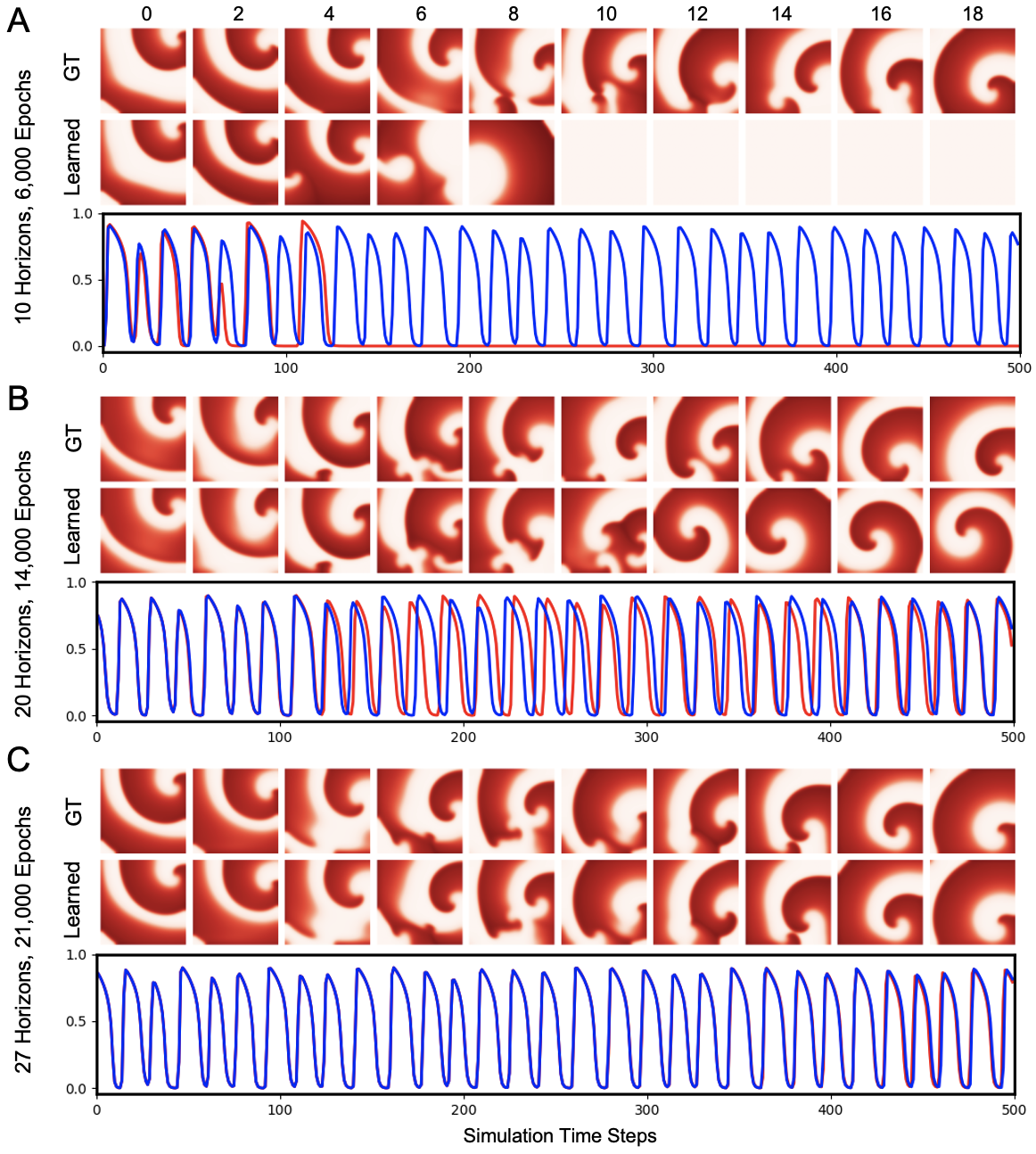}
  \caption{
Agreement between learned and original ground-truth (GT) spiral wave dynamics after gradient-descent-based learning with different observation and learning durations using the rolling multi-horizon learning schedule shown in Fig.~\ref{fig:supplement:learningschedule}D).
After learning with sufficiently many observations and epochs ($\sim$4 rotations), the two dynamics co-evolve congruently over long periods of time (>20 rotations), while with fewer epochs and observations, the learned dynamics diverge quickly, see also Fig.~\ref{fig:results:2D:spiral-divergence-loss}.
Snapshots show dynamics after 2, 4, 6, 8, etc. rotations (1 rotation $\approx$ 15 time steps).
Learning stopped at 0 and the two dynamics are evolved independently afterwards.
The time-series were measured at the center of the simulation domain (blue: original dynamics, red: learned).
\textbf{A} Learning stopped prematurely after 6,000 epochs (10 horizons, 600 epochs each horizon).
The learned dynamics diverge quickly from the ground-truth and coincidentally self-terminate.
\textbf{B} Learning stopped prematurely after 14,000 epochs (20 horizons). While there is still strong initial agreement, the dynamics diverge after 6-7 action potentials.
\textbf{C} Fully learned dynamics after 21,000 epochs (27 horizons). Due to the increasing horizon duration, all model parameters and the initial dynamical state at $t=0$ are recovered so well that the dynamics co-evolve congruently over many rotations, see also Fig.~\ref{fig:results:2D:spiral-divergence-loss}. 
Observation durations range between 1-4 rotations, see also Fig.~\ref{fig:results:2D:spiral-divergence-loss}B).
 }
  \label{fig:results:2D:spiral-divergence-timeseries}
\end{figure*}

We found that the rolling multi-horizon scheme shown in Fig.~\ref{fig:supplement:learningschedule}D) is critical for achieving better dynamical state reconstructions than the state shown in the third panel toward the end of the first horizon (Horizon 1, Epoch 500), see also Fig.~\ref{fig:supplement:learning-setup}C).
The learned initial voltage and refractory fields $(v'_0,r'_0)^1$ at the end of horizon 1 are just coarse approximations and not as continuous as their original counterparts. 
The parameter guesses are poor at the end of horizon 1.
Moreover, without the multi-horizon scheme, the learning saturates: the state and parameters do not improve further with longer horizons and more epochs, but only improve with additional horizons, see also Fig.~\ref{fig:supplement:learning-setup}B,C).
The details of the rolling multi-horizon learning schedule are provided in table \ref{tab:learning-schedule}.
Each horizon in Fig.~\ref{fig:methods:initialcondition} contains 10 observations $\{ \bar{v}_1, \bar{v}_2, ..., \bar{v}_{10} \}$ (spatial patterns). 
The observation loss is computed over the 10 observations, and the dynamics are learned over 600 epochs per horizon. 
At the end of the first 10 horizons, the dynamics were learned over 6,000 epochs total.
In total, the wave is learned over 30 horizons or 70 time steps. 
As each horizon starts with the next time step, the diffusion of the voltage wave provides a smoother, slightly better initial guess for the next horizon, enabling the learning to achieve better and better approximations over the sequence of horizons.
Employing the rolling multi-horizon learning scheme, it is possible to identify the full dynamical state and all 6 model parameters of the Aliev-Panfilov (AP) model, see table \ref{tab:parameters-2D-learned} and Fig.~\ref{fig:barplots}B).
The initial condition is approximated sufficiently well in the first 2-3 horizons, while the parameters need longer to be recovered in subsequent horizons.
The parameters $\{ D, a, k \}$ converge relatively quickly and can be recovered within margins of less than $1\%$, while $\{ \epsilon_0, \mu_1, \mu_2 \}$ require longer and can be recovered only within margins of $5\%$, see table \ref{tab:parameters-2D-learned}.
Importantly, the initial parameter guesses can be substantially off from the ground-truth values, yet can still be recovered within a few percent, see Fig.~\ref{fig:barplots}B).
Because the focal wave eventually dies out at the boundaries, the observation time is limited, and it is difficult to evaluate learning accuracy on data other than the data used for learning. 
Nevertheless, by the end of the learning schedule, the learned and original waves are congruent and visually indistinguishable, and we obtain satisfactory parameter estimations.
Yet, the parameter estimations are not as good as with spiral wave dynamics, see next section.

\subsection{2D Spiral Waves}
\label{sec:results:2Dspiral}

\begin{figure}[htb]
  \centering
  \includegraphics[clip, trim=0.0cm 0.0cm 0.0cm 0.0cm, width=0.42\textwidth]{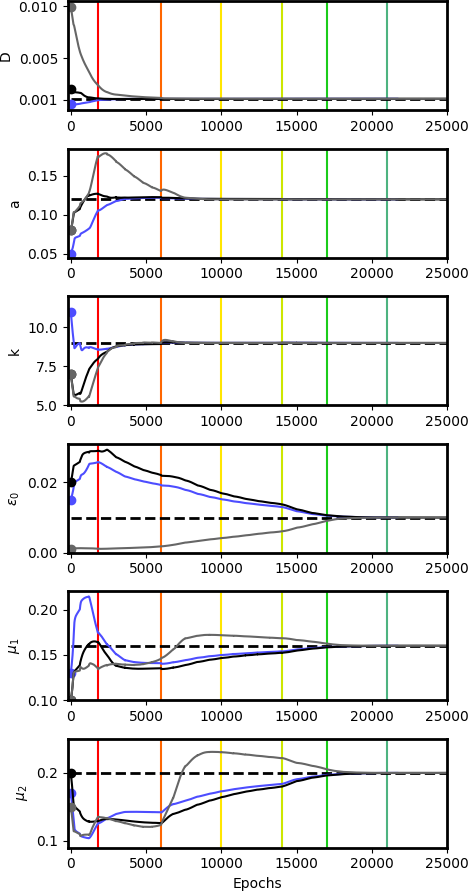}
  \caption{
Convergence of learned Aliev Panfilov model parameters $\{ D, a, k, \epsilon_0, \mu_1, \mu_2 \}$ towards ground-truth values (dashed line) starting with 3 different initial parameter guesses (black, gray, blue) with the dynamics shown in Figs.~\ref{fig:results:2D:spiral-divergence-loss} and \ref{fig:results:2D:spiral-divergence-timeseries}. 
The behavior is the same for all parameter combinations we tested.
Vertical lines same as in Fig.~\ref{fig:results:2D:spiral-divergence-loss}A).
Increasing horizon durations (at 6,000 and 14,000 epochs) induces pivoting and leads to faster convergence.
 }
  \label{fig:results:parameter_history}
\end{figure}

\begin{figure}[htb]
  \centering
  \includegraphics[clip, trim=0.0cm 0.0cm 0.0cm 0.0cm, width=0.44\textwidth]{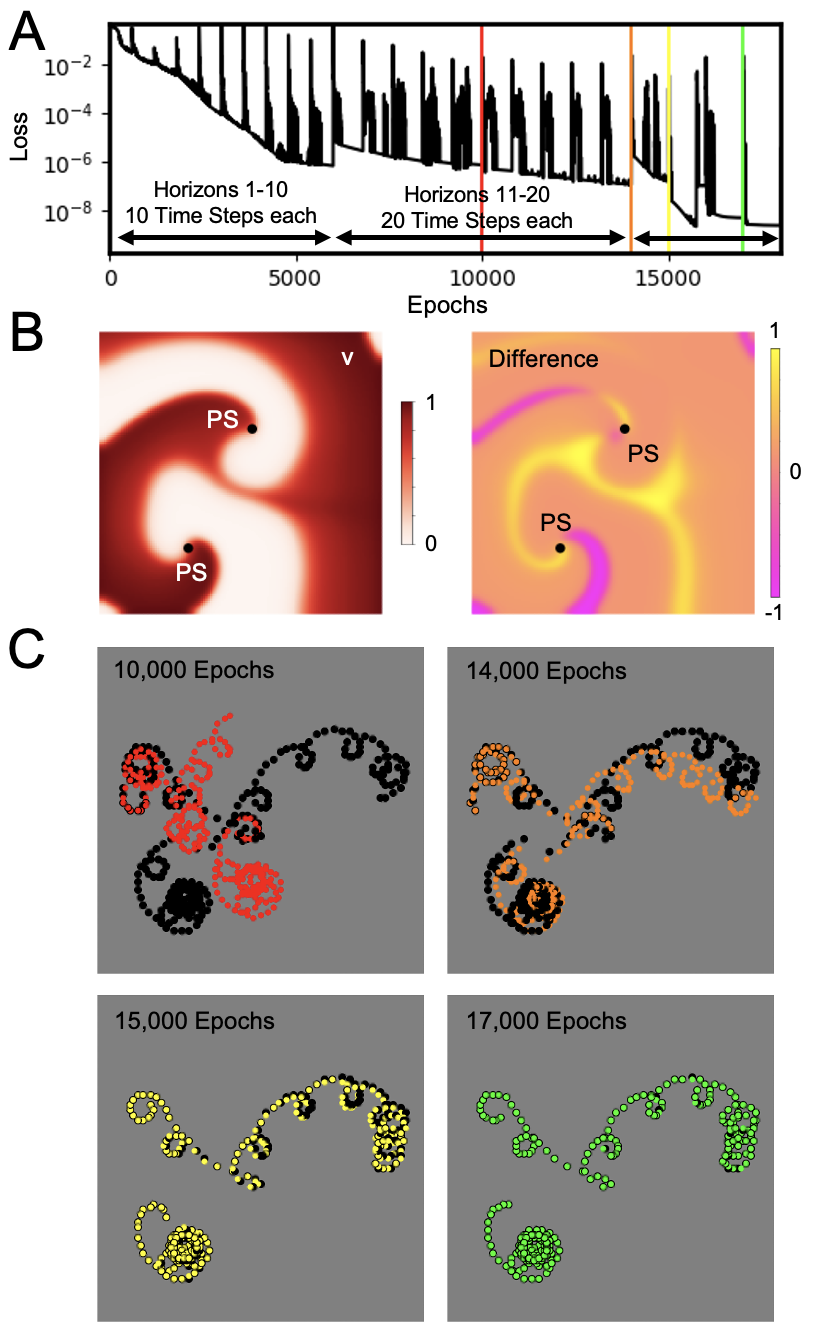}
  \caption{
  Learning and predicting meandering spiral wave dynamics.
\textbf{A} Loss curve during learning with rolling multi-horizon schedule (vertical bars correspond to panels in C). Total of 24 horizons with 10 (horizons 1-10), 20 (horizons 11-20), and 40 time steps (horizons 21-24) each.
\textbf{B} Spiral wave pattern (red: voltage) and difference between ground-truth and learned dynamics (right, learning stopped prematurely) together with phase singularities (PS) marking rotational core (at time step 80).
\textbf{C} Spiral wave core trajectories predicted into the future after learning the dynamics (black: ground-truth, red: 10,000 epochs / 15 horizons, orange: 14,000 epochs / 20 horizons, yellow: 15,000 epochs / 21 horizons, green: 17,000 epochs / 23 horizons) plotted over 150 time steps into the future ($\approx 12$ rotations).
 }
  \label{fig:PS}
\end{figure}

Figs.~\ref{fig:results:2D:spiral-divergence-loss} and \ref{fig:results:2D:spiral-divergence-timeseries} show reconstructions of spiral wave dynamics, which we observed for up to 4-5 rotations.
As in section \ref{sec:results:2Dfocus}, the wave dynamics were simulated using the AP model and then also learned using a differentiable version of the AP model.
The learning started with rough guesses of the initial state, as described in eq.~(\ref{eq:init2D}), and the parameters $\theta_i  = \{ D, a,\epsilon_0, k, \mu_1, \mu_2 \}$.
Importantly, the initial parameter guesses were off substantially from the actual ground-truth values, see blue, gray, and black dots and traces in  Fig.~\ref{fig:results:parameter_history}.
The dynamics in Figs.~\ref{fig:results:2D:spiral-divergence-loss} and \ref{fig:results:2D:spiral-divergence-timeseries} are composed of a meandering, drifting spiral wave exhibiting breathing phenomena, alternans, and occasional breakup.
Next to such dynamics, we also fitted dynamics composed of multiple drifting and interacting spiral waves, see Fig.~\ref{fig:results:sparsity}.
Continuously adapting both the initial state $(v'_0,r'_0)$ and all parameters $\theta_i$ in each epoch yields new dynamical trajectories, as also shown in Figs.~\ref{fig:methods:initialcondition}, \ref{fig:results:2D:spiral-divergence-timeseries}, and \ref{fig:results:sparsity}B).
Subsequently, computing the observation loss over all
time steps and backpropagating gradients through the solver enable finding better and better initial states and parameter values that further minimize the loss.
Fig.~\ref{fig:results:2D:spiral-divergence-loss}A) shows that the loss is continuously decreasing in each horizon, spikes at the beginning of each next horizon, but then continues to decrease further and reaches lower values than in the previous horizon.
We found that this resetting helps achieve lower and lower loss values, see also Figs.~\ref{fig:supplement:learning-setup}C), \ref{fig:results:2D:spiral-divergence-loss}A), \ref{fig:PS}A), and \ref{fig:results:sparsity}.
The multi-horizon learning schedule used in Figs.~\ref{fig:methods:initialcondition}, \ref{fig:results:2D:spiral-divergence-loss}-\ref{fig:PS} and \ref{fig:results:sparsity}, comprises 3 stages: 10 horizons with 10 time steps each, which were learned over 600 epochs each, followed by 10 horizons with 20 time steps each, which were learned over 800 epochs each, and lastly 10 horizons with 40 time steps each, which were learned over 1,000 epochs each, see also table \ref{tab:learning-schedule}.
We found that increasing the horizon duration is critical for obtaining better parameter estimates.
The parameter history curves in Fig.~\ref{fig:results:parameter_history} show the learned parameters over the course of the learning for three repeated runs (black, gray, blue), and how some of the parameters respond to increasing the horizon duration at 6,000 and 14,000 epochs.
While the parameters $\{ D, a, k \}$ generally converge quickly, the parameters $\{ \epsilon_0, \mu_1, \mu_2\}$ take much longer to converge, and their convergence is promoted by increasing the horizon duration.
While initially the learning benefits from shorter horizons, later it is beneficial when the learned and ground-truth dynamics need to match over longer periods.
At first, increasing the horizon duration yields a higher loss.
However, with continued learning, the loss decreases to lower values than if the learning had continued with shorter horizon durations.

The spiral wave dynamics can be learned so well within about 4-5 rotations that it becomes possible to forecast their evolution precisely over 20-30 rotations into the future, see Figs.~\ref{fig:results:2D:spiral-divergence-loss} ,\ref{fig:results:2D:spiral-divergence-timeseries}, and \ref{fig:PS}.
Figs.~\ref{fig:results:2D:spiral-divergence-loss}C) and \ref{fig:results:2D:spiral-divergence-timeseries} show the learned and ground-truth spiral wave dynamics after learning, and how they co-evolve and diverge when the final learned state $(v',r')^*$ and parameters $\{D,\epsilon_0,a,k,\mu_1,\mu_2\}^*$ are used to initiate a new simulation, and this simulation is compared to the corresponding ground-truth simulation over the same period.
The eventual divergence is expected as spiral wave dynamics are chaotic, and small differences in the initial conditions or model parameters generally lead to large accumulated errors over time.
Correspondingly, low divergence indicates very small mismatches between the ground-truth and learned configurations.
The first rows in Fig.~\ref{fig:results:2D:spiral-divergence-loss}C) and panels A,B) in Fig.~\ref{fig:results:2D:spiral-divergence-timeseries} show the divergence (pixel-wise error) when the learning was stopped prematurely after 1-3 rotations (or less than approximately 20 horizons, or 40 time steps, or 14,000 epochs).
The last row in Fig.~\ref{fig:results:2D:spiral-divergence-loss}C) and Fig.~\ref{fig:results:2D:spiral-divergence-timeseries}C) show essentially perfect congruence of the learned and ground-truth spiral wave dynamics with negligible pixel-wise error for the next 10-20 rotations (when learning is performed over 27 horizons, or 21,000 epochs, or 67 time steps, or about 4.5 rotations).
Correspondingly, Fig.~\ref{fig:results:2D:spiral-divergence-loss}B) shows the mean absolute error over time for the different learning durations (green: very low error for 20 rotations), as well as the learning duration depicted as horizontal bars for comparison (learning stops at 0).
The vertical bars in the loss curve in Fig.~\ref{fig:results:2D:spiral-divergence-loss}A) and the parameter histories in Fig.~\ref{fig:results:parameter_history} indicate the epochs associated with the learning durations.
With spiral wave dynamics, all parameters can be recovered within a margin of less than $0.05\%$, see table \ref{tab:parameters-2D-learned}.
The barplots in Fig.~\ref{fig:barplots}A) show the ground-truth and learned parameters along with two different sets of initial parameter guesses. 
In total, three very different initial parameter guesses all converge to the ground-truth parameter values, see Fig.~\ref{fig:results:parameter_history}, with some initial guesses being substantially off (e.g. $D$ being about 10x larger than ground-truth value).
Lastly, Fig.~\ref{fig:PS} shows that it is possible to obtain precise predictions of the spiral wave core's trajectory over more than 10 rotations. Here, the learning was performed over 23 horizons on a different example of spiral wave dynamics (with different parameters).
Based on the examples we studied, it appears that parameters converge faster with richer spiral wave dynamics.
In other words, learning is more effective with, for instance, meandering than with stationary spiral waves.
This finding aligns with the higher parameter-estimate margins observed with focal waves.
In summary, spiral wave dynamics can be learned so well that they can be predicted far into the future, even with substantial initial parameter mismatch, and even though they are chaotic and their evolution is generally hard to predict.

\begin{figure*}[htb]
  \centering
  \includegraphics[clip, trim=0.0cm 0.0cm 0.0cm 0.0cm, width=0.98\textwidth]{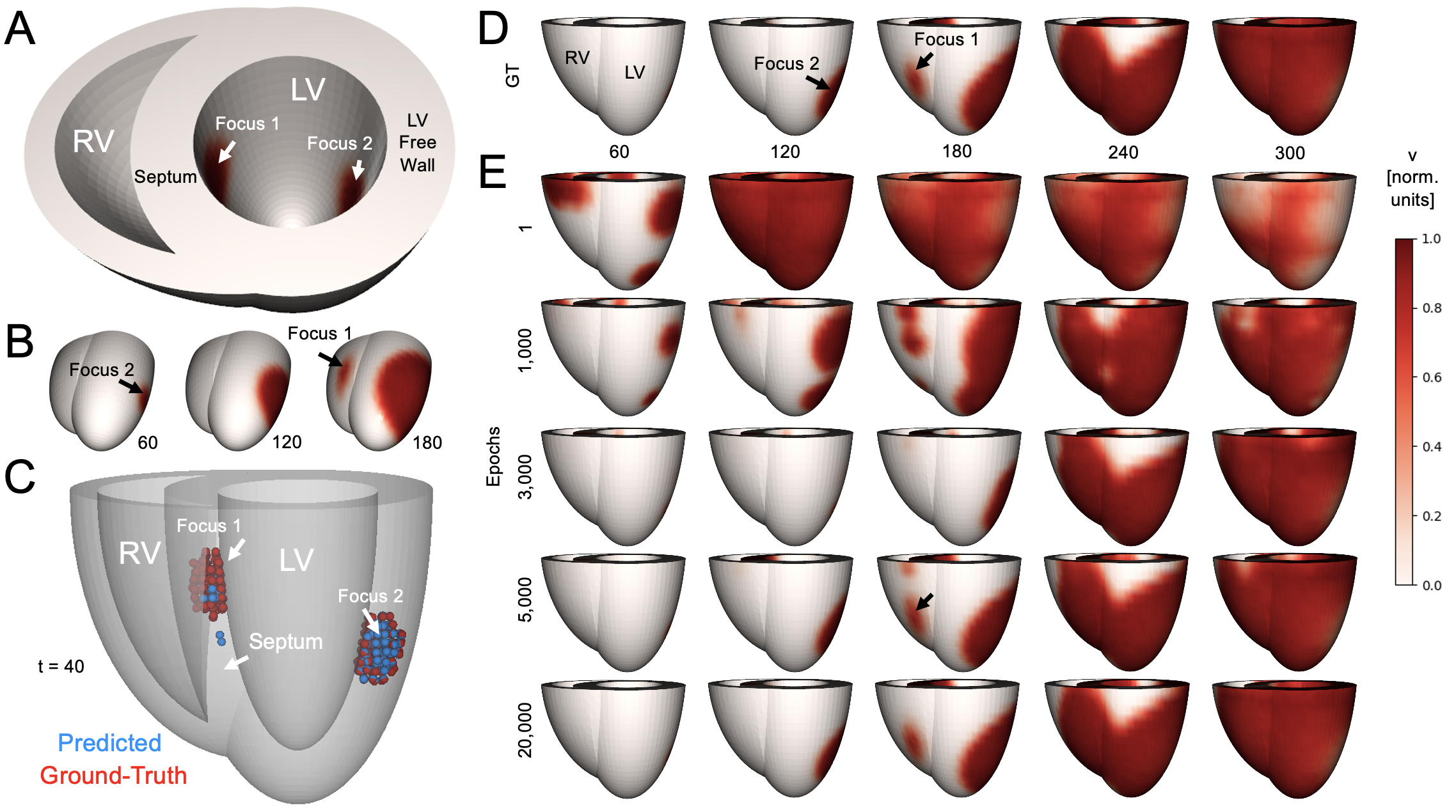}
  \caption{
Learning of transmural bi-focal action potential wave pattern from epicardial observations using differentiable smoothed particle hydrodynamics (SPH) simulations.
\textbf{A} Bi-ventricular simulation geometry. LV: left ventricle, RV: right ventricle. The two foci (red dots marked by white arrows) originate on the endocardial surface of the septum and the LV free wall.
\textbf{B} Epicardial breakthrough of focal waves (apical view). The LV and septal foci break through the surface at around 60 and 180 time steps, respectively (black arrows).
\textbf{C} Early activation sites / focal origins (red: ground-truth, blue: predicted) represented by electrically activated particles with $v>0.5$ at $t=40$ simulation time steps.
\textbf{D} Ground-truth (GT) action potential wave pattern (red: depolarized tissue) on anterior wall.
\textbf{E} Learned action potential wave dynamics after 1, 1,000, 3,000, 5,000, and 20,000 epochs in a single horizon. The observation loss is only computed across the epicardial surface (both RV and LV).
After 20,000 epochs, the ground-truth and learned wave dynamics are indistinguishable across the epicardium, see also Fig.~\ref{fig:results:SPH2}C).
 }
  \label{fig:results:SPH1}
\end{figure*}

\subsection{Recovery of Transmural Action Potential Waves inside Bi-Ventricular Geometry from Epicardial Observations}
\label{sec:results:SPH}

\begin{figure*}[htb]
  \centering
  \includegraphics[clip, trim=0.0cm 0.0cm 0.0cm 0.0cm, width=0.95\textwidth]{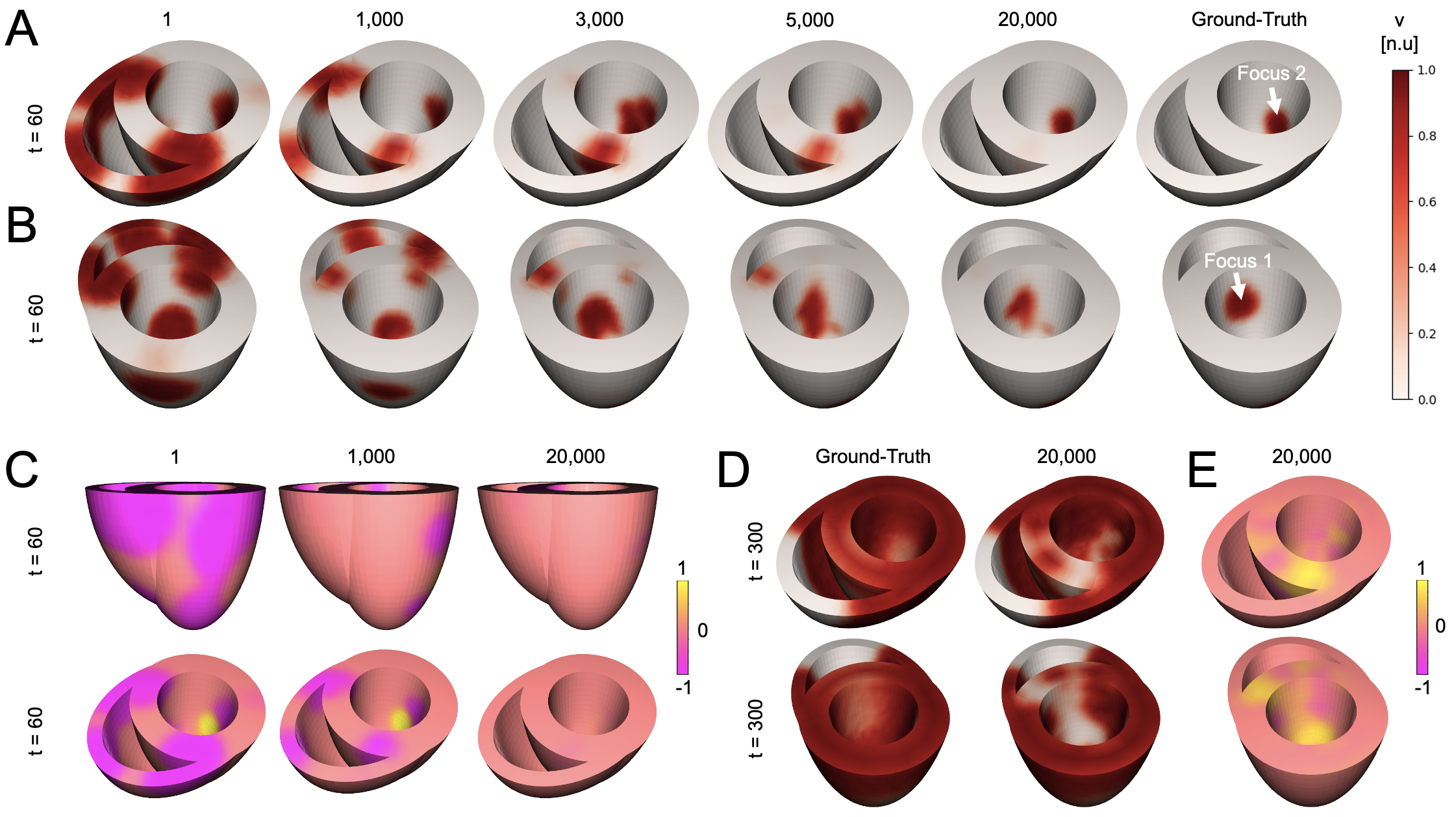}
  \caption{
Localization of endocardial early activation sites from epicardial action potential wave dynamics.
\textbf{A} View onto endocardial left ventricular (LV) free wall at $t=60$ simulation time steps during learning after 1, 1,000, 3,000, 5,000, and 20,000 epochs. The focal point source (Focus 2) emerges on the endocardium after 20,000 epochs, see also Fig.~\ref{fig:results:SPH1}E). Right: Ground-truth (GT).
\textbf{B} Corresponding view onto septum with distorted focal pattern (Focus 1) emerging after 20,000 epochs in proximity to the original source. Right: Ground-truth (GT). Data in Fig.~\ref{fig:results:SPH1}C) derived from panels A,B) at $t=40$.
\textbf{C} Difference between ground-truth (GT) and learned dynamics at $t=60$ becomes negligible after 20,000 epochs on the epicardial surface and the endocardial RV and LV free wall.
\textbf{D} Discrepancies between ground-truth and learned dynamics toward the end of the sequence at $t=300$ on the endocardial surface.
\textbf{E} Substantial differences emerge inside the septum as the learned dynamics evolve (at $t=300$). 
This indicates that activity in deeper tissue is harder to reconstruct and that single-horizon learning cannot fully recover the initial state, c.f. Figs.~\ref{fig:supplement:learningschedule} and \ref{fig:results:SPH4}B).
 }
  \label{fig:results:SPH2}
\end{figure*}

\begin{figure}[htb]
  \centering
  \includegraphics[clip, trim=0.0cm 0.0cm 0.0cm 0.0cm, width=0.48\textwidth]{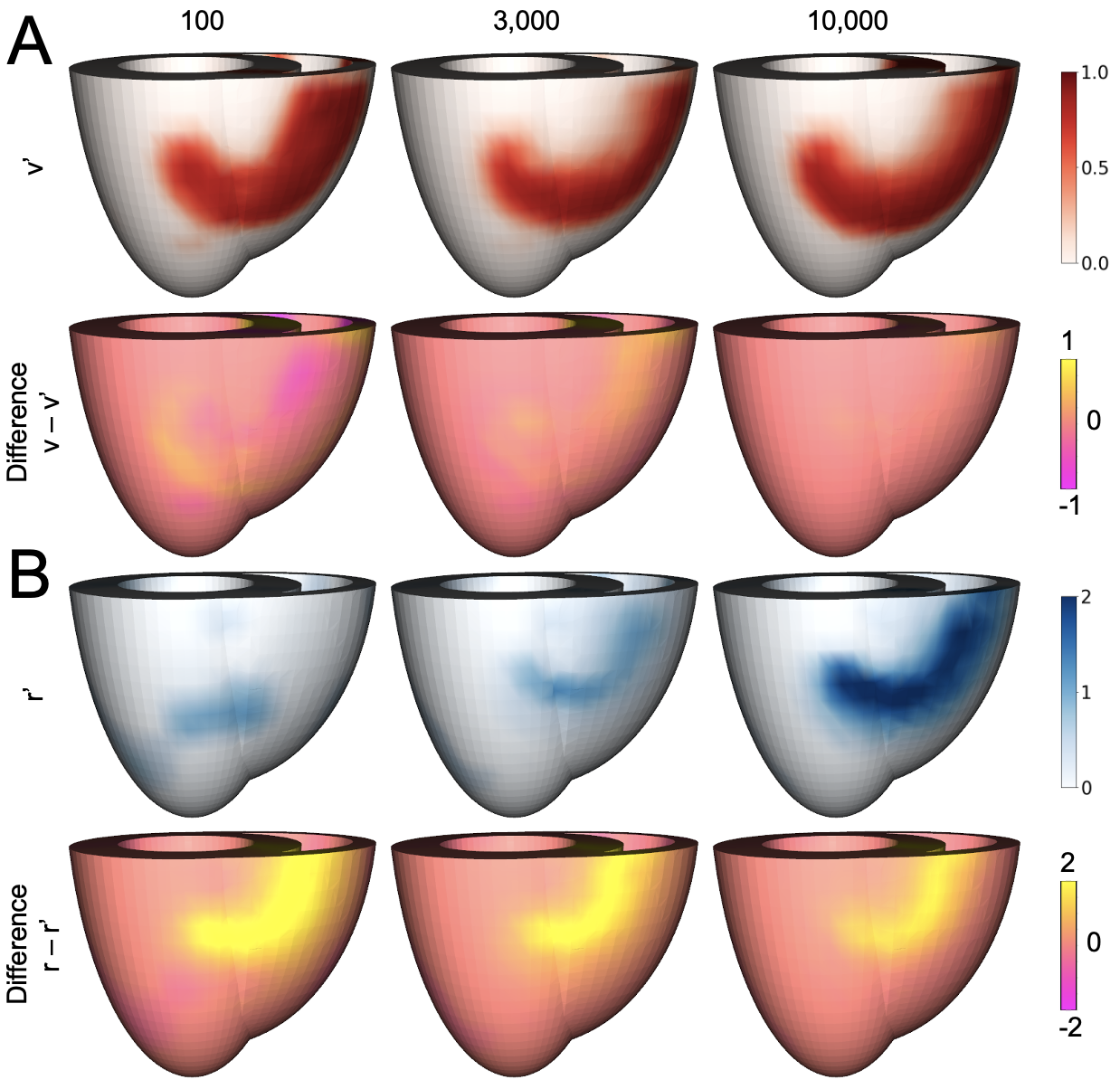}
  \caption{
Recovery of 3D electrical scroll wave from epicardial observations.
\textbf{A} Top: learned voltage variable $v'$ (red: depolarized tissue) on the epicardial surface of the posterior wall at $t=10$ simulation time steps after 100, 3,000, and 10,000 epochs of learning. Bottom: Difference to ground-truth.
\textbf{B} Top: Learned refractory variable $r'$ (blue: refractory tissue) at $t=10$ after 100, 3,000, and 10,000 epochs. Bottom: Difference to ground-truth.
While the voltage variable is directly observable, the refractory dynamics must be recovered because they cannot be directly observed. 
Learning was performed within a single horizon.
 }
  \label{fig:results:SPH3}
\end{figure}

\begin{figure}[htb]
  \centering
  \includegraphics[clip, trim=0.0cm 0.0cm 0.0cm 0.0cm, width=0.48\textwidth]{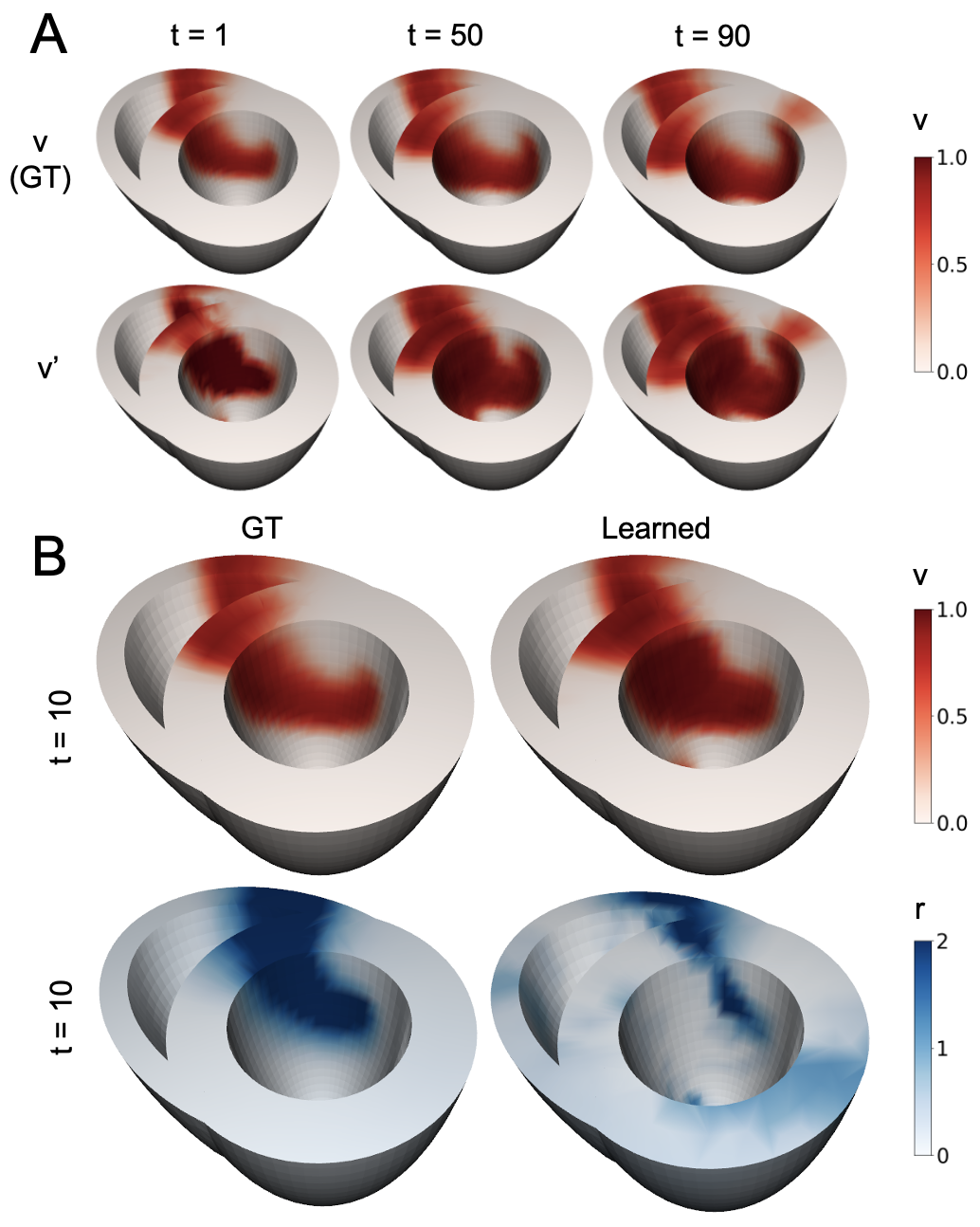}
  \caption{
Endocardial view of learned scroll wave pattern after a single horizon with 20,000 epochs.
\textbf{A} Comparison of ground-truth (GT) voltage $v$ and learned voltage $v'$ over time ($t=1, 50, 90$ simulation time steps). 
The reconstruction accuracy on the endocardium is lower because the voltage variable can only be measured on the epicardium. 
Nevertheless, both $v$ and $v'$ depict a counter-clockwise rotating scroll wave located in the septum and posterior wall, and $v'$ could be refined further during additional learning in subsequent horizons, as demonstrated in section \ref{sec:results:2Dspiral} and Figs.~\ref{fig:results:2D:spiral-divergence-loss} and \ref{fig:results:2D:spiral-divergence-timeseries}.
\textbf{B} Comparison of ground-truth (GT) and learned voltage (red) and refractory (blue) variables early in the sequence ($t=10$). 
While the reconstruction accuracy of $r$ is much better on the epicardium, see Fig.~\ref{fig:results:SPH3}B), it is poor on the endocardium. Nevertheless, the learning has produced a refractory waveback, and this pattern could be refined further as demonstrated in section \ref{sec:results:2Dspiral}.
 }
  \label{fig:results:SPH4}
\end{figure}

Figs.~\ref{fig:results:SPH1} - \ref{fig:results:SPH4} show the reconstruction of transmural action potential wave dynamics from epicardial observations using a single-horizon learning schedule.
The observation loss is calculated across the epicardial surface only, comparing the ground-truth and learned voltage values.
In Fig.~\ref{fig:results:SPH1}, the dynamics correspond to a bi-focal wave pattern that emerges after the application of two stimuli (at $t=0$) on the endocardial surface of the left ventricle (LV) in an idealized bi-ventricular geometry.
One stimulus (Focus 1) was applied to the septum, and the other (Focus 2) was applied a bit lower on the LV free wall, see Figs.~\ref{fig:results:SPH1}A,C) and \ref{fig:results:SPH2}A,B). 
The waves were simulated using the AP model, see section \ref{sec:methods:models}.
The two focal waves propagate outward through the heart walls and eventually emerge on the epicardial surface, see Fig.~\ref{fig:results:SPH1}B,D).
Using our differentiable cardiac SPH simulation, we can learn the three-dimensional transmural dynamics from observations of the wave dynamics across the epicardial surface.
The loss guiding the gradient descent is computed across the spatio-temporal sequence of epicardial action potential wave patterns.
Each epoch, the algorithm modifies the parameters $\{ \theta_1', \theta_2', ... \theta_i'\}$ and initial state $(v_0',r_0')$ at $t=0$ to minimize the loss and match subsequent simulations with the ground-truth dynamics observed on the epicardial surface, see Fig.~\ref{fig:results:SPH1}E).
To initiate the learning, the initial state $(v_0',r_0')$ was set to a random spatial pattern, as shown in Fig.~\ref{fig:methods:sph}B).
We chose this initialization intentionally to avoid making assumptions about potential subsurface patterns, such as the locations of foci.
Instead, with random initialization, the gradient-based algorithm discovers the solution fully automatically from a non-biased random initial state.
In practice, the initialization meant that the first simulations corresponded to a composition of focal and reentrant waves emerging at random locations, see first row in Fig.~\ref{fig:results:SPH1}E).
Note that the learning was performed using a single-horizon scheme, as illustrated in Fig.~\ref{fig:supplement:learningschedule}B), learning dynamics in 300 simulation time steps over 20,000 epochs.

During learning, the learned epicardial voltage wave pattern $v'$ matches the ground-truth voltage wave pattern $v$ early on because the loss is computed across the epicardial surface, see Fig.~\ref{fig:results:SPH1}E).
After 5,000 epochs, both breakthrough patterns originating from foci 1 and 2 are captured well across the entire epicardial surface.
After 20,000 epochs, the learned epicardial wave pattern is visually indistinguishable from the ground-truth wave pattern.
However, it takes longer for the learned dynamics within the heart wall to match the ground-truth dynamics, see Fig.~\ref{fig:results:SPH2}.
After 20,000 epochs, two foci have formed on the endocardial surface, one on the septum and the other on the LV free wall, close to the original ground-truth stimuli locations, see Fig.~\ref{fig:results:SPH2}A,B). 
Fig.~\ref{fig:results:SPH1}C) shows electrically active locations in the early learned states (e.g. at $t=40$, $v>0.5$ per particle), which roughly match the original foci locations.
While the focal pattern in the septum is distorted, it is captured well in the LV free wall, presumably because it is located closer to the epicardial surface.
We obtained similar results when applying a single stimulus either in the LV or the septum.

After only 1 horizon, some of the parameters are recovered reasonably well (e.g. $a$: 1.8\% mismatch), while others still have to converge (e.g. $\mu_1$: 19\% mismatch).
We assume that the foci locations could be refined further by using a multi-horizon scheme, as used in section \ref{sec:results:2Dspiral}.
The learned initial state after one horizon, which is approximately shown in Fig.~\ref{fig:results:SPH2}A,B) at 20,000 epochs, is a much better starting point than the random initialization used at the beginning of the learning. 
Subsequenty, it could be used to start another horizon, and a sequence of horizons would yield much better, refined reconstructions of the foci, similarly as shown for the 2D focal pulse in Fig.~\ref{fig:methods:initialcondition}.
Nevertheless, the multi-horizon schedule is only effective if subsequent horizons are shifted by a few time steps and include additional future time steps, see Fig.~\ref{fig:supplement:learningschedule}D).
Therefore, further refinements would require more data or a high temporal sampling rate, which may be a limitation with focal wave patterns measured across a finite number of samples.
However, if the sole objective is to determine early activation sites, a single horizon may already be sufficient to determine those, as demonstrated in Fig.~\ref{fig:results:SPH1}C).
A multi-horizon scheme is beneficial if the objective is to forecast sustained dynamics, such as spiral or scroll waves, see sections \ref{sec:results:2Dspiral} and \ref{sec:results:3Dbulk}.

We found that the horizon duration over which learning of the intramural wave sources is performed is critical.
The sequence in Figs.~\ref{fig:results:SPH1} and \ref{fig:results:SPH2} includes 300 simulation time steps, which we learned in a single horizon.
We found that, if the sequence included only 200 time steps, and we aimed to recover a single pulse originating in the septum, then the learned dynamics would decay, because there are no waves on the epicardial surface (where the loss is computed) until later in the sequence.
The breakthrough of the action potential wave occurs only at $t=180$ time steps.
This is too late for the gradient-based optimization: the gradient descent scheme has already downregulated the system's excitability, the wave decays, and the optimization gets stuck in a local minimum.
However, with 300 simulation time steps, the pulse is recovered correctly.

Note that in Fig.~\ref{fig:results:SPH2}C), towards the end of the sequence at $t=300$ simulation time steps, the difference between the learned and ground-truth epicardial wave patterns is negligible, while in panels D,E), the endocardial / intramural wave pattern exhibits substantial differences. 
These differences within the heart walls arise from small differences in the learned initial state, which lead to larger divergence and errors over time. 
The finding highlights that, even though the epicardial loss is small, the learned intramural wave pattern is not necessarily accurate and might be a degenerate solution.
Additional horizons, boundary conditions imposed on the dynamics, or further assumptions, such as a limited number of wave sources, could help to alleviate such issues.

Next to focal patterns, we also reconstructed a 3D electrical scroll wave within the bi-ventricular geometry, see Figs.~\ref{fig:results:SPH3} and \ref{fig:results:SPH4}.
As with the focal wave, the learning was performed only within a single horizon.
Fig.~\ref{fig:results:SPH3}A) shows that the scroll wave shape can be recovered across the epicardium almost immediately (<100 epochs), because the voltage pattern is directly observable. 
On the other hand, the refractory dynamics cannot be directly observed and must be learned.
Accordingly, panel B) shows the refractory variable as it is learned over the epochs until it forms the refractory waveback.
Note that the panels show the scroll wave early in the sequence at $t=10$.
At later time steps and at fewer epochs (< 3000 epochs) the wave disintegrates similarly as the focal waves shown in Fig.~\ref{fig:results:SPH1}E).
After 10,000 epochs, the scroll wave becomes a stable, rotating pattern on the epicardium.
As in the focal case, it takes longer for the learned intramural dynamics to match the ground-truth dynamics, see Fig.~\ref{fig:results:SPH4}.
After 20,000 epochs, the learned voltage variable $v'$ forms a scroll wave-like shape on the endocardium that roughly matches the original scroll wave, even in the septum, and rotates at the same speed in the same direction.
Nevertheless, the learned refractory variable $r'$ on the endocardium is only a coarse approximation of the ground-truth pattern, and would require further refinement in subsequent horizons.
Again, the learning would benefit from a multi-horizon learning schedule, as it is demonstrated in the next section \ref{sec:results:3Dbulk}.

\subsection{Recovery of 3D Scroll Wave Dynamics from Surface Measurements in 3D Bulk}
\label{sec:results:3Dbulk}

\begin{figure}[htb]
  \centering
  \includegraphics[clip, trim=0.0cm 0.0cm 0.0cm 0.0cm, width=0.48\textwidth]{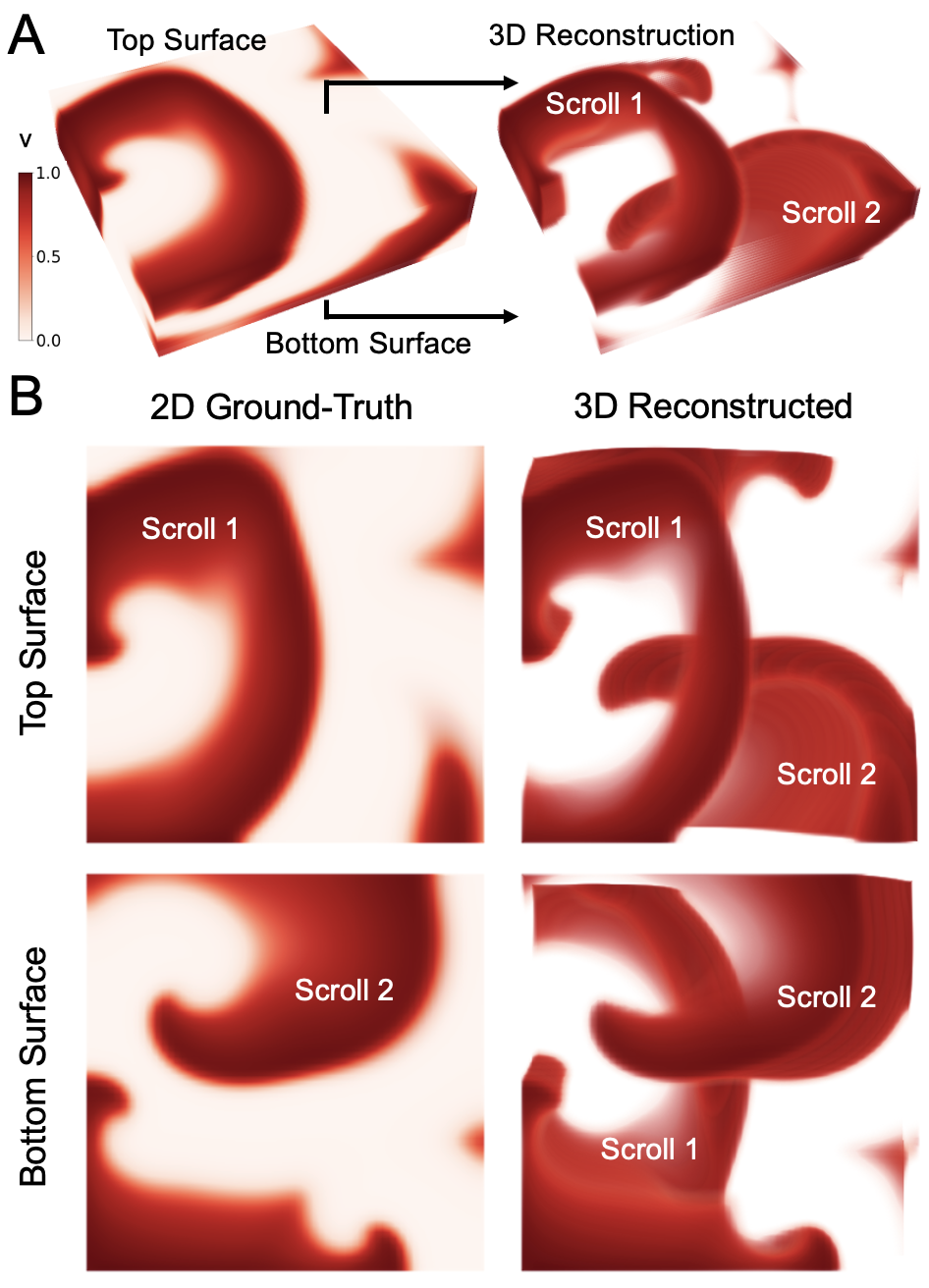}
  \caption{
Reconstruction of 3D scroll wave dynamics inside a bulk/slab from observations of its top and bottom surfaces. 
\textbf{A} Comparison of ground-truth voltage $v$ (left, opaque) across surface and reconstructed 3D voltage $v'$ dynamics (right, translucent) throughout volume.
The configuration mimics simultaneous epi- and endocardial measurements of fibrillatory action potential wave dynamics in the atrial or ventricular wall.
\textbf{B} View of top and bottom surfaces with fully dissociated wave patterns (left) and corresponding views of reconstructed volumetric 3D scroll wave pattern (right).
Overall, the parameters can be estimated with a residual error of less than $0.5\%$, see table \ref{tab:parameters-3d-bulk-learned}, and the reconstructed dynamics are visually indistinguishable from the ground-truth, see Fig.~\ref{fig:results:scroll}.
 }
  \label{fig:results:scroll-top-bottom}
\end{figure}

\begin{figure}[htb]
  \centering
  \includegraphics[clip, trim=0.0cm 0.0cm 0.0cm 0.0cm, width=0.48\textwidth]{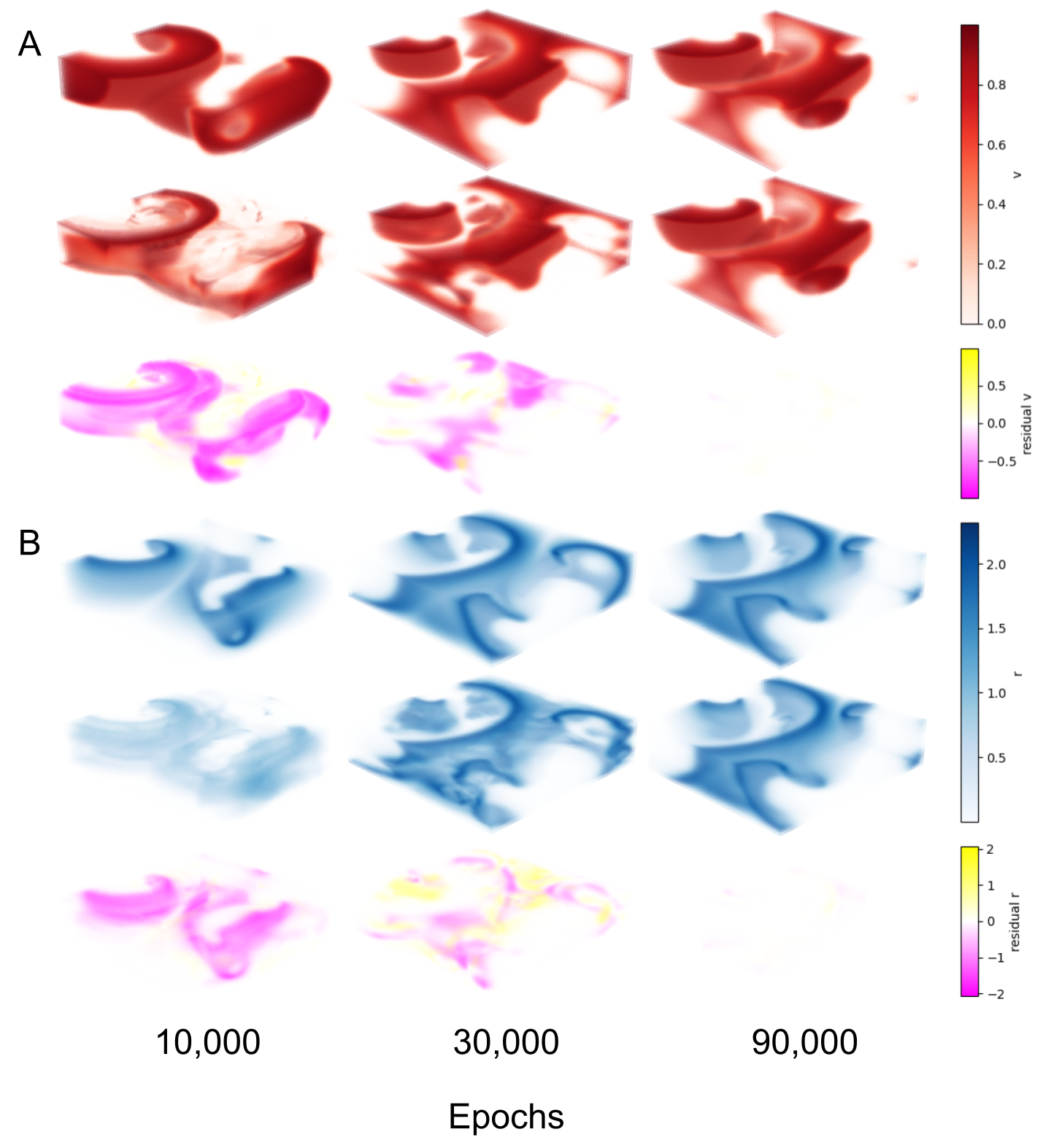}
  \caption{
Learning of scroll wave dynamics by observing 3-4 rotations across the bulk's top and bottom surfaces.
\textbf{A} Ground-truth 3D scroll wave dynamics (red: voltage, first row), recovery of the scroll wave pattern during learning over $55$ horizons shown at $10,000$, $30,000$, and $90,000$ epochs (second row), and residuals between ground-truth and learned voltage $v$ (third row).
The residual MSE per voxel is negligible in the order of $10^{-4} - 10^{-3}$ in the interior of the bulk.
Learning AP scroll waves with the AP model yields near-perfect reconstructions.
\textbf{B} Corresponding panels for the refractory variable $r$.
  }
  \label{fig:results:scroll}
\end{figure}

Figs.~\ref{fig:results:scroll-top-bottom} and \ref{fig:results:scroll} demonstrate the recovery of 3D scroll wave dynamics in a $64 \times 64 \times 16$ bulk medium with isotropic conduction, in which only the top and bottom surfaces are observed.
The observation loss is computed exclusively across the two $64 \times 64 \times 1$ top and bottom surface layers, comparing the ground-truth and learned voltage values $v$ and $v'$, while the interior $14$ layers remain entirely unobserved (bulk is opaque) throughout training, see Fig.~\ref{fig:results:scroll-top-bottom}A).
Because the interior initial condition cannot be directly measured, it was initialized as a linearly weighted average of the two observed surfaces, with weights proportional to proximity: the center layer receives equal weight from both surfaces ($0.5/0.5$), while a layer at depth 3 is weighted $0.25/0.75$, and so forth.
As in previous sections, the refractory variable was estimated on the surface according to eq.~(\ref{eq:init2D}).
The parameters $\theta = \{D, a, \epsilon_0, k, \mu_1, \mu_2\}$ were initialized between $6-10\%$ off from the true values.

The ground-truth system was simulated with the Aliev-Panfilov (AP) model and was initialized using a scroll wave with a slightly tilted vortex filament between the top and bottom surfaces, and then evolved forward by 1,000 time steps, producing the more complex chaotic dynamics shown in Fig.~\ref{fig:results:scroll}A) (top row).
The chaotic episode was then used as the ground-truth for learning with the first chaotic state as the ground-truth initial condition.
The training window included $0$--$90$ time units, corresponding to approximately $3$--$4$ scroll wave rotations, with each rotation taking approximately $26$ time units. 
During this training window, the top and bottom surfaces were dissociated, each surface exhibiting substantially different wave patterns composed of 1-3 interacting spiral waves, similarly as shown in the left panel in Fig.~\ref{fig:results:scroll-top-bottom}B).
Figs.~\ref{fig:results:scroll-top-bottom} and \ref{fig:results:scroll} show the learned scroll wave dynamics after the learning was completed, and the true and learned dynamics are co-evolved into the future for comparison.
This learning setup is particularly challenging because the surface observations reflect qualitatively very different wave patterns, and the interior dynamics connecting them must be inferred entirely from the two opposing surfaces without direct observation of the 3D dynamics.

We used a multi-horizon learning schedule comprising four "warm-up" horizons of lengths $20$, $25$, $30$, and $35$ time steps (all starting at $t=0$, totaling $8,000$ epochs), followed by $51$ rolling horizons of length $40$ time steps each, with each successive horizon propagated forward by one time step (i.e., $t=0$--$40$, $1$--$41$, $2$--$42$, \ldots), for a total of $110,000$ epochs ($8,000$ warm-up and $102,000$ training), see also table \ref{tab:learning-schedule3Dscroll}.
Each horizon was trained for $2,000$ steps.
Between horizons, the learned system is propagated forward using the learned initial conditions and parameters, and then anchored to the true observed surface voltage values $v$ at the new horizon boundaries.
As in the 2D case, the rolling multi-horizon schedule is critical for learning accurate dynamics: it enables the loss signal from the observed surfaces to be progressively propagated into the unobserved interior as parameter estimates improve.

The multi-horizon training scheme plays an especially important role in recovering the unobserved interior initial conditions.
Supplementary Video 1 visualizes the learned initial conditions across successive horizons, and shows that the interior initial conditions begin as rigid and spatially discontinuous, reflecting little more than the linear interpolation used at initialization. 
However, these patterns gradually smooth into a realistic and physically plausible interior state that closely matches the ground-truth scroll wave pattern.
This progressive refinement is only possible because improved parameter estimates allow the surface anchor values to be accurately reflected back into the interior through forward simulation.
Fig.~\ref{fig:results:scroll} illustrates this convergence: the top row in panel A) shows the ground-truth scroll wave dynamics (red: voltage), the second row shows the recovery of the scroll wave pattern within the differentiable simulation over the course of $55$ horizons, and the third row shows the voxel-wise difference (residuals) between the ground-truth voltage $v$ and learned voltage $v'$.
Panel B) shows the corresponding rows for the refractory variable $r$.
Both panels display snapshots at $10,000$, $30,000$, and $90,000$ training steps, and the learned initial conditions approach the ground-truth at approximately $90,000$ training steps ($45$ horizons).

After training, all learned parameters converged to within $0.003 - 0.3\%$ of the true parameter values, comparable to the near-exact recovery observed in the 2D spiral wave experiments, see table~\ref{tab:parameters-3d-bulk-learned}.
The internal scroll wave dynamics can be reconstructed exactly with residual errors (MSE: mean square error) per voxel in the order of $10^{-4} - 10^{-3}$ in the interior of the bulk.
The rapid convergence in this setting is notable given the limited surface-only observations and the high-dimensional interior state that must be inferred.
To evaluate long-term predictive accuracy, we ran both the learned and ground-truth systems forward for $1,000$ time steps beyond the training window (approximately $40$ spiral rotations), as shown in Supplementary Video 2.
Over this extended period, both systems dissociate into a single meandering spiral wave with a filament that is approximately straight through the bulk and consistently positioned between the top and bottom surfaces.
No significant differences between the learned and true system are observed across the full $1,000$ time steps, demonstrating that the recovered parameters and initial conditions are accurate enough to sustain precise long-range forecasts even for complex three-dimensional scroll wave dynamics observed only at the boundaries.

\begin{figure*}[htb]
  \centering
  \includegraphics[clip, trim=0.0cm 0.0cm 0.0cm 0.0cm, width=0.8\textwidth]{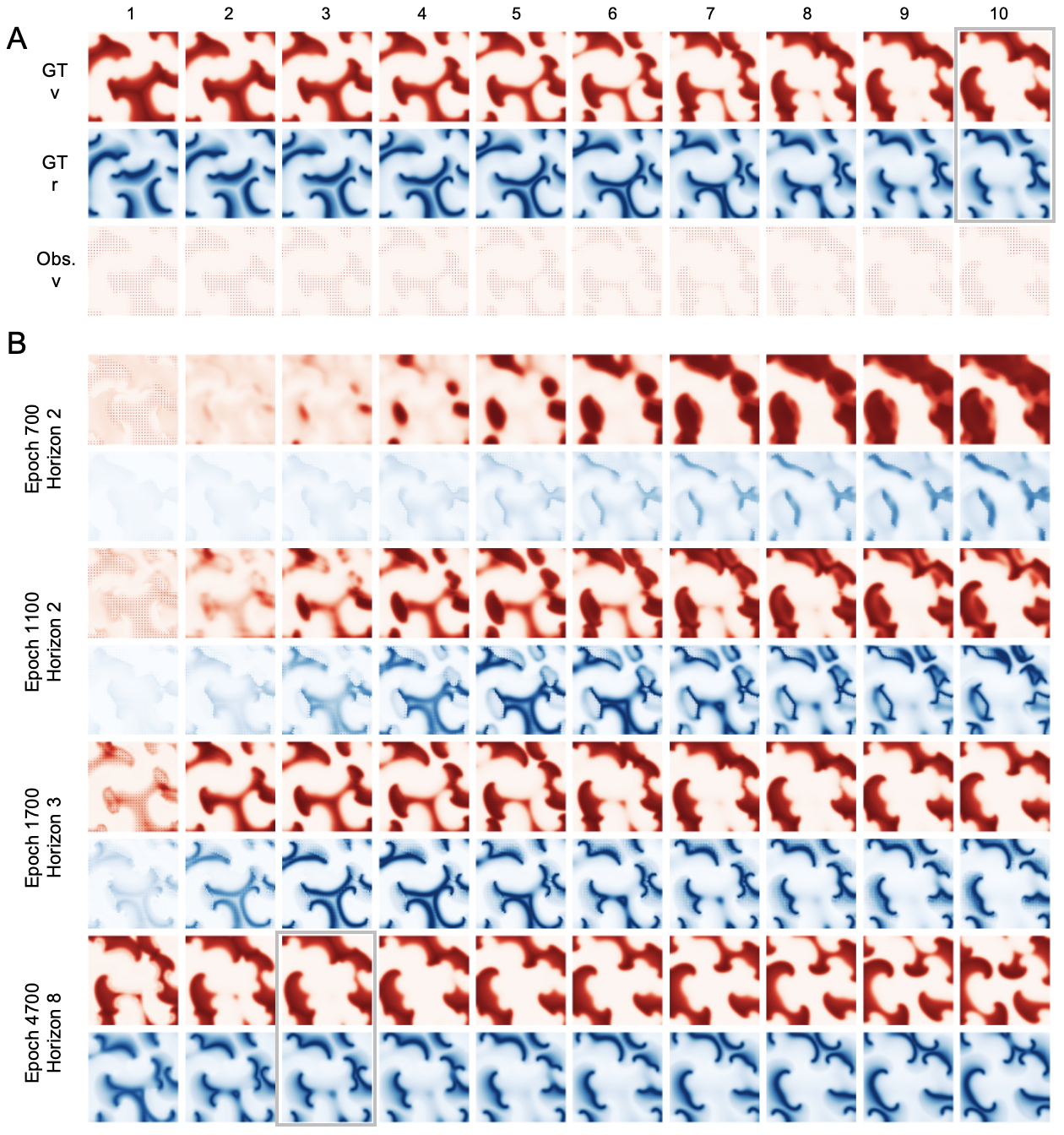}
  \caption{
State and parameter recovery with sparse observations. 
\textbf{A} Ground-truth (GT) multi-spiral wave dynamics showing voltage $v$ (red) and refractory variable $r$ (blue) of the AP model. 
The observations $\bar{v}$ were limited to a grid (subsampled every 4th pixel of the voltage variable, 0 otherwise) to mimic sparse measurement data obtained with a multi-electrode array.
\textbf{B} Learning and recovery of the dynamics over horizons and cumulative number of epochs (600 epochs per horizon in the first 10 horizons).
In each horizon, the observation window is shifted one time step into the future. 
Therefore, the 3rd frame in horizon 8 is the 10th frame in the ground-truth data (gray rectangle).
The final 2nd state in the final epoch of one horizon is used to construct the initial state in the next horizon, as shown in Fig.~\ref{fig:supplement:learningschedule}D), where it is mixed with the observation $\bar{v}$ at every 4th pixel.
While the state estimation is mostly completed within $\sim 5,000$ epochs, the parameter estimation converges and provides sufficiently accurate parameter estimates after $\sim 40,0000$ epochs, see Fig.~\ref{fig:barplots}C) and table \ref{tab:parameters-2D-learned}.
With sparse measurements, recovery only succeeds by adding a smoothness loss for the refractory variable.
 }
  \label{fig:results:sparsity}
\end{figure*}

\subsection{Sparse Measurements}
\label{sec:results:sparse}

Fig.~\ref{fig:results:sparsity} demonstrates that 2D spiral wave dynamics can be recovered from sparse measurements or observations. 
Panel A) shows the ground-truth (GT) multi-spiral wave dynamics (voltage variable: red, refractory variable: blue) obtained with the Aliev-Panfilov (AP) model together with the sparse observations of the voltage variable $\bar{v}$.
Here, we measured voltage with 'electrodes' distributed on a regular grid, subsampling the full-resolution dynamics at every 4th pixel, or at $1/4$ of the original resolution, or across $32 \times 32$ electrodes.
The data mimics measurements obtained with multi-electrode arrays.
The initial condition was set up as in previous sections, using eq.~(\ref{eq:init2D}), but only for pixels located at the electrode positions. 
All other pixels were initially set to zero.
During learning, the observation loss was only computed at the electrode locations.
At the beginning of each new horizon, the learned voltage variable $v'$ was updated with voltage observations $\bar{v}$ from the electrode locations.

To counteract artifacts associated with the sparsity, we used a smoothness loss that enforces smooth spatial gradients in the field of the learned refractory variable $r'$, as described in section \ref{sec:methods:learn:loss}. 
The smoothness loss was introduced in addition to the observation loss and other regularization losses, and affected only the refractory variable.
We set $\alpha=0.01$ to avoid inhibiting or completely smoothing out the reaction-diffusion dynamics.
With small $\alpha$, we inhibited only strong discontinuities in $r'$ between pixels associated with the grid and the other pixels, while the dynamics themselves remained unaffected.
Without the smoothness loss, the dynamics could not be learned at all; instead, we obtained strong spatial discontinuities in the learned dynamics, resembling the electrode grid pattern.

Fig.~\ref{fig:results:sparsity}B) shows that even complicated multi-spiral wave dynamics can be learned with sparse observations.
The learned spiral wave pattern is visually indistinguishable from the ground-truth spiral wave pattern within 8 horizons or fewer than 5,000 epochs.
Nevertheless, learning with sparse measurements takes longer: the same dynamics can be fully learned within about 20,000 epochs with full-resolution data, but learning with sparse observations requires more epochs to achieve similar accuracies.
Table \ref{tab:parameters-2D-learned} provides a comparison of the residual error of the learned parameters with full resolution learning versus with sparse observations, see also Fig.~\ref{fig:barplots}C).
In particular, the errors for the parameters $\{ \epsilon_0, \mu_1, \mu_2 \}$ stay in the single to double digit percent range after 20,000 epochs. 
By comparison, at full resolution, the errors are negligible after 20,000 epochs for the same dynamics.
However, the errors continue to decrease further with continued learning, and eventually, if evolved into the future, the reconstructed sparse dynamics co-evolve congruently with the ground-truth dynamics over many rotations, similarly as shown in Figs.~\ref{fig:results:2D:spiral-divergence-loss} and \ref{fig:results:2D:spiral-divergence-timeseries} in section \ref{sec:results:2Dspiral}.

\subsection{Cross-Model Fitting}
\label{sec:results:crossmodel}
\begin{figure}[htb]
  \centering
  \includegraphics[clip, trim=0.0cm 0.0cm 0.0cm 0.0cm, width=0.48\textwidth]{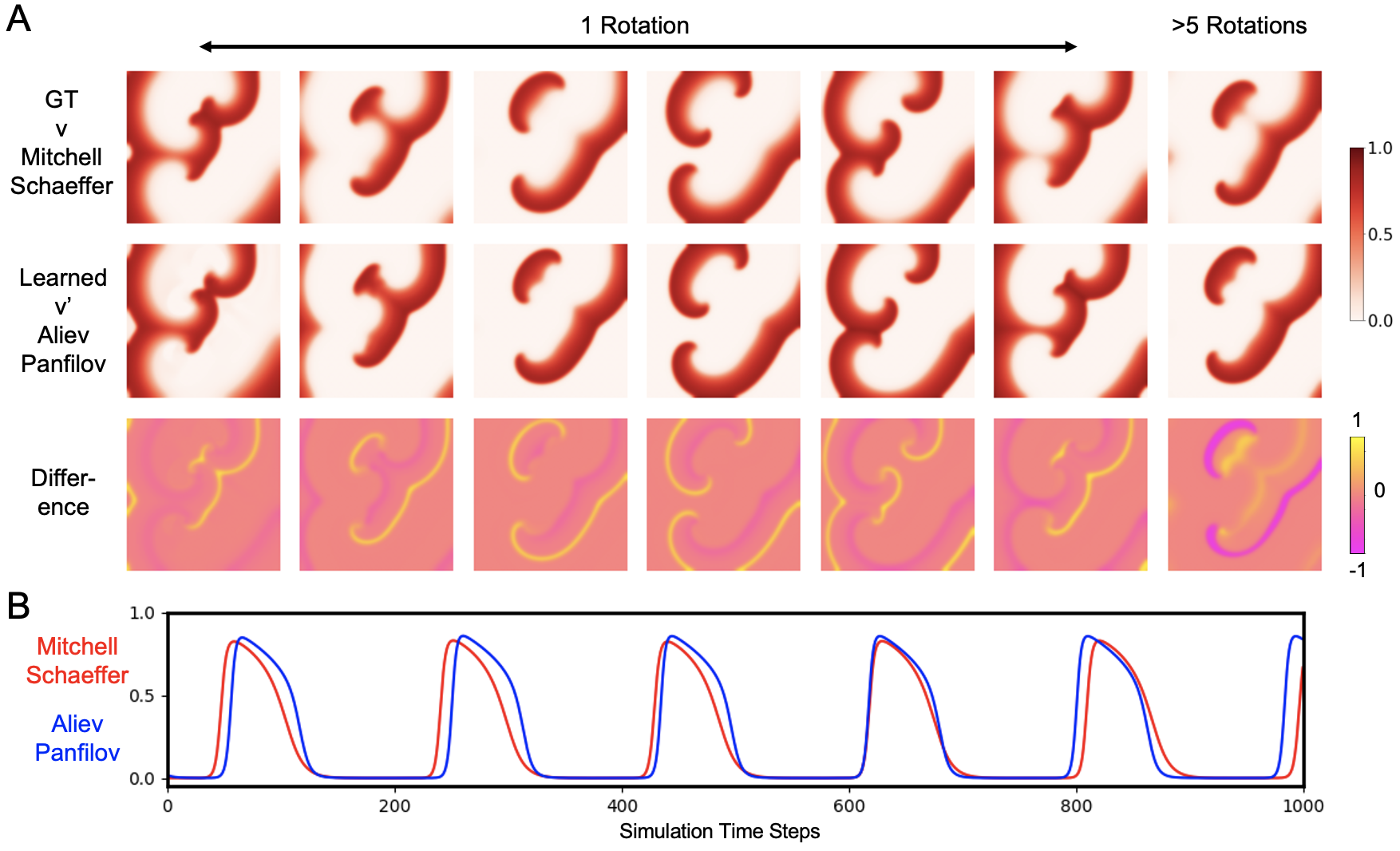}
  \caption{
Cross-model learning: fitting the Aliev-Panfilov (AP) model to spiral wave data generated with the Mitchell-Schaeffer (MS) model.
\textbf{A} Comparison of ground-truth (GT) MS voltage data (top) versus fitted AP simulation (center) and pixel-wise difference (bottom). While the MS wave pattern's topology is captured well, there are residual mismatches between the MS and the fitted AP dynamics visible at the wave fronts and backs.
\textbf{B} Trace of the MS (red) and AP (blue) models sampled from a single location of the 2D dynamics.
 }
  \label{fig:results:MS-AP}
\end{figure}

In sections \ref{sec:results:2Dfocus}-\ref{sec:results:sparse}, we assumed that the model equations are known and fitted a differentiable version of the Aliev-Panfilov (AP) model to spatio-temporal data generated with the AP model.
Here, we simulated spiral wave dynamics with the Mitchell-Schaeffer (MS) model and subsequently fitted the AP model to the MS data, see Fig.~\ref{fig:results:MS-AP}.
Fig.~\ref{fig:results:MS-AP}A) shows a comparison of ground-truth (GT) MS multi-spiral wave data and the learned or fitted AP simulation after the learning process was completed, and the two patterns co-evolve independently.
Qualitatively, the MS wave pattern is captured well by the AP model, but residual discrepancies remain due to intrinsically different characteristics of the two models, see also the comparison of the traces in panel B).
Accordingly, the pixel-wise difference map in panel A) (bottom) shows mismatches at the wave fronts and backs.
Importantly, to be able to fit one model to another, we first had to fit a sequence of action potentials in a single cell (0D), then action potential waves in a cable (1D) ring (with periodic boundary conditions), before fitting the 2D spatio-temporal dynamics with the obtained parameter estimates. 
Before performing the 2D fit, we also used a DDPM, see section \ref{sec:methods:initialization}, to generate an appropriate initial condition for the AP model.
All fits were performed fully automatically using gradient-based optimization enabled by the differentiable simulations.
Direct 2D spatio-temporal fitting, as in sections \ref{sec:results:2Dfocus}-\ref{sec:results:sparse}, with arbitrary initial parameter guesses, did not succeed.

\subsection{Fitting Spiral Waves in a Cardiac Monolayer Cell Culture}
\label{sec:results:experiments}
\begin{figure}[htb]
  \centering
  \includegraphics[clip, trim=0.0cm 0.0cm 0.0cm 0.0cm, width=0.48\textwidth]{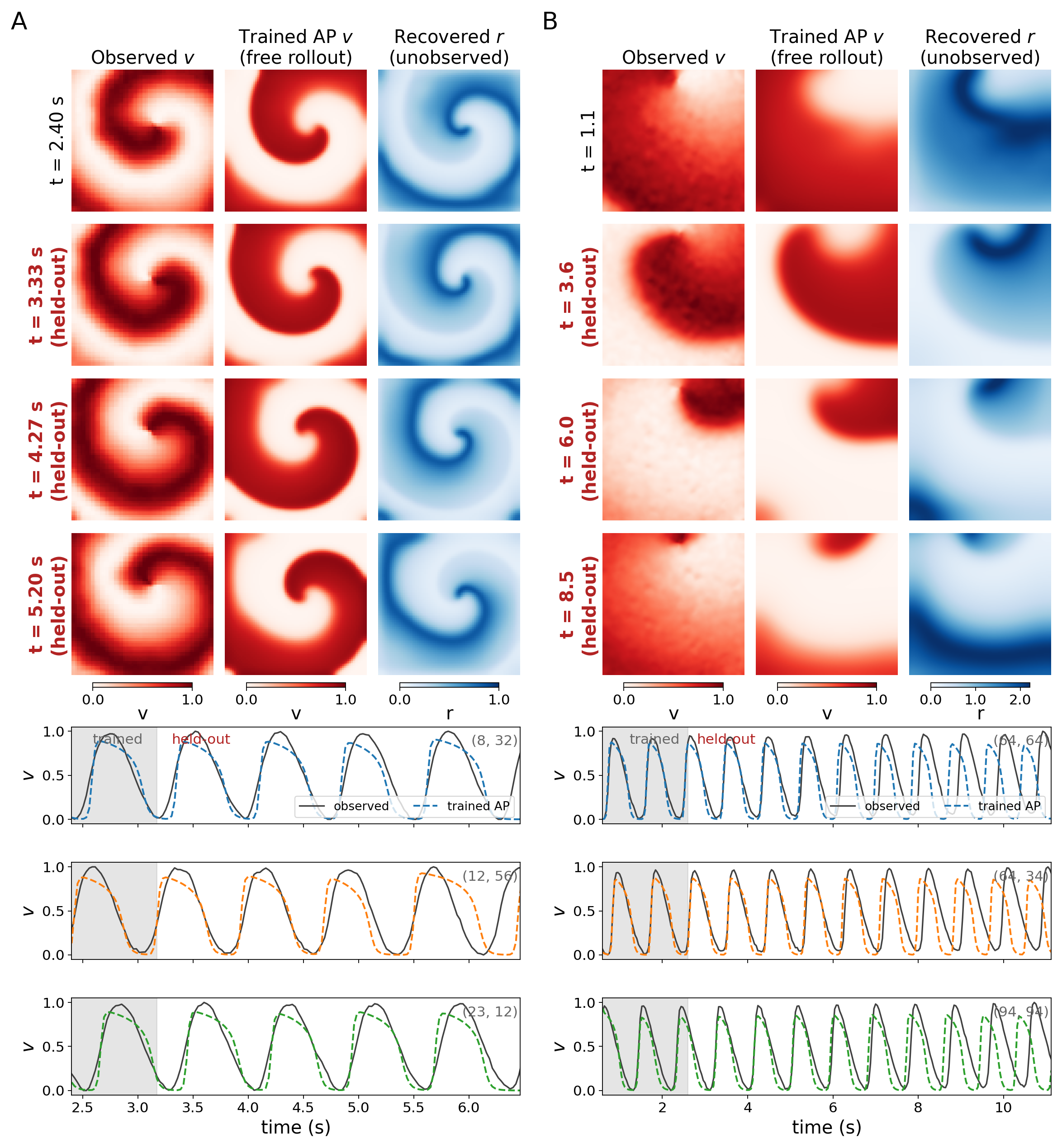}
  \caption{
Automated fitting to a spiral wave in two different monolayer cell cultures using differentiable cardiac electrophysiology simulations.
\textbf{A} Left: Counter-clockwise rotating calcium spiral wave (which we use as a proxy for voltage, red: depolarized, white: resting, normalized units) in the petri dish imaged using calcium-sensitive fluorescent dye. Center and Right: Fitted spiral (red: excitatory variable, blue: refractory variable) in the Aliev-Panfilov (AP) model at the end of the learning within 1-2 rotations (approx.). A diffusion model (DDPM) was used to generate an appropriate initial condition for the AP model.
\textbf{B} Left: Clockwise rotating voltage spiral wave (red: depolarized, white: resting, normalized units) in the petri dish imaged using voltage-sensitive fluorescent dye. Center and Right: Fitted spiral (red: excitatory variable, blue: refractory variable) in the Aliev-Panfilov (AP) model at the end of the learning, see also Supplementary Video 3. Learning occurs within 2 spiral rotations (approx.). A perceptual loss using the V-JEPA framework and a larger simulation domain enable fitting with noise, even when the rotor core is located close to the field-of-view boundary.
 }
  \label{fig:results:cellculture}
\end{figure}

Fig.~\ref{fig:results:cellculture} shows two fits of the AP model to optical mapping data of two spiral waves in two separate experiments: one showing a counter-clockwise and the other a clockwise rotating spiral wave.
The data was obtained in two different cardiac monolayer cell cultures, see section \ref{sec:methods:cellculture}.
The first spiral wave in panel A) was imaged using calcium-sensitive dye, which we used as a proxy for a voltage wave.
The second spiral wave in panel B) is a voltage or action potential spiral wave imaged using voltage-sensitive dye.
In both cases, the spiral wave dynamics were learned over 1-2 rotations using a differentiable AP model, after which the original spiral wave in the cell culture and the learned simulated spiral wave then co-evolve congruently for about 4-5 rotations before the simulated AP spiral slowly diverges.
In one case the simulated spiral wave is slightly slower and in the other case slightly faster.
Both the original and simulated spiral waves rotate in the same counter-clockwise or clockwise directions around a phase singularity located either at the center or near the boundary.
Both real and simulated spiral waves have a comparable action potential duration or wavelength.
Both regimes in panels A) and B) are very different single spiral wave regimes.
The fits were performed fully automatically using a multi-horizon learning schedule, directly with the 2D data.
The imaging data was pixel-wise normalized, and contains noise and mild imaging artifacts.

Importantly, we found that action potential wave patterns in experimental recordings cannot simply be fitted using a pixel-wise observation loss as with the simulated data.
To fit the PDE model to the experimental data in panel A), the DDPM approach described at the end of section \ref{sec:methods:initialization} needed to be applied during initialization and repeatedly at the beginning of each new horizon during the first few horizons to reproject the learned initial state back onto the AP manifold.
This scheme provided the initial state and reset or constrained subsequent learned initial states to avoid degenerate states, while the gradient-based learning refined these states to match the observations.
To fit the PDE model to the experimental data in panel B), three adaptations were required:
(1) First, because the phase singularity of the rotating spiral wave may lie near the edge of the field of view, the PDE was integrated on a spatially expanded domain whose boundaries were positioned automatically so that the spiral tip, which was detected as the centroid of the top-1\% temporal-mean $|\nabla v|$, fell within the central third of the simulation grid.
Initial conditions in the expanded margin were filled by linearly blending from the nearest observation-boundary pixel ($w=1$) to a quiescent rest state ($w=0$), and the observation loss was computed only over the original recording footprint by cropping the expanded simulation output.
(2) Second, pixel-wise loss (MSE) alone proved insufficient to guide optimization. 
To address this, we augmented the total loss with a perceptual loss $\mathcal{L}_p$, see also section \ref{sec:methods:learn:loss}.
The total loss for the experimental fit was $\mathcal{L} = \mathcal{L}_{obs} + \mathcal{L}_{reg} + w_p \cdot \mathcal{L}_p$ with $w_p = 2.0$ and the phase loss disabled throughout.
(3) Third, the model parameters were optimized in log-space via the bounded parameterization described in section \ref{sec:methods:diffsim:finitedifferences}, with wider bounds than with the simulated twin AP/AP data to reflect the absence of a ground-truth reference ($D \in [10^{-4}, 0.05]$, $\varepsilon_0 \in [10^{-3}, 0.5]$, $a \in [0.01, 1.0]$, $k \in [2, 20]$, $\mu_1, \mu_2 \in [0.01, 0.5]$).
Training proceeded in two sequential stages: an initial run using a progressively expanding horizon schedule (starting at $[0, 10]$ frames and expanding to spans of 51 frames while sliding forward one frame per horizon, with up to 1200 epochs per horizon), followed by a second run initialized from the learned parameters of the first, applying the same expanding-then-sliding schedule to further refine the fit.

\section{Discussion}
\label{sec:discussion}

We provide a proof-of-principle that differentiable cardiac electrophysiology simulations are an effective tool for fitting biophysical models to spatio-temporal data of action potential wave dynamics. 
Using auto-differentiation techniques and gradient descent, it becomes possible to reproduce various normal and abnormal electrical rhythms, such as focal, spiral, or scroll waves in 2D and 3D tissues from partial, sparse, and noisy measurements.
The differentiable SPH simulations in bi-ventricular geometries demonstrate that intramural wave sources can be identified from epicardial measurements, even when they lie inside the septum, far from the epicardium.
The differentiable 2D and 3D slab simulations demonstrate that even complex reentrant spiral and scroll wave dynamics can be inferred from sparse or surface measurements, respectively.
With simulated data, it is possible to retrieve accurate state and parameter estimates, with assumed general model structure (equations), see sections \ref{sec:results:2Dfocus}-\ref{sec:results:2Dspiral}, and moderate state estimates when the equations are unknown and approximated with distinct model equations, see section \ref{sec:results:crossmodel}.
The {\it in vitro} example in Fig.~\ref{fig:results:cellculture} demonstrates that the technique can be applied to experimental and not just simulated data.
Our results suggest that, in future work, differentiable simulations could be used to recover transmural wave dynamics throughout the heart muscle from optical mapping, multi-electrode array (MEA), or catheter mapping recordings during sinus rhythm, premature atrial or ventricular complexes, or reentrant rhythms, such as atrial or ventricular tachycardia or even fibrillation.

The provided examples illustrate potential future {\it ex vivo}, {\it in vitro}, or clinical imaging applications.
The sparse observations across a grid in section \ref{sec:results:sparse} mimic recordings obtained with a MEA. 
Even sparser, more irregular sampling could correspond to catheter mapping.
The epi- and/or endocardial observations in sections \ref{sec:results:SPH} and \ref{sec:results:3Dbulk} could be obtained with panoramic 3D optical mapping~\cite{Chowdhary2026} or simultaneous dual-surface optical mapping of the epi- and endocardium \cite{Mitrea2009}.
The settings match those found in imaging experiments:
Planar or focal electrical waves can be imaged across the heart surface during ventricular pacing or premature ventricular complexes. 
In Figs.~15-17 in Chowdhary \& Lebert et al.~\cite{Chowdhary2026}, a focal wave was imaged using panoramic voltage-sensitive optical mapping and captured at 500 fps within about 60 milliseconds or 30 video images propagating across the ventricles.
In Fig.~\ref{fig:methods:initialcondition}, we aimed to learn the 2D focal pulse within 70 snapshots and obtained acceptable results using fewer than 40 snapshots.
Correspondingly, in Figs. 18 and 19 in Chowdhary \& Lebert et al.~\cite{Chowdhary2026}, reentrant action potential vortex waves were imaged at 500 fps for about 10 seconds.
At this temporal resolution, a single rotation can be imaged in about 30-50 video frames, and each recording contains many rotations.
Correspondingly, with the extended observation time and self-sustained dynamics, we could easily fit a model to multiple rotations and compare forecasts against ground-truth across the heart surface for many more rotations.
Accordingly, the results in Figs.~\ref{fig:results:SPH1}-\ref{fig:results:scroll} suggest that the spatial and temporal resolutions are sufficient such that our method could be used to recover intramural wave dynamics from panoramic optical mapping data. 
For example, it could be possible to reconstruct reentrant wave dynamics during atrial or ventricular fibrillation, or to locate focal waves originating from the Purkinje system during sinus rhythm.
With catheter mapping data, it may be possible to fit regular or repetitive rhythms, such as atrial flutter, provided that the full simulation pipeline, including the electrogram computation, is differentiable.
These are clinically important applications that could enhance the localization and ablation of cardiac arrhythmias, and we aim to pursue further translational studies using differentiable cardiac electrophysiology simulations in future work.

A key finding of our study is that successful application to experimental data requires augmenting differentiable simulations with complementary machine learning approaches, such as the Video Joint-Embedding Predictive Architecture (V-JEPA) perceptual loss or denoising diffusion probabilistic models (DDPMs), see section \ref{sec:results:experiments}.
V-JEPA helps overcome limitations of a purely pixel-based loss, and compensates for noise and spatial mismatches between the model and experimental data.
DDPMs can be used to estimate initial conditions, or smooth intermittent dynamical states and reproject degenerate states back onto the biophysical model's manifold.
These add-ons can be seamlessly integrated with gradient-based optimization, and we found them to be key components with the noisy, artifact-afflicted cell culture data.
Overall, our set of techniques provides a more general framework that enables differentiable simulations to be applied to real-world data and beyond the specific settings considered in other studies.
Our 2D results are closely related to the work by Lettermann et al.~\cite{Lettermann2024}, Kashtanova et al.~\cite{Kashtanova2023}, Berg et al.~\cite{Berg2011}, and Herrero Martin et al. \cite{HerreroMartin2022}.
Lettermann et al.~\cite{Lettermann2024} estimated parameters of the four-variable Bueno-Orovio-Cherry-Fenton~\cite{BuenoOrovioCherryFenton2008} (BOCF) model during spiral wave dynamics using a very similar gradient-based optimization approach and a pixel-based loss, which restricts the approach to simulated pixel-based data.
They subsequently fitted the two-variable Aliev-Panfilov (AP) model to the BOCF data.
More precisely, they estimated the 10 most influential parameters of the 28-parameter BOCF model, assuming the complete initial dynamical state was already known.
The AP model was fitted to the BOCF data without assuming the initial AP model state (because the corresponding state is a priori unknown), but the fitted AP dynamics quickly diverged from the BOCF dynamics, indicating poor state and parameter recovery. 
The results suggest that the much simpler AP model likely cannot reproduce the BOCF dynamics.
By contrast, in our study, we chose two models that are more compatible with each other, utilized our DDPM approach to facilitate the cross-model fitting, and, as a result, the fitted AP dynamics co-evolve quite well with the Mitchell-Schaeffer (MS) dynamics.
Kashtanova et al.~\cite{Kashtanova2023} fitted cardiac action potentials imaged with optical mapping during pacing using a hybrid approach of differentiable physics simulations and deep learning.
Here, we fitted physics-only differentiable physics simulations to optical mapping data of more complex spiral waves.
Instead of a neural network component, we used the V-JEPA perceptual loss.
Our work is also comparable to that of Berg et al.~\cite{Berg2011}, who used a data assimilation approach to identify model states and parameters in 2D excitable media.
While the context is slightly different, Fig.~\ref{fig:results:sparsity} in section \ref{sec:results:sparse} shows results comparable to those in Fig.~1 in Berg et al..
Nevertheless, while they performed classical parameter sweeps to find optimal parameter values, we instead utilized gradient-based optimization, which generally enables us to find optimal parameters more efficiently, particularly in higher-dimensional parameter spaces.
Our work is also related to that of Herrero Martin et al. \cite{HerreroMartin2022} and Chiu et al. \cite{Chiu2025}, who used PINNs for parameter inference with 2D and 3D simulation and optical mapping data. 
While they were able to estimate spatial heterogeneity and global parameters such as action potential durations, they obtained relatively large errors for individual simulation parameters and modest performance with simulated fibrillatory wave patterns.
None of the aforementioned studies were demonstrated with realistic heart shapes or experimental data showing complex fibrillation-like patterns, and our study provides additional insights regarding sparse, partial, and noisy observations.

Compared to prior work on reconstructing 3D scroll wave dynamics from surface observations, our method achieves lower reconstruction error and improved consistency in the unobserved bulk interior. 
In Marcotte et al.~\cite{Marcotte2021}, reconstruction from surface data exhibits persistent interior error, particularly away from observed boundaries, even with stochastic model perturbations. 
The data assimilation approach performs sequential state estimation using an ensemble, where observations of the excitatory variable are assimilated 
over time, but the interior state and model parameters are not directly optimized. 
Similarly, Hoffman et al.~\cite{Hoffman2016} demonstrated that data assimilation based on the Local Ensemble Transform Kalman Filter could recover major features of hidden three-dimensional scroll-wave dynamics from dual-surface observations, but reconstruction errors remained largest in the bulk interior and along wave fronts, particularly as information propagated away from observed surfaces. Hoffman and Cherry~\cite{Hoffman2020} further showed that reconstruction quality is highly sensitive to localization radius, observation density, and model mismatch, with ensemble collapse leading to growing interior errors when uncertainty is underestimated.
In Lebert et al.~\cite{Lebert2023} and Stenger et al.~\cite{Stenger2023}, the reconstructions were performed using convolutional neural networks (CNNs), which produced large errors near the center of the bulk. 
In Baranwal et al.~\cite{Baranwal2024} and Stenger et al.~\cite{Stenger2023}, denoising diffusion probabilistic models improved reconstruction quality relative to CNNs for long observation windows, but remained sensitive to noise and hallucination artifacts. 
Importantly, these purely data-driven approaches reconstructed only the excitatory variable and did not recover hidden state variables or model parameters, making application to experimental imaging data difficult in the absence of paired ground-truth volumetric training data.
In contrast, here we jointly learn the full initial condition and model parameters from observations of only the excitatory variable on the top and bottom surfaces, recovering all interior states for both the excitatory and refractory variables without direct supervision. 
This yields near-zero surface loss ($\sim 10^{-9}$) and low bulk errors ($\sim 10^{-5}$--$10^{-4}$). 
The multi-horizon training scheme was essential, as it enables surface constraints to propagate into the interior through repeated forward simulation, progressively refining both parameters and initial conditions. Unlike prior work, which evaluates performance during assimilation with continuous observational updates, we assess the learned 
system in a predictive setting; the learned dynamics remain stable over 1,000 time steps beyond training (approximately 38 spiral rotations), with residual mean squared errors on the order of $10^{-3}$--$10^{-5}$, demonstrating substantially improved system identification and long-term predictive accuracy.
Together with the fit of the experimental spiral waves in the cell culture, these findings suggest that our approach is potentially able to produce numerical 3D reconstructions of complex wave dynamics, potentially within the fibrillating heart muscle. 
For instance, it could be used to uncover epi- and endocardially dissociated atrial fibrillation dynamics~\cite{Eckstein2010, deGroot2016, Walters2020} or reconstruct scroll waves during ventricular fibrillation~\cite{Davidenko1992, Pertsov1993, Christoph2018, FentonKarma1998, Berenfeld1999, Qu2000, Bernus2007, Mitrea2009}.

Our method appears particularly effective for more complex spiral wave dynamics and less effective for simpler wave patterns, such as focal or stationary spiral waves.
On the one hand, this may be explained by the extent of the observation data: spiral wave dynamics, if sustained, can be observed over longer periods than a single pulse.
On the other hand, more complex dynamics are more expressive, more unique, and might reveal more information about the system's state and model parameters than simpler patterns.
As a result, the loss landscape might be more favorable with more complex wave patterns than with simpler ones.
Future work will need to determine whether this behavior is advantageous for analyzing imaging data of atrial or ventricular fibrillation.
An outstanding challenge is choosing the appropriate model equations when attempting to reconstruct fibrillation.
While it is possible to obtain near-perfect reconstructions of spiral and scroll wave dynamics when the model equations are known, see sections \ref{sec:results:2Dfocus}-\ref{sec:results:sparse}, with model mismatch, see section \ref{sec:results:crossmodel}, or experimental data, see section \ref{sec:results:experiments}, the reconstructions are more challenging. 
More work is required to assess whether the current biophysical models are inadequate or the fitting framework requires further development.
In the future, our approach might enable systematic assessments of how well cardiac electrophysiology models reproduce fibrillation and related dynamics.

\section{Conclusions}
\label{sec:conclusions}

Gradient-based optimization opens new opportunities in computational cardiology and arrhythmia research that extend beyond the training of neural networks.
Differentiable physics simulations describing cardiac electrophysiology are a promising new tool for personalizing computer simulations of the heart with many potential applications in basic cardiovascular research and diagnostics.

\section*{Conflict of Interest}
The authors declare that the research was conducted in the absence of any commercial or financial relationships that could be construed as a potential conflict of interest.

\section*{Data Availability Statement}
The source code will be made available upon publication.

\section*{Funding}
This research was funded by the University of California, San Francisco, the National Institutes of Health (DP2HL168071), and the Sandler Program for Breakthrough Biomedical Research, which is partially funded by the Sandler Foundation (to JC). 
The RTX A5000/6000 GPUs used in this study were donated by the NVIDIA Corporation via the Academic Hardware Grant Program (to JC).
This research was also supported in part by grant NSF PHY-2309135 to the Kavli Institute for Theoretical Physics (KITP) and the Gordon and Betty Moore Foundation Grant No. 2019.02 (to EE and JC).

$\\$

\section*{Author Contributions}
AP developed the finite differences-based differentiable simulations.
SC developed the SPH-based differentiable simulations.
AP and AH conceived the multi-horizon learning schedule.
AP, AH, and JC performed the analysis.
AP developed the DDPM approach.
AH developed the V-JEPA approach.
WL and EE performed the voltage-sensitive cell culture imaging experiment.
JC designed the figures and wrote the manuscript with input from AP, AH, SC, and EE.
All authors read and approved the final version of the manuscript.
SC and JC conceived the work.

\bibliography{references}

@misc{Adam,
      title={Adam: A Method for Stochastic Optimization}, 
      author={Diederik P. Kingma and Jimmy Ba},
      year={2017},
      eprint={1412.6980},
      archivePrefix={arXiv},
      primaryClass={cs.LG},
      url={https://arxiv.org/abs/1412.6980}, 
}

@article{AlievPanfilov1996,
  title   = {A simple two-variable model of cardiac excitation},
  author  = {Aliev, R. R. and Panfilov, A. V.},
  journal = {Chaos, Solitons \& Fractals},
  volume  = {7},
  number  = {3},
  pages   = {293-301},
  year    = {1996},
  doi     = {10.1016/0960-0779(95)00089-5}
}

@article{Alonso2016,
doi = {10.1088/0034-4885/79/9/096601},
url = {https://dx.doi.org/10.1088/0034-4885/79/9/096601},
year = {2016},
month = {aug},
publisher = {IOP Publishing},
volume = {79},
number = {9},
pages = {096601},
author = {Sergio Alonso and Markus Bär and Blas Echebarria},
title = {Nonlinear physics of electrical wave propagation in the heart: a review},
journal = {Reports on Progress in Physics},
}

@article{Baranwal2024,
    author = {Baranwal, Tanish and Lebert, Jan and Christoph, Jan},
    title = {Dreaming of electrical waves: Generative modeling of cardiac excitation waves using diffusion models},
    journal = {APL Machine Learning},
    volume = {2},
    number = {3},
    pages = {036113},
    year = {2024},
    month = {09},
    issn = {2770-9019},
    doi = {10.1063/5.0194391}
}

@article{Berenfeld1999,
title = {Dynamics of Intramural Scroll Waves in Three-dimensional Continuous Myocardium with Rotational Anisotropy},
journal = {Journal of Theoretical Biology},
volume = {199},
number = {4},
pages = {383-394},
year = {1999},
issn = {0022-5193},
doi = {https://doi.org/10.1006/jtbi.1999.0965},
url = {https://www.sciencedirect.com/science/article/pii/S0022519399909657},
author = {Omer Berenfeld and Arkady M Pertsov}
}

@article{Berg2011,
  title   = {Synchronization Based System Identification of an Extended Excitable System},
  author  = {Berg, Sebastian and Luther, Stefan and Parlitz, Ulrich},
  year    = {2011},
  month   = sep,
  journal = {Chaos: An Interdisciplinary Journal of Nonlinear Science},
  volume  = {21},
  number  = {3},
  issn    = {10541500},
  doi     = {10.1063/1.3613921},
  pmid    = {21974639}
}

@article{Bernus2007,
author = {Olivier Bernus and Karthik S. Mukund and Arkady M. Pertsov},
title = {{Detection of intramyocardial scroll waves using absorptive transillumination imaging}},
volume = {12},
journal = {Journal of Biomedical Optics},
number = {1},
publisher = {SPIE},
pages = {014035},
year = {2007},
doi = {10.1117/1.2709661},
URL = {https://doi.org/10.1117/1.2709661}
}

@article{Bot2012,
AUTHOR={Bot, Corina T. and Kherlopian, Armen R. and Ortega, Francis A. and Christini, David J. and Krogh-Madsen, Trine },
TITLE={Rapid Genetic Algorithm Optimization of a Mouse Computational Model: Benefits for Anthropomorphization of Neonatal Mouse Cardiomyocytes},
JOURNAL={Frontiers in Physiology},
VOLUME={Volume 3 - 2012},
YEAR={2012},
URL={https://www.frontiersin.org/journals/physiology/articles/10.3389/fphys.2012.00421},
DOI={10.3389/fphys.2012.00421},
ISSN={1664-042X}
}

@article{BuenoOrovioCherryFenton2008,
  title    = {Minimal model for human ventricular action potentials in tissue},
  journal  = {Journal of Theoretical Biology},
  volume   = {253},
  number   = {3},
  pages    = {544-560},
  year     = {2008},
  issn     = {0022-5193},
  doi      = {10.1016/j.jtbi.2008.03.029},
  url      = {https://www.sciencedirect.com/science/article/pii/S0022519308001690},
  author   = {Alfonso Bueno-Orovio and Elizabeth M. Cherry and Flavio H. Fenton}
}

@article{Cairns2017,
    author = {Cairns, Darby I. and Fenton, Flavio H. and Cherry, E. M.},
    title = {Efficient parameterization of cardiac action potential models using a genetic algorithm},
    journal = {Chaos: An Interdisciplinary Journal of Nonlinear Science},
    volume = {27},
    number = {9},
    pages = {093922},
    year = {2017},
    month = {08},
    issn = {1054-1500},
    doi = {10.1063/1.5000354}
}

@article{Cairns2025,
    author = {Cairns, Darby I. and Comstock, Maxfield Roth and Fenton, Flavio H. and Cherry, Elizabeth M.},
    title = {CardioFit: a WebGL-based tool for fast and efficient parametrization of cardiac action potential models to fit user-provided data},
    journal = {Royal Society Open Science},
    volume = {12},
    number = {8},
    pages = {250048},
    year = {2025},
    month = {08},
    issn = {2054-5703},
    doi = {10.1098/rsos.250048},
    url = {https://doi.org/10.1098/rsos.250048}
}

@article{Camps2025,
title = {Harnessing 12-lead ECG and MRI data to personalise repolarisation profiles in cardiac digital twin models for enhanced virtual drug testing},
journal = {Medical Image Analysis},
volume = {100},
pages = {103361},
year = {2025},
issn = {1361-8415},
doi = {https://doi.org/10.1016/j.media.2024.103361},
url = {https://www.sciencedirect.com/science/article/pii/S136184152400286X},
author = {Julia Camps and Zhinuo Jenny Wang and Ruben Doste and Lucas Arantes Berg and Maxx Holmes and Brodie Lawson and Jakub Tomek and Kevin Burrage and Alfonso Bueno-Orovio and Blanca Rodriguez}
}

@article{Chen2012,
  author={Chen, Fulong and Chu, Angdi and Yang, Xuefei and Lei, Yao and Chu, Jizheng},
  journal={IEEE Transactions on Biomedical Engineering}, 
  title={Identification of the Parameters of the Beeler–Reuter Ionic Equation With a Partially Perturbed Particle Swarm Optimization}, 
  year={2012},
  volume={59},
  number={12},
  pages={3412-3421},
  doi={10.1109/TBME.2012.2216265}}

@InProceedings{Chiu2025,
author="Chiu, Ching-En
and Roy, Aditi
and Cechnicka, Sarah
and Gupta, Ashvin
and Pinto, Arieh Levy
and Galazis, Christoforos
and Christensen, Kim
and Mandic, Danilo
and Varela, Marta",
title="Physics-Informed Neural Networks Can Accurately Model Cardiac Electrophysiology in 3D Geometries and Fibrillatory Conditions",
booktitle="Statistical Atlases and Computational Models of the Heart. Workshop, CMRxRecon and MBAS Challenge Papers.",
year="2025",
publisher="Springer Nature Switzerland",
address="Cham",
pages="98--109",
isbn="978-3-031-87756-8"
}

@article{Chowdhary2026,
  author = {Chowdhary, Shrey and Lebert, Jan and Dickman, Shai and Manetta, Mason and Gordon, Charles and Christoph, Jan},
  title = {Panoramic voltage-sensitive optical mapping of contracting hearts using cooperative multiview motion tracking with 12 cameras},
  year    = {2026},
  journal = {The Journal of Physiology},
  volume  = {n/a},
  number  = {n/a},
  doi = {https://doi.org/10.1113/JP290533}
}

@article{Christoph2018,
  title   = {Electromechanical Vortex Filaments during Cardiac Fibrillation},
  author  = {J. Christoph and M. Chebbok and C. Richter and J. Schr\"oder-Schetelig and P. Bittihn and S. Stein and I. Uzelac and F. H. Fenton and G. Hasenfuss and R. Jr. Gilmour and S. Luther},
  doi     = {10.1038/nature26001},
  journal = {Nature},
  volume  = {555},
  pages   = {667 - 672},
  year    = {2018}
}

@article{Davidenko1992,
  author  = {Davidenko, Jorge M. and Pertsov, Arcady V. and Salomonsz, Remy and Baxter, William and Jalife, José},
  title   = {Stationary and drifting spiral waves of excitation in isolated cardiac muscle},
  journal = {Nature},
  volume  = {355},
  pages   = {349-351},
  year    = {1992},
  doi     = {10.1038/355349a0}
}

@article{deGroot2016,
  author  = {Natasja de Groot  and Lisette van der Does  and Ameeta Yaksh  and Eva Lanters  and Christophe Teuwen  and Paul Knops  and Pieter van de Woestijne  and Jos Bekkers  and Charles Kik  and Ad Bogers  and Maurits Allessie },
  title   = {Direct Proof of Endo-Epicardial Asynchrony of the Atrial Wall During Atrial Fibrillation in Humans},
  journal = {Circulation: Arrhythmia and Electrophysiology},
  volume  = {9},
  number  = {5},
  pages   = {e003648},
  year    = {2016},
  doi     = {10.1161/CIRCEP.115.003648}
}

@article{Dokos2004,
title = {Parameter estimation in cardiac ionic models},
journal = {Progress in Biophysics and Molecular Biology},
volume = {85},
number = {2},
pages = {407-431},
year = {2004},
note = {Modelling Cellular and Tissue Function},
issn = {0079-6107},
doi = {https://doi.org/10.1016/j.pbiomolbio.2004.02.002},
url = {https://www.sciencedirect.com/science/article/pii/S0079610704000306},
author = {Socrates Dokos and Nigel H. Lovell}
}

@article{Doste2019,
author = {Doste, Ruben and Soto-Iglesias, David and Bernardino, Gabriel and Alcaine, Alejandro and Sebastian, Rafael and Giffard-Roisin, Sophie and Sermesant, Maxime and Berruezo, Antonio and Sanchez-Quintana, Damian and Camara, Oscar},
title = {A rule-based method to model myocardial fiber orientation in cardiac biventricular geometries with outflow tracts},
journal = {International Journal for Numerical Methods in Biomedical Engineering},
volume = {35},
number = {4},
pages = {e3185},
doi = {https://doi.org/10.1002/cnm.3185},
url = {https://onlinelibrary.wiley.com/doi/abs/10.1002/cnm.3185},
eprint = {https://onlinelibrary.wiley.com/doi/pdf/10.1002/cnm.3185},
note = {e3185 cnm.3185},
year = {2019}
}

@article{Doste2026,
title = {An automated computational pipeline for generating large-scale cohorts of patient-specific ventricular models in electromechanical in silico trials},
journal = {Computer Methods and Programs in Biomedicine},
volume = {279},
pages = {109290},
year = {2026},
issn = {0169-2607},
doi = {https://doi.org/10.1016/j.cmpb.2026.109290},
url = {https://www.sciencedirect.com/science/article/pii/S0169260726000581},
author = {Ruben Doste and Julia Camps and Zhinuo Jenny Wang and Lucas Arantes Berg and Maxx Holmes and Hannah Smith and Marcel Beetz and Lei Li and Abhirup Banerjee and Vicente Grau and Blanca Rodriguez}
}

@article{Eckstein2010,
  author  = {Eckstein, Jens and Maesen, Bart and Linz, Dominik and Zeemering, Stef and van Hunnik, Arne and Verheule, Sander and Allessie, Maurits and Schotten, Ulrich},
  title   = {{Time course and mechanisms of endo-epicardial electrical dissociation during atrial fibrillation in the goat}},
  journal = {Cardiovascular Research},
  volume  = {89},
  number  = {4},
  pages   = {816-824},
  year    = {2010},
  month   = {10},
  issn    = {0008-6363},
  doi     = {10.1093/cvr/cvq336}
}

@misc{Eing2026,
      title={Video Joint-Embedding Predictive Architectures for Facial Expression Recognition}, 
      author={Lennart Eing and Cristina Luna-Jiménez and Silvan Mertes and Elisabeth André},
      year={2026},
      eprint={2601.09524},
      archivePrefix={arXiv},
      primaryClass={cs.CV},
      url={https://arxiv.org/abs/2601.09524}, 
}

@misc{equinox,
      title={Equinox: neural networks in JAX via callable PyTrees and filtered transformations}, 
      author={Patrick Kidger and Cristian Garcia},
      year={2021},
      eprint={2111.00254},
      archivePrefix={arXiv},
      primaryClass={cs.LG},
      url={https://arxiv.org/abs/2111.00254}, 
}

@phdthesis{kidger2021,
  title = {On Neural Differential Equations},
  author = {Patrick Kidger},
  year = {2021},
  school = {University of Oxford}
}

@article{FentonKarma1998,
  author    = {Fenton, Flavio and Karma, Alain},
  journal   = {Chaos: An Interdisciplinary Journal of Nonlinear Science},
  number    = {1},
  pages     = {20--47},
  publisher = {AIP},
  title     = {Vortex dynamics in three-dimensional continuous myocardium with fiber rotation: Filament instability and fibrillation},
  volume    = {8},
  year      = {1998},
  doi       = {10.1063/1.166311}
}

@article{Gepstein1997,
  author  = {Lior Gepstein  and Gal Hayam  and Shlomo A. Ben-Haim },
  title   = {A Novel Method for Nonfluoroscopic Catheter-Based Electroanatomical Mapping of the Heart},
  journal = {Circulation},
  volume  = {95},
  number  = {6},
  pages   = {1611-1622},
  year    = {1997},
  doi     = {10.1161/01.CIR.95.6.1611}
}

@article{Gillette2021,
title = {A Framework for the generation of digital twins of cardiac electrophysiology from clinical 12-leads ECGs},
journal = {Medical Image Analysis},
volume = {71},
pages = {102080},
year = {2021},
issn = {1361-8415},
doi = {https://doi.org/10.1016/j.media.2021.102080},
url = {https://www.sciencedirect.com/science/article/pii/S1361841521001262},
author = {Karli Gillette and Matthias A.F. Gsell and Anton J. Prassl and Elias Karabelas and Ursula Reiter and Gert Reiter and Thomas Grandits and Christian Payer and Darko Štern and Martin Urschler and Jason D. Bayer and Christoph M. Augustin and Aurel Neic and Thomas Pock and Edward J. Vigmond and Gernot Plank}
}

@article{Heinson2023,
author = {Yuli W. Heinson and Julie L. Han and Emilia Entcheva},
title = {{Portable low-cost macroscopic mapping system for all-optical cardiac electrophysiology}},
volume = {28},
journal = {Journal of Biomedical Optics},
number = {1},
publisher = {SPIE},
pages = {016001},
year = {2023},
doi = {10.1117/1.JBO.28.1.016001},
URL = {https://doi.org/10.1117/1.JBO.28.1.016001}
}

@article{HerreroMartin2022,
  doi       = {10.3389/fcvm.2021.768419},
  url       = {https://doi.org/10.3389/fcvm.2021.768419},
  year      = {2022},
  month     = feb,
  publisher = {Frontiers Media {SA}},
  volume    = {8},
  author    = {Clara Herrero Martin and Alon Oved and Rasheda A. Chowdhury and Elisabeth Ullmann and Nicholas S. Peters and Anil A. Bharath and Marta Varela},
  title     = {{EP}-{PINNs}: Cardiac Electrophysiology Characterisation Using Physics-Informed Neural Networks},
  journal   = {Frontiers in Cardiovascular Medicine}
}

@article{Hoffman2016,
    author = {Hoffman, M. J. and LaVigne, N. S. and Scorse, S. T. and Fenton, F. H. and Cherry, E. M.},
    title = {Reconstructing three-dimensional reentrant cardiac electrical wave dynamics using data assimilation},
    journal = {Chaos: An Interdisciplinary Journal of Nonlinear Science},
    volume = {26},
    number = {1},
    pages = {013107},
    year = {2016},
    month = {01},
    issn = {1054-1500},
    doi = {10.1063/1.4940238},
    url = {https://doi.org/10.1063/1.4940238}
}

@article{Hoffman2020,
  doi       = {10.1098/rsta.2019.0388},
  url       = {https://doi.org/10.1098/rsta.2019.0388},
  year      = {2020},
  month     = may,
  publisher = {The Royal Society},
  volume    = {378},
  number    = {2173},
  pages     = {20190388},
  author    = {Matthew J. Hoffman and Elizabeth M. Cherry},
  title     = {Sensitivity of a data-assimilation system for reconstructing three-dimensional cardiac electrical dynamics},
  journal   = {Philosophical Transactions of the Royal Society A: Mathematical,  Physical and Engineering Sciences}
}

@misc{Hu2020,
      title={DiffTaichi: Differentiable Programming for Physical Simulation}, 
      author={Yuanming Hu and Luke Anderson and Tzu-Mao Li and Qi Sun and Nathan Carr and Jonathan Ragan-Kelley and Frédo Durand},
      year={2020},
      eprint={1910.00935},
      archivePrefix={arXiv},
      primaryClass={cs.LG},
      url={https://arxiv.org/abs/1910.00935}, 
}

@article {Huang2015,
	Title = {A Photostable Silicon Rhodamine Platform for Optical Voltage Sensing},
	Author = {Huang, Yi-Lin and Walker, Alison S and Miller, Evan W},
	DOI = {10.1021/jacs.5b06644},
	Number = {33},
	Volume = {137},
	Month = {August},
	Year = {2015},
	Journal = {Journal of the American Chemical Society},
	ISSN = {0002-7863},
	Pages = {10767—10776},
	URL = {https://europepmc.org/articles/PMC4666802}
}

@misc{jax,
  author        = {Bradbury, J. and Frostig, R. and Hawkins, P. and Johnson, M. J.  and Leary, C. and Maclaurin, D. and Necula, G. and Paszke, A. and VanderPlas, J. and Wanderman-Milne, S. and Zhang, Q.},
  title         = {JAX: Composable transformations of Python+NumPy programs},
  year          = {2018}
}

@article{Kashtanova2023,
author = {Kashtanova, Victoriya  and Pop, Mihaela  and Ayed, Ibrahim  and Gallinari, Patrick  and Sermesant, Maxime },
title = {Simultaneous data assimilation and cardiac electrophysiology model correction using differentiable physics and deep learning},
journal = {Interface Focus},
volume = {13},
number = {6},
pages = {20230043},
year = {2023},
doi = {10.1098/rsfs.2023.0043}
}

@article{Klimas2020,
title = {Multimodal on-axis platform for all-optical electrophysiology with near-infrared probes in human stem-cell-derived cardiomyocytes},
journal = {Progress in Biophysics and Molecular Biology},
volume = {154},
pages = {62-70},
year = {2020},
note = {Novel optics-based approaches for cardiac electrophysiology},
issn = {0079-6107},
doi = {https://doi.org/10.1016/j.pbiomolbio.2019.02.004},
url = {https://www.sciencedirect.com/science/article/pii/S0079610718302797},
author = {Aleksandra Klimas and Gloria Ortiz and Steven C. Boggess and Evan W. Miller and Emilia Entcheva}
}

@article{LaVigne2017,
    author = {LaVigne, Nicholas S. and Holt, Nathan and Hoffman, Matthew J. and Cherry, Elizabeth M.},
    title = {Effects of model error on cardiac electrical wave state reconstruction using data assimilation},
    journal = {Chaos: An Interdisciplinary Journal of Nonlinear Science},
    volume = {27},
    number = {9},
    pages = {093911},
    year = {2017},
    month = {08},
    issn = {1054-1500},
    doi = {10.1063/1.4999603}
}

@article{Lebert2023,
  title     = {Reconstruction of three-dimensional scroll waves in excitable media from two-dimensional observations using deep neural networks},
  author    = {Lebert, Jan and Mittal, Meenakshi and Christoph, Jan},
  journal   = {Phys. Rev. E},
  volume    = {107},
  issue     = {1},
  pages     = {014221},
  numpages  = {14},
  year      = {2023},
  month     = {Jan},
  publisher = {American Physical Society},
  doi       = {10.1103/PhysRevE.107.014221},
  url       = {https://link.aps.org/doi/10.1103/PhysRevE.107.014221}
}

@article{Lettermann2024,
  title     = {Tutorial: a beginner's guide to building a representative model of dynamical systems using the adjoint method},
  author    = {Lettermann, Leon and Jurado, Alejandro and Betz, Timo and W{\"o}rg{\"o}tter, Florentin and Herzog, Sebastian},
  journal   = {Commun. Phys.},
  publisher = {Springer Science and Business Media LLC},
  volume    =  {7},
  number    =  {1},
  year      =  {2024}
}

@article{Liu2023,
author = {Liu, Wei and Han, Julie L. and Tomek, Jakub and Bub, Gil and Entcheva, Emilia},
title = {Simultaneous Widefield Voltage and Dye-Free Optical Mapping Quantifies Electromechanical Waves in Human Induced Pluripotent Stem Cell-Derived Cardiomyocytes},
journal = {ACS Photonics},
volume = {10},
number = {4},
pages = {1070-1083},
year = {2023},
doi = {10.1021/acsphotonics.2c01644},
URL = {https://doi.org/10.1021/acsphotonics.2c01644},
eprint = {https://doi.org/10.1021/acsphotonics.2c01644}
}

@article{Loewe2015,
AUTHOR={Loewe, Axel  and Wilhelms, Mathias  and Schmid, Jochen  and Krause, Mathias J.  and Fischer, Fathima  and Thomas, Dierk  and Scholz, Eberhard P.  and Dössel, Olaf  and Seemann, Gunnar },
TITLE={Parameter Estimation of Ion Current Formulations Requires Hybrid Optimization Approach to Be Both Accurate and Reliable},
JOURNAL={Frontiers in Bioengineering and Biotechnology},
VOLUME={Volume 3 - 2015},
YEAR={2016},
DOI={10.3389/fbioe.2015.00209},
ISSN={2296-4185}
}

@article{Lombardo2016,
    doi = {10.1371/journal.pcbi.1005060},
    author = {Lombardo, Daniel M. AND Fenton, Flavio H. AND Narayan, Sanjiv M. AND Rappel, Wouter-Jan},
    journal = {PLOS Computational Biology},
    publisher = {Public Library of Science},
    title = {Comparison of Detailed and Simplified Models of Human Atrial Myocytes to Recapitulate Patient Specific Properties},
    year = {2016},
    month = {08},
    volume = {12},
    url = {https://doi.org/10.1371/journal.pcbi.1005060},
    pages = {1-15},
    number = {8}
}

@inproceedings{Lugmayr2022,
  author    = {Lugmayr, Andreas and Danelljan, Martin and Romero, Andres and Yu, Fisher and Timofte, Radu and Van Gool, Luc},
  title     = {RePaint: Inpainting using Denoising Diffusion Probabilistic Models},
  booktitle = {Proceedings of the IEEE/CVF Conference on Computer Vision and Pattern Recognition (CVPR)},
  pages     = {11451--11461},
  year      = {2022},
  doi       = {10.1109/cvpr52688.2022.01117}
}

@article{Marcotte2021,
    author = {Marcotte, Christopher D. and Fenton, Flavio H. and Hoffman, Matthew J. and Cherry, Elizabeth M.},
    title = {Robust data assimilation with noise: Applications to cardiac dynamics},
    journal = {Chaos: An Interdisciplinary Journal of Nonlinear Science},
    volume = {31},
    number = {1},
    pages = {013118},
    year = {2021},
    month = {01},
    issn = {1054-1500},
    doi = {10.1063/5.0033539},
    url = {https://doi.org/10.1063/5.0033539}
}

@article{Marcotte2023,
    author = {Marcotte, C. D. and Hoffman, M. J. and Fenton, F. H. and Cherry, E. M.},
    title = {Reconstructing cardiac electrical excitations from optical mapping recordings},
    journal = {Chaos: An Interdisciplinary Journal of Nonlinear Science},
    volume = {33},
    number = {9},
    pages = {093141},
    year = {2023},
    month = {09},
    issn = {1054-1500},
    doi = {10.1063/5.0156314}
}

@article{Mendez2024,
title = {Reconstructing ventricular cardiomyocyte dynamics and parameter estimation using data assimilation},
journal = {Biophysical Journal},
volume = {123},
number = {23},
pages = {4050-4066},
year = {2024},
issn = {0006-3495},
doi = {https://doi.org/10.1016/j.bpj.2024.10.018},
url = {https://www.sciencedirect.com/science/article/pii/S0006349524006908},
author = {Mario J. Mendez and Elizabeth M. Cherry and Gregory S. Hoeker and Steven Poelzing and Seth H. Weinberg}
}

@inproceedings{Meng2022,
  author    = {Meng, Chenlin and He, Yutong and Song, Yang and Jiaming, Song and Wu, Jiajun and Zhu, Jun-Yan and Ermon, Stefano},
  title     = {SDEdit: Guided Image Synthesis and Editing with Stochastic Differential Equations},
  booktitle = {International Conference on Learning Representations (ICLR)},
  year      = {2022},
  url       = {https://openreview.net/forum?id=aBsCjcPu_tE}
}

@article{MitchellSchaeffer2003,
	author = {Mitchell, Colleen C. and Schaeffer, David G.},
	date = {2003/09/01},
	doi = {10.1016/S0092-8240(03)00041-7},
	id = {Mitchell2003},
	isbn = {1522-9602},
	journal = {Bulletin of Mathematical Biology},
	number = {5},
	pages = {767--793},
	title = {A two-current model for the dynamics of cardiac membrane},
	url = {https://doi.org/10.1016/S0092-8240(03)00041-7},
	volume = {65},
	year = {2003}
}

@inproceedings{Mitrea2009,
  author    = {Mitrea, Bogdan G. and Wellner, Marcel and Pertsov, Arkady M.},
  booktitle = {2009 Annual International Conference of the IEEE Engineering in Medicine and Biology Society},
  title     = {Monitoring intramyocardial reentry using alternating transillumination},
  year      = {2009},
  volume    = {},
  number    = {},
  pages     = {4194-4197},
  doi       = {10.1109/IEMBS.2009.5334048}
}

@article{MonteiroDaRocha2016,
title = {Deficient cMyBP-C protein expression during cardiomyocyte differentiation underlies human hypertrophic cardiomyopathy cellular phenotypes in disease specific human ES cell derived cardiomyocytes},
journal = {Journal of Molecular and Cellular Cardiology},
volume = {99},
pages = {197-206},
year = {2016},
issn = {0022-2828},
doi = {https://doi.org/10.1016/j.yjmcc.2016.09.004},
url = {https://www.sciencedirect.com/science/article/pii/S0022282816303492},
author = {Andre {Monteiro da Rocha} and Guadalupe Guerrero-Serna and Adam Helms and Carly Luzod and Sergey Mironov and Mark Russell and José Jalife and Sharlene M. Day and Gary D. Smith and Todd J. Herron}
}

@article{Neic2017,
title = {Efficient computation of electrograms and ECGs in human whole heart simulations using a reaction-eikonal model},
journal = {Journal of Computational Physics},
volume = {346},
pages = {191-211},
year = {2017},
issn = {0021-9991},
doi = {https://doi.org/10.1016/j.jcp.2017.06.020},
url = {https://www.sciencedirect.com/science/article/pii/S0021999117304655},
author = {Aurel Neic and Fernando O. Campos and Anton J. Prassl and Steven A. Niederer and Martin J. Bishop and Edward J. Vigmond and Gernot Plank}
}

@article{Pertsov1993,
  author  = {Pertsov, A. M. and Davidenko, J. M. Salomonsz, R. and Baxter, W. T. and Jalife, José},
  title   = {Spiral waves of excitation underlie reentrant activity in isolated cardiac muscle},
  journal = {Circulation Research},
  volume  = {72},
  pages   = {631-650},
  year    = {1993},
  doi     = {10.1161/01.res.72.3.631}
}

@article{Pezzutto2020,
    author = {Pezzuto, Simone and Prinzen, Frits W and Potse, Mark and Maffessanti, Francesco and Regoli, François and Caputo, Maria Luce and Conte, Giulio and Krause, Rolf and Auricchio, Angelo},
    title = {Reconstruction of three-dimensional biventricular activation based on the 12-lead electrocardiogram via patient-specific modelling},
    journal = {EP Europace},
    volume = {23},
    number = {4},
    pages = {640-647},
    year = {2020},
    month = {11},
    issn = {1099-5129},
    doi = {10.1093/europace/euaa330},
    url = {https://doi.org/10.1093/europace/euaa330},
    eprint = {https://academic.oup.com/europace/article-pdf/23/4/640/36916991/euaa330.pdf},
}

@article{Pezzutto2022,
title = {Learning cardiac activation maps from 12-lead ECG with multi-fidelity Bayesian optimization on manifolds},
journal = {IFAC-PapersOnLine},
volume = {55},
number = {20},
pages = {175-180},
year = {2022},
note = {10th Vienna International Conference on Mathematical Modelling MATHMOD 2022},
issn = {2405-8963},
doi = {https://doi.org/10.1016/j.ifacol.2022.09.091},
url = {https://www.sciencedirect.com/science/article/pii/S2405896322012794},
author = {Simone Pezzuto and Paris Perdikaris and Francisco Sahli Costabal}
}

@inbook{pytorch, 
author = {Paszke, Adam and Gross, Sam and Massa, Francisco and Lerer, Adam and Bradbury, James and Chanan, Gregory and Killeen, Trevor and Lin, Zeming and Gimelshein, Natalia and Antiga, Luca and Desmaison, Alban and K\"{o}pf, Andreas and Yang, Edward and DeVito, Zach and Raison, Martin and Tejani, Alykhan and Chilamkurthy, Sasank and Steiner, Benoit and Fang, Lu and Bai, Junjie and Chintala, Soumith}, 
title = {PyTorch: an imperative style, high-performance deep learning library}, 
year = {2019}, 
publisher = {Curran Associates Inc.}, 
address = {Red Hook, NY, USA}, 
booktitle = {Proceedings of the 33rd International Conference on Neural Information Processing Systems}, 
articleno = {721}, 
 pages = {8024--8035},
numpages = {12}
}

@article{Qu2000,
title = {Scroll Wave Dynamics in a Three-Dimensional Cardiac Tissue Model: Roles of Restitution, Thickness, and Fiber Rotation},
journal = {Biophysical Journal},
volume = {78},
number = {6},
pages = {2761-2775},
year = {2000},
issn = {0006-3495},
doi = {https://doi.org/10.1016/S0006-3495(00)76821-4},
url = {https://www.sciencedirect.com/science/article/pii/S0006349500768214},
author = {Zhilin Qu and Jong Kil and Fagen Xie and Alan Garfinkel and James N. Weiss}
}

@article{Rappel2022,
title = {The physics of heart rhythm disorders},
journal = {Physics Reports},
volume = {978},
pages = {1-45},
year = {2022},
issn = {0370-1573},
doi = {https://doi.org/10.1016/j.physrep.2022.06.003},
author = {Wouter-Jan Rappel}
}

@article{Rheaume2023,
    author = {Rheaume, E and Velasco-Perez, H and Cairns, D and Comstock, M and Rheaume, E and Kaboudian, A and Uzelac, I and Cherry, E and Fenton, F H},
    title = {A Modified Fitzhugh-Nagumo Model that Reproduces the Action Potential and Dynamics of the Ten Tusscher et al. Cardiac Model in Tissue},
    journal = {Computing in Cardiology 2023},
    volume = {50},
    number = {},
    pages = {},
    year = {2023},
    month = {},
    issn = {2325-887X},
    doi = {10.22489/CinC.2023.424}
}

@inproceedings{jax-md,
author = {Schoenholz, Samuel S. and Cubuk, Ekin D.},
title = {JAX, M.D. a framework for differentiable physics},
year = {2020},
isbn = {9781713829546},
publisher = {Curran Associates Inc.},
address = {Red Hook, NY, USA},
booktitle = {Proceedings of the 34th International Conference on Neural Information Processing Systems},
articleno = {959},
numpages = {14},
location = {Vancouver, BC, Canada},
series = {NIPS '20}
}

@InProceedings{Seemann2009,
author="Seemann, Gunnar and Lurz, S. and Keller, D. U. J. and Weiss, D. L. and Scholz, E. P. and D{\"o}ssel, O.",
editor="Vander Sloten, Jos
and Verdonck, Pascal
and Nyssen, Marc
and Haueisen, Jens",
title="Adaption of Mathematical Ion Channel Models to measured data using the Particle Swarm Optimization",
booktitle="4th European Conference of the International Federation for Medical and Biological Engineering",
year="2009",
publisher="Springer Berlin Heidelberg",
address="Berlin, Heidelberg",
pages="2507--2510",
isbn="978-3-540-89208-3"
}

@article{Stenger2023,
  author  = {Stenger,R.  and Herzog,S.  and Kottlarz,I.  and Rüchardt,B.  and Luther,S.  and Wörgötter,F.  and Parlitz,U. },
  title   = {Reconstructing in-depth activity for chaotic 3D spatiotemporal excitable media models based on surface data},
  journal = {Chaos: An Interdisciplinary Journal of Nonlinear Science},
  volume  = {33},
  number  = {1},
  pages   = {013134},
  year    = {2023},
  doi     = {10.1063/5.0126824},
  url     = {https://doi.org/10.1063/5.0126824},
  eprint  = {https://doi.org/10.1063/5.0126824}
}

@Article{Syed2005,
author={Syed, Z. and Vigmond, E. and Nattel, S. and Leon, L. J.},
title={Atrial cell action potential parameter fitting using genetic algorithms},
journal={Medical and Biological Engineering and Computing},
year={2005},
month={Oct},
day={01},
volume={43},
number={5},
pages={561-571},
issn={1741-0444},
doi={10.1007/BF02351029},
url={https://doi.org/10.1007/BF02351029}
}

@InProceedings{Thomas2025,
author="Thomas, Benjamin J. and Goodbrake, Christian and Meyer, Kenneth and Sacks, Michael S.",
editor="Chabiniok, Radom{\'i}r
and Zou, Qing
and Hussain, Tarique
and Nguyen, Hoang H.
and Zaha, Vlad G.
and Gusseva, Maria",
title="High Speed Cardiac Simulations Using the JAX Framework",
booktitle="Functional Imaging and Modeling of the Heart",
year="2025",
publisher="Springer Nature Switzerland",
address="Cham",
pages="275--281",
isbn="978-3-031-94559-5"
}

@misc{Thuerey2021,
      title={Physics-based Deep Learning}, 
      author={N. Thuerey and B. Holzschuh and P. Holl and G. Kohl and M. Lino and Q. Liu and P. Schnell and F. Trost},
      year={2025},
      eprint={2109.05237},
      archivePrefix={arXiv},
      primaryClass={cs.LG},
      url={https://arxiv.org/abs/2109.05237}, 
}

@article{Walters2020,
  title    = {Site-Specific Epicardium-to-Endocardium Dissociation of Electrical Activation in a Swine Model of Atrial Fibrillation},
  journal  = {JACC: Clinical Electrophysiology},
  volume   = {6},
  number   = {7},
  pages    = {830-845},
  year     = {2020},
  issn     = {2405-500X},
  doi      = {https://doi.org/10.1016/j.jacep.2020.04.015},
  url      = {https://www.sciencedirect.com/science/article/pii/S2405500X20302802},
  author   = {Tomos E. Walters and Geoffrey Lee and Adam Lee and Richard Sievers and Jonathan M. Kalman and Edward P. Gerstenfeld}
}

@article{Winchenbach2026,
title = {diffSPH: Differentiable smoothed particle hydrodynamics for hybrid machine learning solutions in fluid mechanics},
journal = {Journal of Computational Physics},
volume = {555},
pages = {114769},
year = {2026},
issn = {0021-9991},
doi = {https://doi.org/10.1016/j.jcp.2026.114769},
url = {https://www.sciencedirect.com/science/article/pii/S0021999126001191},
author = {Rene Winchenbach and Nils Thuerey}
}

@article{Zhang2021,
  title    = {{SPHinXsys}: An open-source multi-physics and multi-resolution library based on smoothed particle hydrodynamics},
  journal  = {Computer Physics Communications},
  volume   = {267},
  pages    = {108066},
  year     = {2021},
  issn     = {0010-4655},
  doi      = {https://doi.org/10.1016/j.cpc.2021.108066},
  url      = {https://www.sciencedirect.com/science/article/pii/S0010465521001788},
  author   = {Chi Zhang and Massoud Rezavand and Yujie Zhu and Yongchuan Yu and Dong Wu and Wenbin Zhang and Jianhang Wang and Xiangyu Hu}
}

@article{Zhang2021b,
  author  = {Zhang, Chi and Wang, Jianhang and Rezavand, Massoud and Wu, Dong and Hu, Xiangyu},
  doi     = {10.1016/j.cma.2021.113847},
  issn    = {0045-7825},
  journal = {Computer Methods in Applied Mechanics and Engineering},
  month   = aug,
  pages   = {113847},
  title   = {An Integrative Smoothed Particle Hydrodynamics Method for Modeling Cardiac Function},
  volume  = {381},
  year    = {2021}
}

\clearpage

\newpage

\section*{Supplementary Information}
\label{sec:supplement}
\subsection*{Supplementary Videos}
Supplementary Videos will be provided at: \url{https://cardiacvision.ucsf.edu/videos/diffsim/}.

\subsection{Supplementary Tables}

\begin{table}[!htbp]
  \begin{tabular}{@{}c|cccccc@{}}\toprule
    Param.    									& $D$  		& $k$ 	& $a$ 	& $\epsilon_0$ 	& $\mu_1$ 	& $\mu_2$ 	\\ \midrule
     Fig.~\ref{fig:methods:initialcondition}                  		&  0.0011  	& 8.0  	& 0.1500  	& 0.02		& 0.15 		& 0.15  		\\ 
    Figs.~\ref{fig:results:2D:spiral-divergence-loss}, \ref{fig:results:2D:spiral-divergence-timeseries}            	&  0.0011   	& 9.0 	& 0.1200 	& 0.01 		& 0.16   		& 0.2 		\\
    Fig.~\ref{fig:PS}         								& 0.0010  	 & 6.0 	& 0.0500  	& 0.09  	& 0.10   	& 0.15 		\\
    Figs.~\ref{fig:results:SPH1}-\ref{fig:results:SPH4}        		& $D_{iso}$ 	 & 8.0	& 0.0500 	& 0.02 	& 0.20  	& 0.30 		\\
    Figs.~\ref{fig:results:scroll-top-bottom}, \ref{fig:results:scroll}  & 0.0011  	 & 8.0 	& 0.1035  	& 0.02	& 0.15	& 0.15		\\
    Fig.~\ref{fig:results:sparsity}         						&  0.0011   & 8.0  	&  0.1000	& 0.02	& 0.16 	& 0.12   		 \\ \bottomrule
  \end{tabular}
  \caption{
    Parameters of Aliev-Panfilov~\cite{AlievPanfilov1996} (AP) model used to simulate electrical action potential wave patterns in different sections of this study. 
    }
  \label{tab:parameters}
\end{table}

\begin{table}[!htbp]
  \begin{tabular}{@{}c|ccccc@{}}\toprule
    					& Horizon  	&  Horizons	& Duration	& Epochs 	& Epochs	 \\ 
					& Number		&  (n)		& 			& $n_e$ 		& Total 		\\ \midrule
    Stage 1 			& 1-10   	 	& 10    		& 10  		& 600  		& 6,000  	 	\\ \midrule
    Stage 2				& 11-20   		& 10			& 20 			& 800 		& 14,000 		 \\ \midrule
    Stage 3        			& 21-30   		& 10			& 40 			& 1,000		& 24,000 		\\ \bottomrule
  \end{tabular}
  \caption{
    Parameters of learning schedule used in Figs.~\ref{fig:methods:initialcondition}, \ref{fig:results:2D:spiral-divergence-loss}-\ref{fig:PS} and \ref{fig:results:sparsity}. The schedule comprises 3 subsequent stages with 10 horizons each and 600, 800, and 1,000 epochs per horizon. The horizon duration increases from 10 to 20 to 40 time steps in each stage, and each horizon is shifted by 1 time step ($\Delta t = 1$).
After 10, 20, and 30 horizons, the learning was performed over 20, 40, and 70 time steps, respectively.
At the end of the full schedule, the learning was performed over 24,000 epochs.
The schedule can, in principle, vary and include more or fewer stages, horizons, horizon durations, or epochs, see also table \ref{tab:learning-schedule3Dscroll}.
In Figs.~\ref{fig:results:2D:spiral-divergence-loss}, \ref{fig:results:2D:spiral-divergence-timeseries} and \ref{fig:PS} the learning was stopped prematurely (e.g. after Stage 2 or 14,000 epochs).
    }
  \label{tab:learning-schedule}
\end{table}

\begin{table}[!htbp]
  \begin{tabular}{@{}c|ccccc@{}}\toprule
    					& Horizon  	&  Hor.	& Duration	& Epochs 	& Epochs	 \\ 
					& Number		&  (n)		& 			& $n_e$ 		 & Total 		\\ \midrule
    W1	 			& 1	   	 	& 1    		& 20  		& 2,000  		& 2,000  	 	\\ \midrule
    W2				& 2   			& 1			& 25 			& 2,000 		& 4,000 		 \\ \midrule
    W3				& 3	   		& 1			& 30 			& 2,000 		& 6,000 		 \\ \midrule
    W4				& 4   			& 1			& 35 			& 2,000 		& 8,000 		 \\ \midrule
    Rolling        			& 5-55   		& 51			& 40 			& 2,000		& 110,000 		\\ \bottomrule
  \end{tabular}
  \caption{
    Parameters of learning schedule used in Figs.~\ref{fig:results:scroll-top-bottom} and \ref{fig:results:scroll}. 
    The schedule comprises 4 subsequent "warm-up" horizons (W1-W4), which refine the initial condition, followed by a rolling multi-horizon schedule comprising 51 horizons, which learn the dynamics, see also section \ref{sec:results:3Dbulk}.
The horizon duration increases during warm-up and then stays constant during the rolling horizons.
Each horizon is learned over 2,000 epochs. 
The 51 multi-horizons are learned over 102,000 epochs.
At the end of the full schedule, the learning was performed over 110,000 epochs.
    }
  \label{tab:learning-schedule3Dscroll}
\end{table}

\begin{table}[!htbp]
  \begin{tabular}{@{}c|cccccc@{}}\toprule
    Fig.										& $D$  	& $a$ 	& $k$ 	& $\epsilon_0$ & $\mu_1$ & $\mu_2$ \\ \midrule
    \ref{fig:methods:initialcondition}   					& 0.0011   & 0.1500  & 8.0000  	& 0.0200  & 0.1500  	& 0.1500   \\
                                                                 				& 0.0011   & 0.1493  & 8.0118  	& 0.0195  & 0.1524  	& 0.1572    \\
											& 0.05\%   & 0.45\% & 0.15\%  	& 2.70\%  & 1.61\%  	& 4.77\%   \\ \midrule
    \ref{fig:results:2D:spiral-divergence-loss},\ref{fig:results:2D:spiral-divergence-timeseries}         	& 0.0011   & 0.1200	& 9.0000 	& 0.0100 	& 0.1600 	& 0.2000	 \\
            										& 0.0011   & 0.1200	& 9.0000 	& 0.0100 	& 0.1600 	& 0.2000 	\\
										        & 0.00\%   & 0.00\%	& 0.00\% 	& 0.02\% 	& 0.00\% 	& 0.00\%	 \\ \midrule
   \ref{fig:results:sparsity}         						& 0.0011   & 0.1000	& 8.0000 	& 0.0200 	& 0.1600 	& 0.1200 	\\
            										& 0.0011   & 0.0998	& 7.9564 	& 0.0185 	& 0.1637 	& 0.1333 	\\
										        & 1.22\%   & 0.15\%	& 0.54\% 	& 7.64\% 	& 2.32\% 	& 11.10\% \\ \bottomrule
  \end{tabular}
  \caption{
    Ground-truth parameters (top), learned parameters (center), and error (bottom) for the different 2D wave dynamics shown in Figs.~\ref{fig:methods:initialcondition}, \ref{fig:results:2D:spiral-divergence-loss} and \ref{fig:results:sparsity}.
    All results were obtained with 20,000 epochs and the same multi-horizon learning schedule. 
    The errors are negligible with spiral wave dynamics (Fig.~\ref{fig:results:2D:spiral-divergence-loss}), but can be up to several percent with sparse or limited observations. 
    The errors in Fig.~\ref{fig:results:sparsity} would decrease further with additional learning.
    }
  \label{tab:parameters-2D-learned}
\end{table}

\begin{table}[!htbp]
  \begin{tabular}{@{}c|cccccc@{}}\toprule
    Fig.                               & $D$      & $a$      & $k$      & $\epsilon_0$ & $\mu_1$  & $\mu_2$  \\ \midrule
    \ref{fig:results:scroll}      & 0.0011   & 0.1035   & 8.0000   & 0.0200       & 0.1500   & 0.1500   \\
                                        & 0.0011   & 0.1036   & 8.0003   & 0.0200       & 0.1498   & 0.1496   \\
                                        & 0.04\%   & 0.06\%   & 0.00\%   & 0.20\%       & 0.15\%   & 0.27\%   \\ \bottomrule
  \end{tabular}
  \caption{
    Ground-truth parameters (top), learned parameters (center), and error (bottom) for 3D scroll wave dynamics in a bulk with surface-only observations, as shown in Fig.~\ref{fig:results:scroll}.
    }
  \label{tab:parameters-3d-bulk-learned}
\end{table}

\clearpage

\setcounter{figure}{0}
\renewcommand{\thefigure}{S\arabic{figure}}

\subsection{Supplementary Figures}

\begin{figure}[!htbp]
  \centering
  \includegraphics[clip, trim=0.0cm 0.0cm 0.0cm 0.0cm, width=0.95\columnwidth]{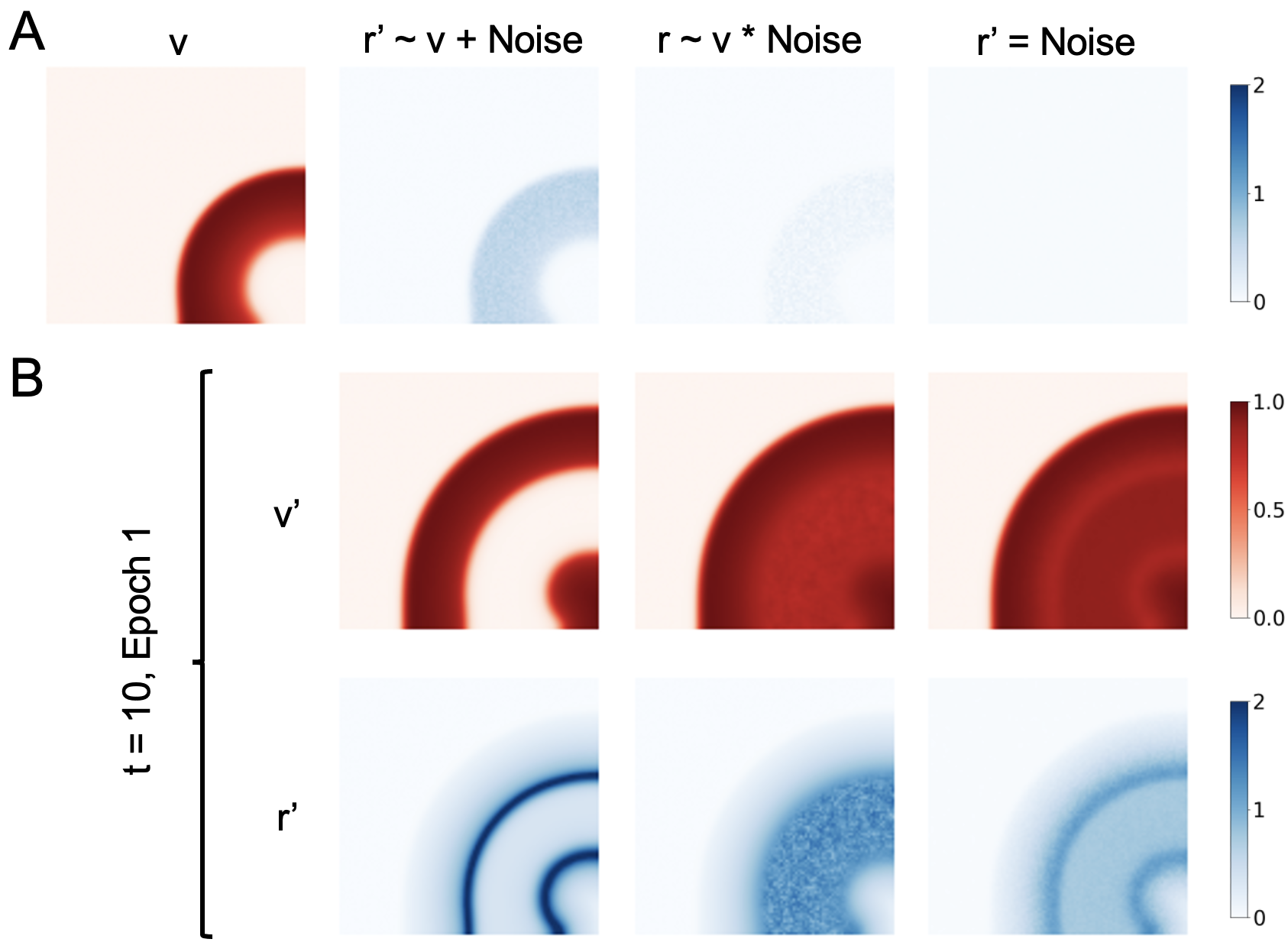}
  \caption{
\textbf{A} Focal voltage wave pattern $v$ (red) with initial guesses for the learned refractory pattern $r'$ (blue). Left to right: $r' = v + \sigma$, $r'=v \cdot \sigma$ and $r'=\sigma$ with Gaussian noise $\sigma$. The first option is the best-performing initial condition, see also Fig.~\ref{fig:supplement:learning-setup}A). Constant fields $r'=a, a \in \mathbb{R}$ or $r'=0$ lead to numerical instability.
\textbf{B} Initial conditions evolved at the end of Epoch 1 in Horizon 1 with 10 time steps.
 }
  \label{fig:supplement:2D-initialcondition}
\end{figure}

\begin{figure}[!htbp]
  \centering
  \includegraphics[clip, trim=0.0cm 0.0cm 0.0cm 0.0cm, width=0.92\columnwidth]{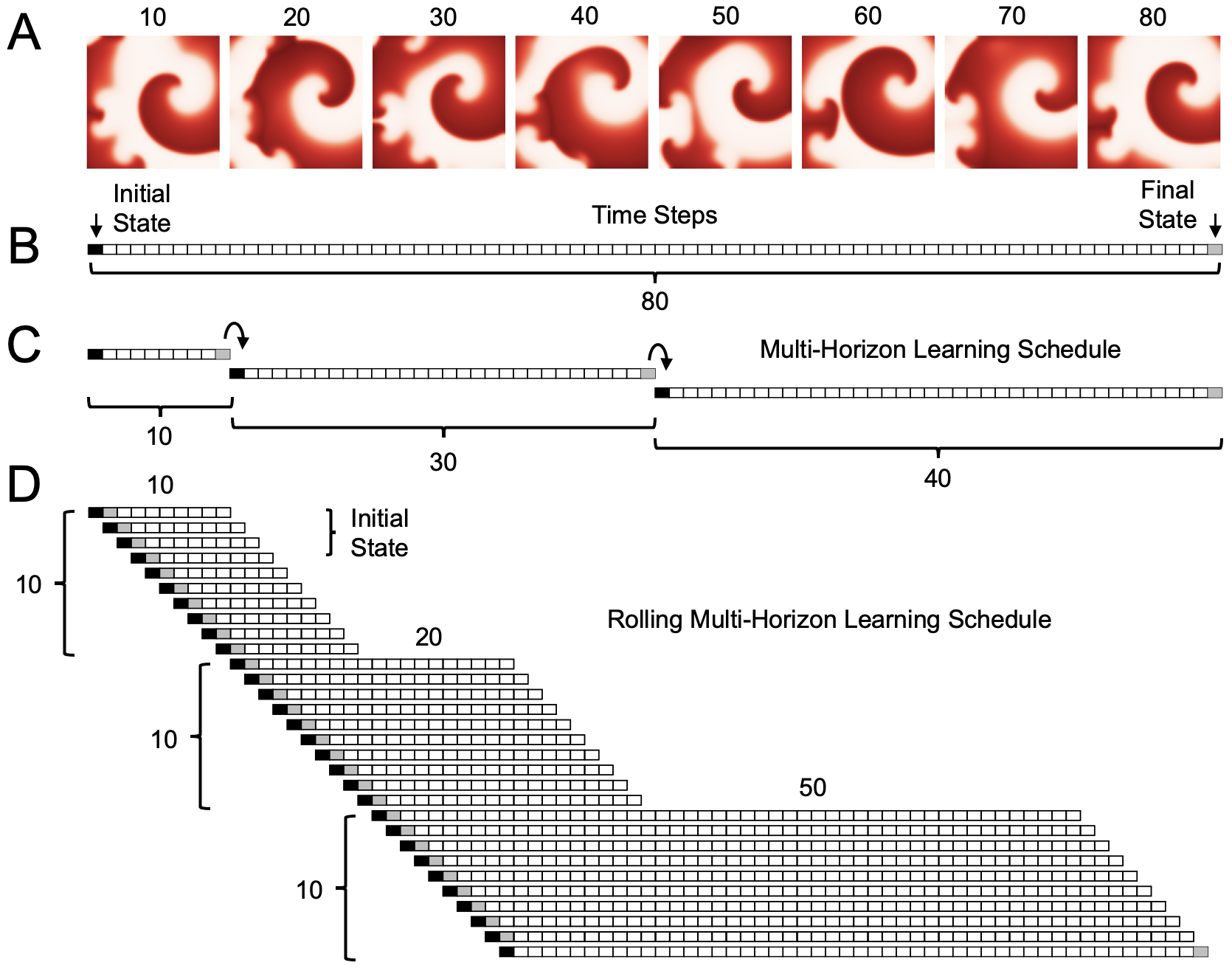}
  \caption{
Learning schedule for gradient-descent-based state and parameter estimation of spatio-temporal electrical wave dynamics in excitable media.
\textbf{A} Example of simulated spiral wave dynamics observed over about 3 rotations, the observations comprising 80 time steps.
\textbf{B} Learning over a single horizon that includes all 80 time steps, starting with an initial guess of the initial condition / first dynamical state (black), with the task to obtain an estimate of all model parameters and the final dynamical state at the final time step (gray).
\textbf{C} Multi-horizon learning schedule consisting of 3 subsequent horizons with different lengths (example).
In each horizon, the last learned state (gray) is used to construct the first state of the next horizon (black).
\textbf{D} Multi-horizon learning schedule with rolling, overlapping horizons of different lengths, altogether covering the 80 time steps (example). Here, the second learned time step is used to initiate the next horizon.}
  \label{fig:supplement:learningschedule}
\end{figure}

\begin{figure}[!htbp]
  \centering
  \includegraphics[clip, trim=0.0cm 0.0cm 0.0cm 0.0cm, width=0.92\columnwidth]{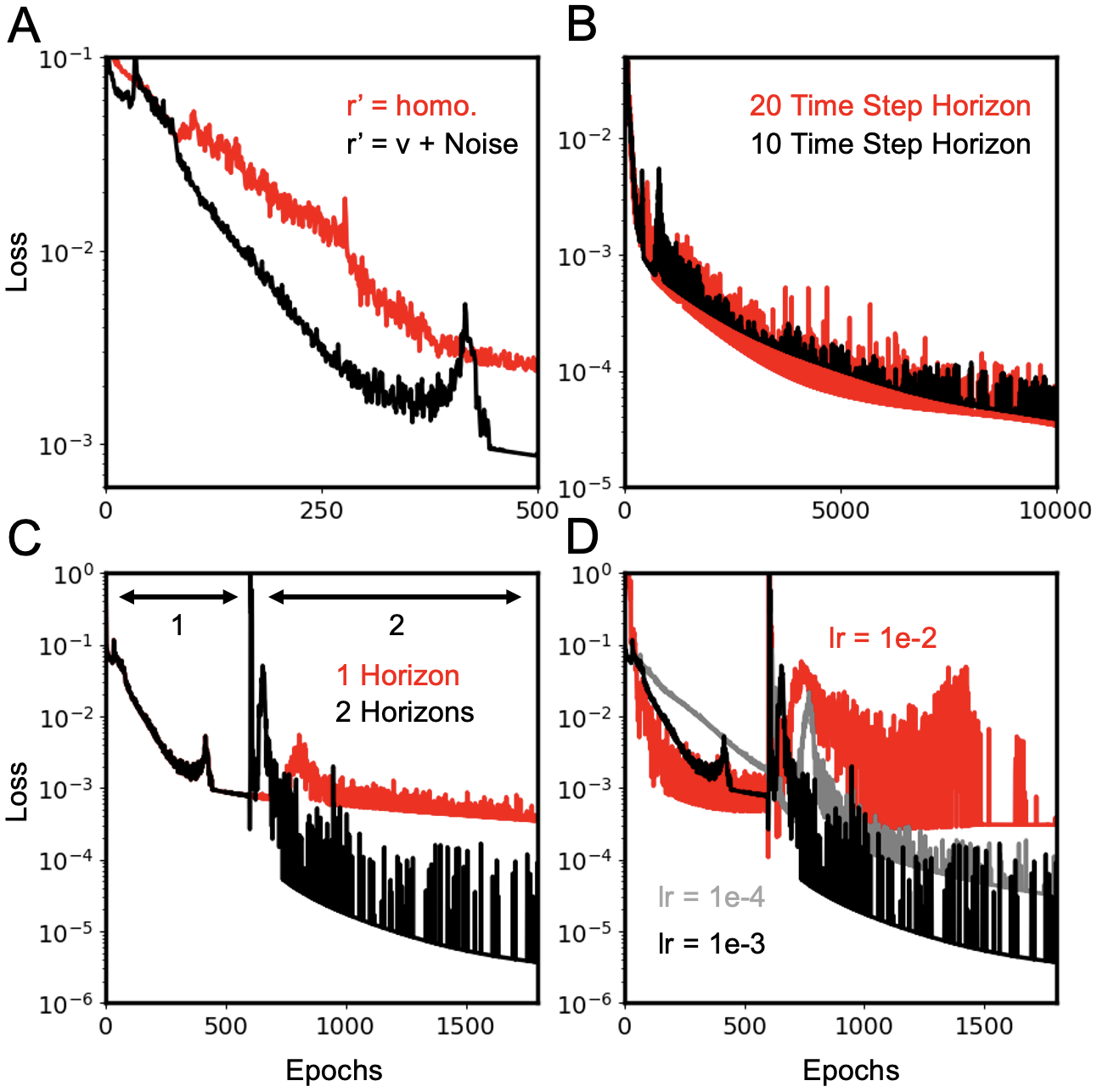}
  \caption{
Effect of learning schedule onto loss curve with data shown in Fig.~\ref{fig:methods:initialcondition}.
\textbf{A} Steeper loss curve with $r'\sim v$ or focal-shaped (black) vs. homogeneous noisy (red) initialization of $r'$, see also Fig.~\ref{fig:supplement:2D-initialcondition}A).
\textbf{B} Single-horizon learning over horizons with 20 (red) or 10 (black) time steps. Loss never reaches levels $< 10^{-5}$.
\textbf{C} Single-horizon (red) vs. 2-horizon (black) learning schedule with 10 time step horizons. Loss reaches levels  $< 10^{-5}$ after fewer than 1500 epochs in horizon 2.
\textbf{D} 2-horizon learning schedules with constant learning rates of $0.01$ (red), $0.0001$ (gray), and $0.001$ (black, ideal).
 }
  \label{fig:supplement:learning-setup}
\end{figure}

\begin{figure}[!htbp]
  \centering
  \includegraphics[clip, trim=0.0cm 0.0cm 0.0cm 0.0cm, width=0.9\columnwidth]{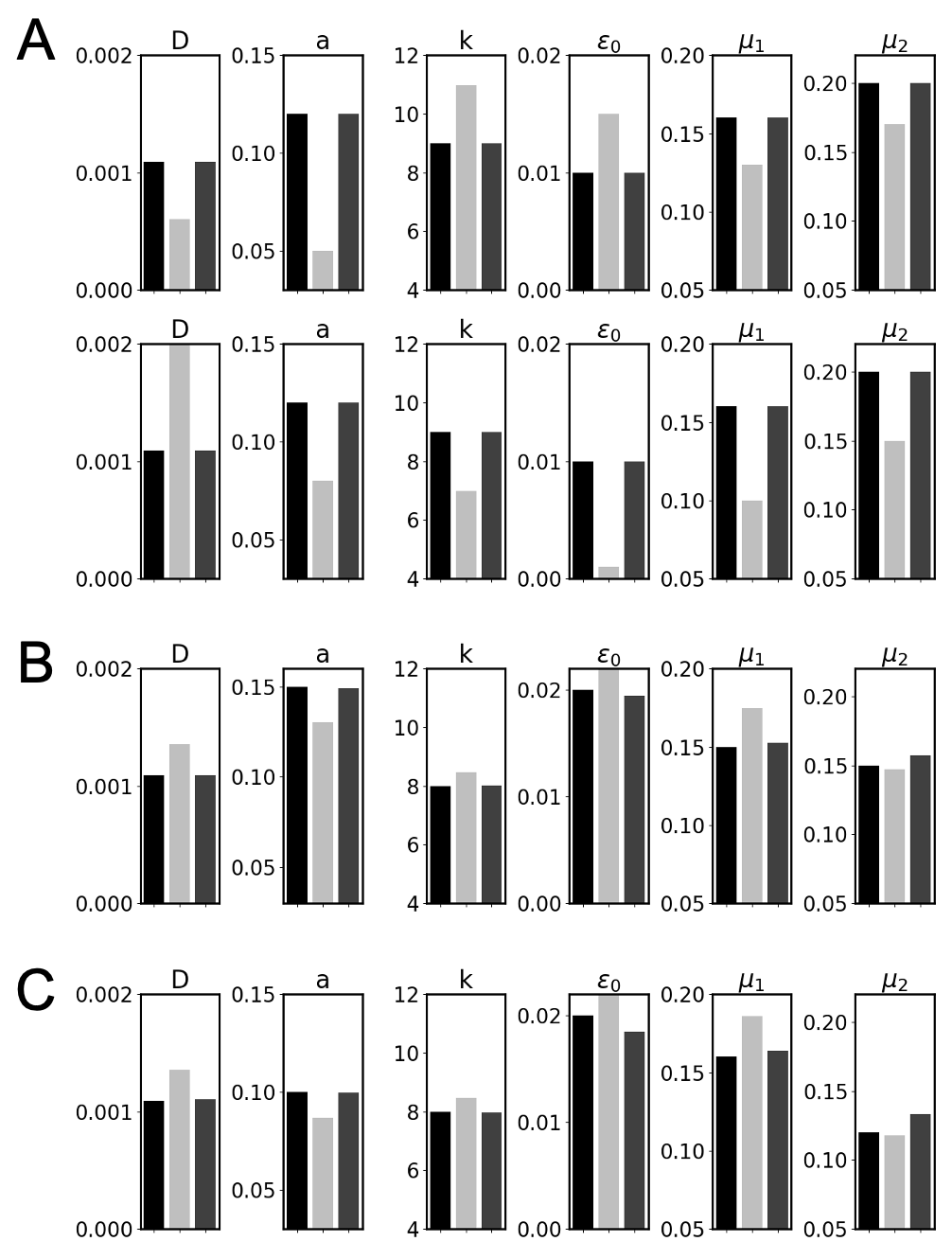}
  \caption{
Parameters (black: ground-truth, light gray: initial guess, dark gray: learned) for different wave dynamics.
\textbf{A} Spiral wave dynamics shown in Figs.~\ref{fig:results:2D:spiral-divergence-loss} and \ref{fig:results:2D:spiral-divergence-timeseries} fully learned (2 separate runs) after 70 observation time steps ($\sim 4.5$ rotations) within 30 horizons learned over 24,000 epochs.
\textbf{B} Focal wave shown in Fig.~\ref{fig:methods:initialcondition}. While parameters $\{ D, a, k \}$ have converged, $\{ \epsilon_0, \mu_1, \mu_2 \}$ have not fully converged after 80 observation time steps or 40 horizons or 34,000 epochs. 
\textbf{C} Sparse multi-spiral waves shown in Fig.~\ref{fig:results:sparsity}. As in C), the parameters $\{ \epsilon_0, \mu_1, \mu_2 \}$ have not fully converged after 40 horizons, indicating that longer observation times are necessary. In both cases, the parameter history curves indicate that the learned parameters would eventually converge. See Table \ref{tab:parameters-2D-learned} for values and errors.
 }
  \label{fig:barplots}
\end{figure}

\clearpage

\end{document}